\documentclass[aps,preprintnumbers,amsmath,amssymb,superscriptaddress,floatfix,nofootinbib]{revtex4}
\usepackage{lipsum}
\usepackage{cancel}
\usepackage{slashed}
\usepackage{graphicx}
\usepackage{epsfig}
\usepackage{comment}
\usepackage{epstopdf}
\usepackage{amsmath}
\usepackage{amsfonts}
\usepackage{braket}
\usepackage{amssymb}
\usepackage{setspace}
\usepackage{amsfonts,color}
\usepackage{ragged2e}
\usepackage[labelfont=bf]{caption}
\usepackage{natbib}
\usepackage{subfig}
\usepackage{color}
\usepackage{bbold}
\usepackage{bm}
\usepackage{mathrsfs}
\usepackage{url}
\usepackage{rotating}
\usepackage{pdflscape}
\usepackage{diagbox}
\usepackage{ulem}
\usepackage[dvipsnames]{xcolor}
\allowdisplaybreaks[1]
\usepackage{multirow}
\usepackage{hhline}
\usepackage{makecell}
\usepackage{float}
\usepackage{hyperref}

\begin{document}

\title{The $1/N_c$ Operator Analysis of the Combined Octet and Decuplet Baryons Contact Interactions in SU(3) Chiral Effective Field Theory}

\author{Chindanai Bubpatate}
\email{chinnai@kkumail.com}
\affiliation{Khon Kaen Particle Physics and Cosmology Theory Group (KKPaCT), Department of Physics, Faculty of Science, Khon Kaen University,123 Mitraphap Rd., Khon Kaen, 40002, Thailand}

\author{Dominador F. Vaso, Jr.}
\email{dominadorjr.vaso@g.msuiit.edu.ph}
\affiliation{Department of Physics, Mindanao State University - Iligan Institute of Technology, Iligan City, 9200, Philippines}
\affiliation{Khon Kaen Particle Physics and Cosmology Theory Group (KKPaCT), Department of Physics, Faculty of Science, Khon Kaen University,123 Mitraphap Rd., Khon Kaen, 40002, Thailand} 
\author{Daris Samart}
\email{darisa@kku.ac.th, corresponding author}
\affiliation{Khon Kaen Particle Physics and Cosmology Theory Group (KKPaCT), Department of Physics, Faculty of Science, Khon Kaen University,123 Mitraphap Rd., Khon Kaen, 40002, Thailand}

\date{\today}

\begin{abstract}
In this work, we construct the non-derivative four-point interactions for Octet and Decuplet baryons in the SU(3) Chiral Effective Field Theory (ChEFT) framework, and there are 104 coupling constant terms. The non-relativistic expansion of the baryon fields has been considered up to the {Next-to-Leading Order} (NLO) of the three-momentum expansion. We find 28 and 106 Low-Energy Constants (LECs) for {Leading Order} (LO) and NLO, respectively. Using the Hartree Hamiltonian of the $1/N_c$ expansion of the operator product up to {Next-to-Next-to-Leading Order} (NNLO), we can reduce the free parameters (LECs) of the ChEFT from 134 down to 24 up to NLO of the three-momentum expansion. Moreover, we will discuss the implications of the $1/N_c$ sum rules in $\Omega\Omega$ and $\Omega N$ scatterings, where the future results from lattice QCD can be used to test our sum rules. 
\end{abstract}

\maketitle
\section{Introduction}\label{sec-1}

Understanding baryon-baryon interactions at low energies is crucial to exploring phenomena in nuclear physics, hypernuclei, and astrophysics, particularly the internal dynamics of neutron stars. At these energy scales, Quantum Chromodynamics (QCD), the fundamental theory of strong interactions, becomes highly non-perturbative, rendering the conventional methods of standard quantum field theory inadequate. Chiral effective field theory (ChEFT), based on effective field theory principles and spontaneous chiral symmetry breaking of QCD, provides a systematic framework for describing hadronic interactions at energies inapplicable to perturbative methods. ChEFT exploits the symmetry patterns of QCD to construct low-energy effective Lagrangians, using baryons and mesons as effective degrees of freedom, thus facilitating reliable calculations and systematic improvements via expansions in small momenta and pion masses \cite{Epelbaum_2009,Machleidt_2011,Haidenbauer_2013,Weinberg:1978kz,Meissner_1993}.

ChEFT has achieved significant successes in describing nucleon-nucleon ($NN$) interactions with remarkable accuracy, systematically incorporating loop corrections and higher-order terms through a robust power-counting scheme \cite{Epelbaum_2009,Machleidt_2011,Haidenbauer_2013,Polinder_2006,Kaiser_1997,Savage_1997,Kaiser_1998,Epelbaoum_1998,Epelbaum_2000,Entem_2003,Entem_2015,Ren_2018,Krebs:2007rh,Epelbaum_2008,Piarulli_2015}. Recent extensions of these methods to hyperon-nucleon ($YN$) and hyperon-hyperon ($YY$) interactions have become essential for understanding hypernuclear physics and astrophysical phenomena, particularly the equation of state for dense neutron-star matter \cite{Gerstung_2020,Nogga:1999bq,Haidenbauer:2019boi,Haidenbauer_2023,Kn_ll_2023,Li:2017zwn,Petschauer:2020urh,Beane:2012ey,Li:2016paq}. Despite these advances, substantial challenges persist due to the proliferation of unknown Low-Energy Constants (LECs) that complicate precise predictions \cite{Haidenbauer_2023}. Recently, explicit formulations of decuplet baryon interactions within ChEFT have been considered, with the aim of enhancing theoretical predictions and facilitating comparisons with lattice QCD and experimental searches for exotic dibaryon states \cite{haiden2017scattdecupletbary}.

Initially proposed by 't Hooft and significantly developed by Witten \cite{tHooft:1973alw,EdWITTEN1979,Witten:1980sp}, the large-$N_c$ limit in QCD ($N_c$ is the number of colors) provides a useful method to reduce the complexity of the low-energy QCD phenomena by imposing constraints on the parameters within effective field theories derived from QCD. In such a limit, the spin-1/2 and spin-3/2 baryons become degeneracy states. The large-$N_c$ framework advantages the simplification of QCD dynamics in the limit of large color numbers, allowing expansions in the inverse powers of $N_c$. This method significantly reduces the number of independent parameters, that is, LECs in effective theories, identifies universal features of hadronic interactions, and provides the potential theoretical predictions \cite{Jenkins:1990jv,Dashen_1995,Liu_2019,richardson2024}. Recent large-$N_c$ analysis have successfully simplified nucleon-nucleon and hyperon-nucleon interaction models, clarifying the interaction of spin and flavor structures and demonstrating consistency with lattice QCD and experimental data \cite{Savage:2011xk,Beane_2011,Vonk_2025}. However, a unified large-$N_c$ approach that systematically combines baryon interactions with octets and decuplets within the SU(3) ChEFT framework remains unexplored. 

Motivated by these considerations, this study aims to generalize existing analyses by explicitly incorporating decuplet baryon degrees of freedom alongside octet baryons within ChEFT and large-$N_c$ frameworks. Particularly, we investigate two-body baryon interactions involving all possible configurations of the octet and decuplet baryons, construct potentials using Next-to-Next-to-Leading-Order (NNLO) Hamiltonians derived from the SU(3) chiral Lagrangian via non-relativistic reduction. In the latter, we systematically organize and derive the $1/N_c$ operator product expansion of the baryon-baryon potentials within SU(3) flavor symmetry up to $1/N_c^2$ order. Finally, one can reduce the number of independent LECs through the large-$N_c$ sum rules. These sum rules arise from the matching spin and flavor structures between potentials derived from the $1/N_c$ Hartree Hamiltonian and the SU(3) chiral Lagrangian. Consequently, this approach considerably decreases the number of independent coupling constants, significantly improving the predictive capacity and efficiency of ChEFT. 
 
This work provides detailed investigations of non-relativistic expansions from the relativistic chiral Lagrangians, extensive $1/N_c$ sum rules, and demonstrates significant parameter reductions achieved by including the decuplet baryon degrees of freedom. Furthermore, our results could be used to compare them with existing lattice QCD and experimental scattering data, highlighting the practical relevance and utility of our refined theoretical framework.

This article is organized as follows. The minimal derivative four-point interactions of the octet and decuplet baryons in ChEFT are constructed and the baryon-baryon potentials with the non-relativistic expansion are worked out in the section \ref{sec-2}. In section \ref{sec-3}, we set up the $1/N_c$ operator production expansion for constructing the $1/N_c$ Hartree Hamiltonian. Then, the $1/N_c$ sum rules are obtained by matching the spin-flavor structures between the baryon-baryon potentials from the SU(3) chiral Lagrangians and the $1/N_c$ Hartree Hamiltonian. Section \ref{sec-4}, we discuss the physical consequences of the $1/N_c$ sum rule's implication on $ \Omega N $ and $ \Omega\Omega $ scattering, respectively. Finally, we close this article with a discussion and conclusion in section \ref{sec-5}.

\section{The baryon-baryon potential from the chiral effective Lagrangians}\label{sec-2}

In this section, we construct the minimal derivative chiral Lagrangian for four-point interactions involving octet and decuplet baryons within SU(3) ChEFT, by systematically accounting for all relevant invariant flavors and Lorentz structures. The resulting minimal set of SU(3) chiral Lagrangians in the non-derivative limit is expressed in terms of particle fields, following the formalism in Ref. \cite{haiden2017scattdecupletbary}. We find
\begin{alignat}{2}
    \notag\mathcal{L}_{BBBB}=&\,\,\,\sum_{i=1}^5g_{00,1}^{i1}\sum_{a,b,c,d=1}^3(\Bar{B}_{ab}\Gamma^{i}B_{bc})(\Bar{B}_{cd}\Gamma^{i}B_{da})\\[1em]
    &+\sum_{i=1}^5g_{00,1}^{i2}\sum_{a,b,c,d=1}^3(\Bar{B}_{ab}\Gamma^{i}B_{cd})(\Bar{B}_{bc}\Gamma^{i}B_{da})\notag\\[1em]
        &+\sum_{i=1}^5g_{00,1}^{i3}\sum_{a,b,c,d=1}^3(\Bar{B}_{ab}\Gamma^{i}B_{ba})(\Bar{B}_{cd}\Gamma^{i}B_{dc})\label{LBBBB}\\[.5cm]
        \mathcal{L}_{DBBB}=&\,\,\,g_{01,1}^{41}\sum_{a,b,c,d,e,f=1}^3\epsilon_{abc}(\Bar{\Delta}_{ade}^{\mu}B_{db})(\Bar{B}_{fc}\gamma_\mu\gamma^5 B_{ef})+h.c.\notag\\[1em]
        &+g_{01,1}^{42}\sum_{a,b,c,d,e,f=1}^3\epsilon_{abc}(\Bar{\Delta}_{ade}^{\mu}B_{fb})(\Bar{B}_{dc}\gamma_\mu\gamma^5 B_{ef})+h.c.\\[.5cm]
        \mathcal{L}_{DBDB}=&\,\,\,\sum_{i=1}^5g_{11,1}^{i1}\sum_{a,b,c,d,e=1}^3(\Bar{\Delta}_{abc}^\mu\Gamma^{i}\Delta_{\mu,abc})(\Bar{B}_{de}\Gamma^{i}B_{ed})\notag\\[1em]
        &+\sum_{i=1}^5g_{11,1}^{i2}\sum_{a,b,c,d,e=1}^3(\Bar{\Delta}_{abc}^\mu\Gamma^{i}\Delta_{\mu,abd})(\Bar{B}_{ce}\Gamma^{i}B_{ed})\notag\\[1em]
        &+\sum_{i=1}^5g_{11,1}^{i3}\sum_{a,b,c,d,e=1}^3(\Bar{\Delta}_{abc}^\mu\Gamma^{i}\Delta_{\mu,abd})(\Bar{B}_{ed}\Gamma^{i}B_{ce})\notag\\[1em]
        &+\sum_{i=1}^5g_{11,1}^{i4}\sum_{a,b,c,d,e=1}^3(\Bar{\Delta}_{abc}^\mu\Gamma^{i}\Delta_{\mu,abe})(\Bar{B}_{bd}\Gamma^{i}B_{ce})\\[.5cm]
        \mathcal{L}_{DDBB}=&\,\,\,\sum_{i=1}^5g_{02,1}^{i1}\sum_{a,b,c,d,e,f,g,h=1}^3\epsilon_{abc}\epsilon_{def}(\Bar{\Delta}_{adg}^\mu\Gamma^{i}B_{gc})(\Bar{\Delta}_{\mu,beh}\Gamma^{i}B_{hf})+h.c.\notag\\[1em]
        &+\sum_{i=3}^5g_{02,2}^{i1}\sum_{a,b,c,d,e,f,g,h=1}^3\epsilon_{abc}\epsilon_{def}(\Bar{\Delta}_{adg}^\mu\Gamma^{i,\nu}B_{gc})(\Bar{\Delta}_{\nu,beh}\Gamma^{i}_{\mu}B_{hf})+h.c\\[.5cm]
        \mathcal{L}_{DDDB}=&\,\,\,g_{12,1}^{41}\sum_{a,b,c,d,e,f,g=1}^{3}\epsilon_{abc}(\Bar{\Delta}^{\mu}_{ade}\gamma^{\nu}\gamma^5\Delta_{\mu,def})(\Bar{\Delta}_{\nu,beh}B_{hf})+h.c.\notag\\[1em]
        &+g_{12,2}^{41}\sum_{a,b,c,d,e,f,g=1}^{3}\epsilon_{abc}(\Bar{\Delta}^{\mu}_{ade}\Delta^\nu_{def})(\Bar{\Delta}_{\mu,beh}\gamma_{\nu}\gamma^5B_{hf})+h.c.\notag\\[1em]
        &+g_{12,3}^{41}\sum_{a,b,c,d,e,f,g=1}^{3}\epsilon_{abc}(\Bar{\Delta}^{\mu}_{ade}\Delta^\nu_{def})(\Bar{\Delta}_{\nu,beh}\gamma_{\mu}\gamma^5B_{hf})+h.c.\\[.5cm]
        \mathcal{L}_{DDDD}=&\,\,\,\sum_{i=1}^{5}g_{22,1}^{i1}\sum_{a,b,c,d,e,f=1}^{3}(\Bar{\Delta}^{\mu}_{abc}\Gamma^{i}\Delta_{\mu,abc})(\Bar{\Delta}_{def}^\nu\Gamma^{i}\Delta_{\nu,def})\notag\\[1em]
        &+\sum_{i=1}^{5}g_{22,2}^{i1}\sum_{a,b,c,d,e,f=1}^{3}(\Bar{\Delta}^{\mu}_{abc}\Gamma^{i}\Delta^{\nu}_{abc})(\Bar{\Delta}_{\mu,def}\Gamma^{i}\Delta_{\nu,def})\notag\\[1em]
        &+\sum_{i=1}^{5}g_{22,3}^{i1}\sum_{a,b,c,d,e,f=1}^{3}(\Bar{\Delta}^{\mu}_{abc}\Gamma^{i}\Delta^{\nu}_{abc})(\Bar{\Delta}_{\nu,def}\Gamma^{i}\Delta
        _{\mu,def})\notag\\[1em]
        &+\sum_{i=3}^{5}g_{22,4}^{i1}\sum_{a,b,c,d,e,f=1}^{3}(\Bar{\Delta}^{\mu}_{abc}\Gamma^{i,\nu}\Delta^{\rho}_{abc})(\Bar{\Delta}_{\nu,def}\Gamma^{i}_{\mu}\Delta_{\rho,def})\notag\\[1em]
        &+\sum_{i=3}^{5}g_{22,5}^{i1}\sum_{a,b,c,d,e,f=1}^{3}(\Bar{\Delta}^{\mu}_{abc}\Gamma^{i,\nu}\Delta^{\rho}_{abc})(\Bar{\Delta}_{\rho,def}\Gamma^{i}_{\mu}\Delta_{\nu,def})\notag\\[1em]
        &+\sum_{i=3}^{5}g_{22,6}^{i1}\sum_{a,b,c,d,e,f=1}^{3}(\Bar{\Delta}^{\mu}_{abc}\Gamma^{i,\nu}\Delta^{\rho}_{abc})(\Bar{\Delta}_{\mu,def}\Gamma^{i}_{\rho}\Delta_{\nu,def})\notag\\[1em]
        &+\sum_{i=3}^{5}g_{22,7}^{i1}\sum_{a,b,c,d,e,f=1}^{3}(\Bar{\Delta}^{\mu}_{abc}\Gamma^{i,\nu}\Delta^{\rho}_{abc})(\Bar{\Delta}_{\nu,def}\Gamma^{i}_{\rho}\Delta_{\mu,def})\notag\\[1em]
        &+\sum_{i=1}^{5}g_{22,1}^{i2}\sum_{a,b,c,d,e,f=1}^{3}(\Bar{\Delta}^{\mu}_{abc}\Gamma^{i}\Delta_{\mu,abd})(\Bar{\Delta}^{\nu}_{def}\Gamma^{i}\Delta_{\nu,cef})\notag\\[1em]
        &+\sum_{i=1}^{5}g_{22,2}^{i2}\sum_{a,b,c,d,e,f=1}^{3}(\Bar{\Delta}^{\mu}_{abc}\Gamma^{i}\Delta^{\nu}_{,abd})(\Bar{\Delta}_{\mu,def}\Gamma^{i}\Delta_{\nu,cef})\notag\\[1em]
        &+\sum_{i=1}^{5}g_{22,3}^{i2}\sum_{a,b,c,d,e,f=1}^{3}(\Bar{\Delta}^{\mu}_{abc}\Gamma^{i}\Delta^{\nu}_{,abd})(\Bar{\Delta}_{\nu,def}\Gamma^{i}\Delta_{\mu,cef})\notag\\[1em]
        &+\sum_{i=3}^{5}g_{22,4}^{i2}\sum_{a,b,c,d,e,f=1}^{3}(\Bar{\Delta}^{\mu}_{abc}\Gamma^{i,\nu}\Delta^{\rho}_{abd})(\Bar{\Delta}_{\nu,def}\Gamma^{i}_{\mu}\Delta_{\rho,cef})\notag\\[1em]
        &+\sum_{i=3}^{5}g_{22,5}^{i2}\sum_{a,b,c,d,e,f=1}^{3}(\Bar{\Delta}^{\mu}_{abc}\Gamma^{i,\nu}\Delta^{\rho}_{abd})(\Bar{\Delta}_{\rho,def}\Gamma^{i}_{\mu}\Delta_{\nu,cef})\notag\\[1em]
        &+\sum_{i=3}^{5}g_{22,6}^{i2}\sum_{a,b,c,d,e,f=1}^{3}(\Bar{\Delta}^{\mu}_{abc}\Gamma^{i,\nu}\Delta^{\rho}_{abd})(\Bar{\Delta}_{\mu,def}\Gamma^{i}_{\rho}\Delta_{\nu,cef})\notag\\[1em]
        &+\sum_{i=3}^{5}g_{22,7}^{i2}\sum_{a,b,c,d,e,f=1}^{3}(\Bar{\Delta}^{\mu}_{abc}\Gamma^{i,\nu}\Delta^{\rho}_{abd})(\Bar{\Delta}_{\nu,def}\Gamma^{i}_{\rho}\Delta_{\mu,cef}).\label{LDDDD}
\end{alignat}\\
\textcolor{black}{The coupling constants, $g_{xy,z}^{i\alpha}$ of the chiral Lagrangians are described below. The subscript $x, y$ represents the number of incoming and outgoing decuplet baryons in the initial and final state for each Lagrangian, and $z$ is {contraction configuration notation} between two bilinear baryons. $i$ and $\alpha$ denote the kind of Lorentz (spin) and flavor structures. The non-derivative Lagrangian, which describes 6 various contact interactions in this work, yields 104 chiral coupling constant parameters. These 6 kinds of interaction and their corresponding number of independent coupling parameters are listed as follows: 15 ($BB \rightarrow BB$), 2 ($BB \rightarrow DB$), 20 ($DB \rightarrow DB$), 8 ($BB \rightarrow DD$), 3 ($DB \rightarrow DD$), and 54 ($DD \rightarrow DD$).} {In this representation}, the octet and decuplet baryons are represented in the fundamental representations as $B_{ab}=\frac{1}{\sqrt{2}}\sum_{m=1}^8\big(\lambda^{m}\big)_{ab}B^{m}$ and $\Delta^{\mu}_{abc}$, respectively. The Latin indices {($a,\,b,\,c,\,\cdots = 1,\,2,\,3\,\cdots$)} are the fundamental indices of the SU(3) flavor symmetry. The $\Gamma_i$ are Clifford algebra elements,
\begin{align}\label{cliffordalge}
    \Gamma^1=\mathbb{1},\,\,\,\,\,\Gamma^2=\gamma^5,\,\,\,\,\,\Gamma^3={\gamma^\mu},\,\,\,\,\,\Gamma^4=\gamma^\mu\gamma_5,\,\,\,\,\,\Gamma^5=\sigma^{\mu\nu}.
\end{align}

\indent {To derive the interaction potential from non-derivative chiral Lagrangians, we begin with the contributions at LO and NLO in the baryon sector.} The non-relativistic expansion of the baryon bilinear ($\Bar{u}\Gamma u$) in terms of the inverse baryon mass, corresponding to the $(Q/M)$ order in \cite{PETSCHAUER20131}, is systematically implemented (see Appendix \ref{A-1-NR-Expansion} for additional details). Subsequently,the flavor structures are constructed according to the alignment of their SU(3) flavor symmetry indices. As a result, we obtain the contact interaction potentials for octet-octet, octet-decuplet, and decuplet-decuplet baryon systems up to second order of the small momentum. Therefore, the general expressions for these potentials are given as follows:
\begin{alignat}{2}
    V_{BBBB}=&\,\,\,\biggl\{c_{S}^{(1)}+c_{T}^{(1)}(\vec{\sigma}_{1}\cdot\vec{\sigma}_2)+c_{5,BBBB}^{(1)}p_-^2+c_{6,BBBB}^{(1)}p_+^2+\bigl[c_{7,BBBB}^{(1)}p_-^2+c_{8,BBBB}^{(1)}p_+^2\bigr](\vec{\sigma}_{1}\cdot\vec{\sigma}_2)\notag\\[.5em]
        &\quad+c_{9,BBBB}^{(1)}(\vec{\sigma}_{1}+\vec{\sigma}_2)\cdot i(\vec{p}_+\times\vec{p}_-)+c_{10,BBBB}^{(1)}(\vec{\sigma}_1\cdot\vec{p}_-)(\vec{\sigma}_2\cdot\vec{p}_-)+c_{11,BBBB}^{(1)}(\vec{\sigma}_1\cdot\vec{p}_+)(\vec{\sigma}_2\cdot\vec{p}_+)\biggr\}\notag\\[.5em]
        &\quad\times\biggl\{\frac{1}{3}\delta^{(a'_1)(a_1)}\delta^{(a'_2)(a_2)}+\frac{1}{2}d^{(a'_1)(a_1)e}d^{e(a'_2)(a_2)}+\frac{1}{2}id^{(a'_1)(a_1)e}f^{e(a'_2)(a_2)}+\frac{1}{2}if^{(a'_1)(a_1)e}d^{e(a'_2)(a_2)}\notag\\[.5em]
        &\hspace{1cm}-\frac{1}{2}f^{(a'_1)(a_1)e}f^{e(a'_2)(a_2)}\biggr\}\notag\\[1em]
        &+\biggl\{c_{S}^{(2)}+c_{T}^{(2)}(\vec{\sigma}_{1}\cdot\vec{\sigma}_2)+c_{5,BBBB}^{(2)}p_-^2+c_{6,BBBB}^{(2)}p_+^2+\bigl[c_{7,BBBB}^{(2)}p_-^2+c_{8,BBBB}^{(2)}p_+^2\bigr](\vec{\sigma}_{1}\cdot\vec{\sigma}_2)\notag\\[.5em]
        &\quad+c_{9,BBBB}^{(2)}(\vec{\sigma}_{1}+\vec{\sigma}_2)\cdot i(\vec{p}_+\times\vec{p}_-)+c_{10,BBBB}^{(2)}(\vec{\sigma}_1\cdot\vec{p}_-)(\vec{\sigma}_2\cdot\vec{p}_-)+c_{11,BBBB}^{(2)}(\vec{\sigma}_1\cdot\vec{p}_+)(\vec{\sigma}_2\cdot\vec{p}_+)\biggr\}\notag\\[.5em]
        &\quad\times\biggl\{\frac{1}{3}\delta^{(a'_1)(a'_2)}\delta^{(a_1)(a_2)}+\frac{1}{2}d^{(a'_1)(a'_2)e}d^{e(a_1)(a_2)}+\frac{1}{2}id^{(a'_1)(a'_2)e}f^{e(a_1)(a_2)}+\frac{1}{2}if^{(a'_1)(a'_2)e}d^{e(a_1)(a_2)}\notag\\[.5em]
        &\hspace{1cm}-\frac{1}{2}f^{(a'_1)(a'_2)e}f^{e(a_1)(a_2)}\biggr\}\notag\\[1em]
        &+\biggl\{c_{S}^{(3)}+c_{T}^{(3)}(\vec{\sigma}_{1}\cdot\vec{\sigma}_2)+c_{5,BBBB}^{(3)}p_-^2+c_{6,BBBB}^{(3)}p_+^2+\bigl[c_{7,BBBB}^{(3)}p_-^2+c_{8,BBBB}^{(3)}p_+^2\bigr](\vec{\sigma}_{1}\cdot\vec{\sigma}_2)\notag\\[.5em]
        &\quad+c_{9,BBBB}^{(3)}(\vec{\sigma}_{1}+\vec{\sigma}_2)\cdot i(\vec{p}_+\times\vec{p}_-)+c_{10,BBBB}^{(3)}(\vec{\sigma}_1\cdot\vec{p}_-)(\vec{\sigma}_2\cdot\vec{p}_-)+c_{11,BBBB}^{(3)}(\vec{\sigma}_1\cdot\vec{p}_+)(\vec{\sigma}_2\cdot\vec{p}_+)\biggr\}\notag\\[.5em]
        &\quad\times\biggl\{\delta^{(a'_1)(a_1)}\delta^{(a'_2)(a_2)}\biggr\},\label{4B}\\[1cm]
        V_{BBDB}=&\,\,\,\biggl\{c_{2,BBDB}^{(1)}(\vec{S}_{1}^{\dagger}\cdot\vec{\sigma}_{2})+\bigl[c_{7,BBDB}^{(1)}p_{-}^{2}+c_{8,BBDB}^{(1)}p_{+}^{2}\bigr](\vec{S}_{1}^{\dagger}\cdot\vec{\sigma}_{2})+c_{9,BBDB}^{(1)}\vec{S}_1^{\dagger}\cdot i(\vec{p}_{+}\times\vec{p}_{-})\notag\\[.5em]
        &\quad+c_{10,BBDB}^{(1)}(\vec{S}_{1}^{\dagger}\cdot\vec{p}_{-})(\vec{\sigma}_{2}\cdot\vec{p}_{-})+c_{11,BBDB}^{(1)}(\vec{S}_{1}^{\dagger}\cdot\vec{p}_{+})(\vec{\sigma}_{2}\cdot\vec{p}_{+})\biggr\}\notag\\[.5em]
        &\quad\times\biggl\{\frac{1}{3\sqrt{2}}\Lambda^{(a_1), i'_1j'_1k'_1}_{(e)}\delta^{(a_2)(a'_2)}+\frac{1}{2\sqrt{2}}(\Lambda^{(a_1), i'_1j'_1k'_1}_{(e)}d^{a_2a'_2e}+i\Lambda^{(a_1), i'_1j'_1k'_1}_{(e)}f^{a_2a'_2e})\biggr\}\notag\\[1em]
        &+\biggl\{c_{2,BBDB}^{(2)}(\vec{S}_{1}^{\dagger}\cdot\vec{\sigma}_{2})+\bigl[c_{7,BBDB}^{(2)}p_{-}^{2}+c_{8,BBDB}^{(2)}p_{+}^{2}\bigr](\vec{S}_{1}^{\dagger}\cdot\vec{\sigma}_{2})+c_{9,BBDB}^{(2)}\vec{S}_1^{\dagger}\cdot i(\vec{p}_{+}\times\vec{p}_{-})\notag\\[.5em]
        &\quad+c_{10,BBDB}^{(2)}(\vec{S}_{1}^{\dagger}\cdot\vec{p}_{-})(\vec{\sigma}_{2}\cdot\vec{p}_{-})+c_{11,BBDB}^{(2)}(\vec{S}_{1}^{\dagger}\cdot\vec{p}_{+})(\vec{\sigma}_{2}\cdot\vec{p}_{+})\biggr\}\notag\\[.5em]
        &\quad\times\biggl\{\frac{1}{3\sqrt{2}}\Lambda^{(a'_2), i'_1j'_1k'_1}_{(e)}\delta^{(a_2)(a_1)}-\frac{1}{2\sqrt{2}}d^{a_2a_1e}\Lambda^{(a'_2), i'_1j'_1k'_1}_{(e)}-\frac{1}{2\sqrt{2}}if^{a_2a_1e}\Lambda^{(a'_2), i'_1j'_1k'_1}_{(e)}\biggr\},\label{3B1D}\\[1cm]
        V_{DBDB}=&\,\,\,\biggl\{c_{1,DBDB}^{(1)}+c_{2,DBDB}^{(1)}(S^{\dagger}\vec{\sigma}_{1}S\cdot\vec{\sigma}_{2})+c_{5,DBDB}^{(1)}p_-^2+c_{6,DBDB}^{(1)}p_+^2\notag\\[.5em]
        &\quad+\bigl[c_{7,DBDB}^{(1)}p_-^2+c_{8,DBDB}^{(1)}p_+^2\bigr](S^{\dagger}\vec{\sigma}_{1}S\cdot\vec{\sigma}_{2})+c_{9,DBDB}^{(1)}(S^{\dagger}\vec{\sigma}_{1}S+\vec{\sigma}_{2})\cdot i(\vec{p}_{+}\times\vec{p}_{-})\notag\\[.5em]
        &\quad+c_{10,DBDB}^{(1)}(S^{\dagger}\vec{\sigma}_{1}S\cdot\vec{p}_-)(\vec{\sigma}_{2}\cdot\vec{p}_-)+c_{11,DBDB}^{(1)}(S^{\dagger}\vec{\sigma}_{1}S\cdot\vec{p}_+)(\vec{\sigma}_{2}\cdot\vec{p}_+)\biggr\}\notag\\[.5em]
        &\quad\times\biggl\{\delta^{i_1j_1k_1}_{i'_1j'_1k'_1}\delta^{(a_2)(a'_2)}\biggr\}\notag\\[1em]
        &+\biggl\{c_{1,DBDB}^{(2)}+c_{2,DBDB}^{(2)}(S^{\dagger}\vec{\sigma}_{1}S\cdot\vec{\sigma}_{2})+c_{5,DBDB}^{(2)}p_-^2+c_{6,DBDB}^{(2)}p_+^2\notag\\[.5em]
        &\quad+\bigl[c_{7,DBDB}^{(2)}p_-^2+c_{8,DBDB}^{(2)}p_+^2\bigr](S^{\dagger}\vec{\sigma}_{1}S\cdot\vec{\sigma}_{2})+c_{9,DBDB}^{(2)}(S^{\dagger}\vec{\sigma}_{1}S+\vec{\sigma}_{2})\cdot i(\vec{p}_{+}\times\vec{p}_{-})\notag\\[.5em]
        &\quad+c_{10,DBDB}^{(2)}(S^{\dagger}\vec{\sigma}_{1}S\cdot\vec{p}_-)(\vec{\sigma}_{2}\cdot\vec{p}_-)+c_{11,DBDB}^{(2)}(S^{\dagger}\vec{\sigma}_{1}S\cdot\vec{p}_+)(\vec{\sigma}_{2}\cdot\vec{p}_+)\biggr\}\notag\\[.5em]
        &\quad\times\biggl\{\frac{1}{3}\delta^{i_1j_1k_1}_{i'_1j'_1k'_1}\delta^{(a'_2)(a_2)}+\frac{1}{2}\Lambda^{(e), i_1j_1k_1}_{i'_1j'_1k'_1}d^{a'_2a_2e}+\frac{1}{2}\Lambda^{(e), i_1j_1k_1}_{i'_1j'_1k'_1}if^{a'_2a_2e}\biggr\}\notag\\[1em]
        &+\biggl\{c_{1,DBDB}^{(3)}+c_{2,DBDB}^{(3)}(S^{\dagger}\vec{\sigma}_{1}S\cdot\vec{\sigma}_{2})+c_{5,DBDB}^{(3)}p_-^2+c_{6,DBDB}^{(3)}p_+^2\notag\\[.5em]
        &\quad+\bigl[c_{7,DBDB}^{(3)}p_-^2+c_{8,DBDB}^{(3)}p_+^2\bigr](S^{\dagger}\vec{\sigma}_{1}S\cdot\vec{\sigma}_{2})+c_{9,DBDB}^{(3)}(S^{\dagger}\vec{\sigma}_{1}S+\vec{\sigma}_{2})\cdot i(\vec{p}_{+}\times\vec{p}_{-})\notag\\[.5em]
        &\quad+c_{10,DBDB}^{(3)}(S^{\dagger}\vec{\sigma}_{1}S\cdot\vec{p}_-)(\vec{\sigma}_{2}\cdot\vec{p}_-)+c_{11,DBDB}^{(3)}(S^{\dagger}\vec{\sigma}_{1}S\cdot\vec{p}_+)(\vec{\sigma}_{2}\cdot\vec{p}_+)\biggr\}\notag\\[.5em]
        &\quad\times\biggl\{\frac{1}{3}\delta^{i_1j_1k_1}_{i'_1j'_1k'_1}\delta^{(a_2)(a'_2)}+\frac{1}{2}\Lambda^{(e), i_1j_1k_1}_{i'_1j'_1k'_1}d^{a_2a'_2e}+\frac{1}{2}\Lambda^{(e), i_1j_1k_1}_{i'_1j'_1k'_1}if^{a_2a'_2e}\biggr\}\notag\\[1em]
        &+\biggl\{c_{1,DBDB}^{(4)}+c_{2,DBDB}^{(4)}(S^{\dagger}\vec{\sigma}_{1}S\cdot\vec{\sigma}_{2})+c_{5,DBDB}^{(4)}p_-^2+c_{6,DBDB}^{(4)}p_+^2\notag\\[.5em]
        &\quad+\bigl[c_{7,DBDB}^{(4)}p_-^2+c_{8,DBDB}^{(4)}p_+^2\bigr](S^{\dagger}\vec{\sigma}_{1}S\cdot\vec{\sigma}_{2})+c_{9,DBDB}^{(4)}(S^{\dagger}\vec{\sigma}_{1}S+\vec{\sigma}_{2})\cdot i(\vec{p}_{+}\times\vec{p}_{-})\notag\\[.5em]
        &\quad+c_{10,DBDB}^{(4)}(S^{\dagger}\vec{\sigma}_{1}S\cdot\vec{p}_-)(\vec{\sigma}_{2}\cdot\vec{p}_-)+c_{11,DBDB}^{(4)}(S^{\dagger}\vec{\sigma}_{1}S\cdot\vec{p}_+)(\vec{\sigma}_{2}\cdot\vec{p}_+)\biggr\}\notag\\[.5em]
        &\quad\times\biggl\{\frac{1}{3}\delta^{i_1j_1k_1}_{i'_1j'_1k'_1}\delta^{(a'_2)(a_2)}+\frac{1}{2}\Lambda^{(e), i_1j_1k_1}_{i'_1j'_1k'_1}d^{ea_2a'_2}+\frac{1}{2}\Lambda^{(e), i_1j_1k_1}_{i'_1j'_1k'_1}if^{ea_2a'_2}\biggr\},\label{2D2B}\\[1cm]
        V_{BBDD}=&\,\,\,\biggl\{c_{2,BBDD}^{(1)}(\vec{S}_1^{\dagger}\cdot\vec{S}_2^{\dagger})+c_{3,BBDD}^{(1)}(S_{1}^{mn\dagger}S_{2}^{mn\dagger})+\bigl[c_{7,BBDD}^{(1)}p_-^2+c_{8,BBDD}^{(1)}p_+^2\bigr](\vec{S}_1^{\dagger}\cdot\vec{S}_2^{\dagger})\notag\\[1em]
        &\quad+c_{10,BBDD}^{(1)}(\vec{S}_{1}^{\dagger}\cdot\vec{p}_-)(\vec{S}_{2}^{\dagger}\cdot\vec{p}_-)+c_{11,BBDD}^{(1)}(\vec{S}_{1}^{\dagger}\cdot\vec{p}_+)(\vec{S}_{2}^{\dagger}\cdot\vec{p}_+)\notag\\[1em]
        &\quad+\bigl[c_{12,BBDD}^{(1)}p_-^2+c_{13,BBDD}^{(1)}p_+^2\bigr](S_1^{mn\dagger}S_2^{mn\dagger})+\bigl[c_{14,BBDD}^{(1)}p_-^ip_-^j+c_{15,BBDD}^{(1)}p_+^ip_+^j\bigr](S_1^{in\dagger}S_2^{jn\dagger})\biggr\}\notag\\[.5em]
        &\quad\times\biggl\{\frac{1}{6}\Lambda^{(a_1)}_{i'_1j'_1k'_1}\Lambda^{(a_2)}_{i'_2j'_2k'_2}+\frac{1}{4}\Lambda^{(a_1), i'_1j'_1k'_1}_{(e)}\Lambda^{(a_2), i'_2j'_2k'_2}_{(e)}\biggr\},\label{2B2D}\\[1cm]
        V_{DBDD}=&\,\,\,\biggl\{c_{2,DBDD}^{(1)}(S^{\dagger}\vec{\sigma}_1S\cdot\vec{S}_2^{\dagger})+c_{3,DBDD}^{(1)}(\Sigma^{mn}_1S_{2}^{mn\dagger})+\bigl[c_{7,DBDD}^{(1)}p_-^2+c_{8,DBDD}^{(1)}p_+^2\bigr](S^{\dagger}\vec{\sigma}_1S\cdot\vec{S}_2^{\dagger})\notag\\[.5em]
        &\quad+c_{9,DBDD}^{(1)}\vec{S}_2^{\dagger}\cdot i(\vec{p}_+\times\vec{p}_-)+c_{10,DBDD}^{(1)}(S^{\dagger}\vec{\sigma}_1S\cdot\vec{p}_-)(\vec{S}_2^{\dagger}\cdot\vec{p}_-)+c_{11,DBDD}^{(1)}(S^{\dagger}\vec{\sigma}_1S\cdot\vec{p}_+)(\vec{S}_2^{\dagger}\cdot\vec{p}_+)\notag\\[.5em]
        &\quad+\bigl[c_{12,DBDD}^{(1)}p_-^2+c_{13,DBDD}^{(1)}p_+^2\bigr](\Sigma^{mn}_1S_{2}^{mn\dagger})+\bigl[c_{14,DBDD}^{(1)}p_-^ip_-^j+c_{15,DBDD}^{(1)}p_+^ip_+^j\bigr](\Sigma^{in}_1S_{2}^{jn\dagger})\biggr\}\notag\\[.5em]
        &\quad\times\biggl\{\frac{1}{3\sqrt{2}}\lambda_{i'_2j'_2k'_2}^{(e)}\delta_{i'_1}^{i_1}\delta_{j'_1}^{j_1}\delta_{k'_1}^{k_1}+\frac{1}{2\sqrt{2}}\Lambda_{i'_1j'_1k'_1}^{(e),i_1j_1k_1}\Lambda_{(e)}^{(a_2),i_2j_2k_2}\delta_{i'_2}^{i_2}\delta_{j'_2}^{j_2}\delta_{k'_2}^{k_2}\biggr\},\label{3D1B}\\[1cm]
        V_{DDDD}=&\,\,\,\biggl\{c_{1,DDDD}^{(1)}+c_{2,DDDD}^{(1)}(S^{\dagger}\vec{\sigma}_1S\cdot S^{\dagger}\vec{\sigma}_2S)+c_{3,DDDD}^{(1)}(\Sigma_1^{mn}\Sigma_2^{mn})+c_{4,DDDD}^{(1)}(\Sigma_1^{mnl}\Sigma_2^{mnl})\notag\\[1em]
        &\quad+c_{5,DDDD}^{(1)}p_-^2+c_{6,DDDD}^{(1)}p_+^2+\bigl[c_{7,DDDD}^{(1)}p_-^2+c_{8,DDDD}^{(1)}p_+^2\bigr](S^{\dagger}\vec{\sigma}_1S\cdot S^{\dagger}\vec{\sigma}_2S)\notag\\[1em]
        &\quad+c_{9,DDDD}^{(1)}(S^{\dagger}\vec{\sigma}_1S+S^{\dagger}\vec{\sigma}_2S)\cdot i(\vec{p}_+\times\vec{p}_-)+c_{10,DDDD}^{(1)}(S^{\dagger}\vec{\sigma}_1S\cdot \vec{p}_-)(S^{\dagger}\vec{\sigma}_2S\cdot \vec{p}_-)\notag\\[1em]
        &\quad+c_{11,DDDD}^{(1)}(S^{\dagger}\vec{\sigma}_1S\cdot \vec{p}_+)(S^{\dagger}\vec{\sigma}_2S\cdot \vec{p}_+)+\bigl[c_{12,DDDD}^{(1)}p_-^2+c_{13,DDDD}^{(1)}p_+^2\bigr](\Sigma^{mn}_1\Sigma^{mn}_2)\notag\\[1em]
        &\quad+\bigl[c_{14,DDDD}^{(1)}p_-^ip_-^j+c_{15,DDDD}^{(1)}p_+^ip_+^j\bigr](\Sigma^{in}_1\Sigma^{jn}_2)+\bigl[c_{16,DDDD}^{(1)}p_-^2+c_{17,DDDD}^{(1)}p_+^2\bigr](\Sigma^{mnl}_1\Sigma^{mnl}_2)\notag\\[1em]
        &\quad+\bigl[c_{18,DDDD}^{(1)}p_-^ip_-^j+c_{19,DDDD}^{(1)}p_+^ip_+^j\bigr](\Sigma^{inl}_1\Sigma^{jnl}_2)\biggr\}\notag\\[.5em]
        &\quad\times\biggl\{\delta^{i_1j_1k_1}_{i'_1j'_1k'_1}\delta^{i_2j_2k_2}_{i'_2j'_2k'_2}\biggr\},\notag\\[1em]
        &+\biggl\{c_{1,DDDD}^{(2)}+c_{2,DDDD}^{(2)}(S^{\dagger}\vec{\sigma}_1S\cdot S^{\dagger}\vec{\sigma}_2S)+c_{3,DDDD}^{(2)}(\Sigma_1^{mn}\Sigma_2^{mn})+c_{4,DDDD}^{(2)}(\Sigma_1^{mnl}\Sigma_2^{mnl})\notag\\[1em]
        &\quad+c_{5,DDDD}^{(2)}p_-^2+c_{6,DDDD}^{(2)}p_+^2+\bigl[c_{7,DDDD}^{(2)}p_-^2+c_{8,DDDD}^{(2)}p_+^2\bigr](S^{\dagger}\vec{\sigma}_1S\cdot S^{\dagger}\vec{\sigma}_2S)\notag\\[1em]
        &\quad+c_{9,DDDD}^{(2)}(S^{\dagger}\vec{\sigma}_1S+S^{\dagger}\vec{\sigma}_2S)\cdot i(\vec{p}_+\times\vec{p}_-)+c_{10,DDDD}^{(2)}(S^{\dagger}\vec{\sigma}_1S\cdot \vec{p}_-)(S^{\dagger}\vec{\sigma}_2S\cdot \vec{p}_-)\notag\\[1em]
        &\quad+c_{11,DDDD}^{(2)}(S^{\dagger}\vec{\sigma}_1S\cdot \vec{p}_+)(S^{\dagger}\vec{\sigma}_2S\cdot \vec{p}_+)+\bigl[c_{12,DDDD}^{(2)}p_-^2+c_{13,DDDD}^{(2)}p_+^2\bigr](\Sigma^{mn}_1\Sigma^{mn}_2)\notag\\[1em]
        &\quad+\bigl[c_{14,DDDD}^{(2)}p_-^ip_-^j+c_{15,DDDD}^{(2)}p_+^ip_+^j\bigr](\Sigma^{in}_1\Sigma^{jn}_2)+\bigl[c_{16,DDDD}^{(2)}p_-^2+c_{17,DDDD}^{(2)}p_+^2\bigr](\Sigma^{mnl}_1\Sigma^{mnl}_2)\notag\\[1em]
        &\quad+\bigl[c_{18,DDDD}^{(2)}p_-^ip_-^j+c_{19,DDDD}^{(2)}p_+^ip_+^j\bigr](\Sigma^{inl}_1\Sigma^{jnl}_2)\biggr\}\notag\\[1em]
        &\quad\times\biggl\{\frac{1}{3}\delta^{i_1j_1k_1}_{i'_1j'_1k'_1}\delta^{i_2j_2k_2}_{i'_2j'_2k'_2}+\frac{1}{2}\Lambda^{(e),\, i_1j_1k_1}_{i'_1j'_1k'_1}\Lambda^{(e),\, i_2j_2k_2}_{i'_2j'_2k'_2}\biggr\},\label{4D}
\end{alignat}
where the spin (transition) matrices $\sigma$, $S$, $\Sigma$, as well as their combinations with multiple indices, are provided in Appendix \ref{A-3-Pauli}.\textcolor{black}{The subscripts 1 and 2 on the spin matrices denote the spin indices of the incoming (outgoing) baryons 1 and 2, respectively. For LECs, $c_{n,\text{t}}^{(\alpha)}$, the superscripts, $\alpha = 1,2,3,4$ refer to the flavor ordinal numbers, while the subscripts refer to the number of potential couplings, $n$ and the configurations of the baryon interaction transition, $\text{t}$. The chiral potentials, which are constructed by non-relativistic expansion in Table \ref{TB1}, provide 134 LECs up to NLO of the three momentum expansion, i.e. $\mathcal{O}((Q/M)^2)$ with small momentum $Q$ and baryon mass $M$ in the large-$N_c$ limit, and their corresponding number of independent coupling parameters are listed as follows:  27 ($BB\rightarrow BB$), 12 ($BB\rightarrow DB$), 36 ($DB \rightarrow DB$), 10 ($BB \rightarrow DD$), 11 ($DB\rightarrow DD$), and 38 ($DD\rightarrow DD$).} The SU(3) flavor structures do not contain redundant terms. The momentum notations in this work are defined as \\
 \begin{equation}
    \begin{split}
        &\vec{p}_+=\frac{1}{2}(\vec{p'}+\vec{p}),\,\,\,\,\,\vec{p}_-=\vec{p'}-\vec{p},\,\,\,\,\,\quad p^2_+=\vec{p}_+\cdot\vec{p}_+, \\[1em]
        &p^2_-=\vec{p}_-\cdot\vec{p}_-,\,\,\,\,\,\quad\vec{p}\times\vec{p'}=\vec{p}_+\times\vec{p}_-,\\[1em]
        &\vec{p}_+\cdot\vec{p}_-=0,
    \end{split} 
\end{equation}
{where} $\vec{p}$ $(\vec{p'})$ denotes the three-momentum for the incoming (outgoing) state in the center of mass frame, and the external momenta are always zero due to the on-shell condition. From the mixed octet-decuplet and decuplet-decuplet states, $i_1j_1k_1$ ($i'_1j'_1k'_1$) and $i_2j_2k_2$ ($i'_2j'_2k'_2$) represent the indices of incoming (outgoing) for the first and second baryons in {the spin-$3/2$ sector} and the additional notations of flavor transition tensors are \cite{asemke2012}, 
\begin{equation}
    \begin{split}
        \Lambda^{a, nop}_{klm}&=\lambda^{(a)}_{nk}\delta_{ol}\delta_{pm}, \quad\Lambda^{a,klm}_{b}=\epsilon_{ijk}\lambda^{(a)}_{li}\lambda^{(b)}_{mj}, \quad\Lambda^{a,b}_{klm}=\epsilon_{ijk}\lambda^{(a)}_{il}\lambda^{(b)}_{jm}.
    \end{split}
\end{equation}
\textcolor{black}{
The coefficients, $c_{n,\text{t}}^{(\alpha)}$ in Eq.(\ref{4B})--(\ref{4D}) are the linear combinations of Lagrangian coupling constants $g_{xy,z}^{i(\alpha)}$ from the Lagrangian via the non-relativistic expansion, $c^{(\alpha)}_{n,{\rm t}}=\sum_{i}A^{i,(\alpha)}_{n,{\rm t}}\,g^{i(\alpha)}_{xy,z}$, where $A^{i,(\alpha)}_{n,\text{t}}$ denotes the coefficient of Lagrangian coupling. These relations are provided in the following table,}
\begin{table}[H]
\color{black}
\centering
    \setlength{\tabcolsep}{6pt} 
    \renewcommand{\arraystretch}{2} 
\begin{tabular}{|c||c|}
\hline
\text{The Potential's LECs} & \multicolumn{1}{l|}{The Lagrangian's coupling constants} \\ \hline
$c_{S}^{(1,2,3)}$ &$g_{00,1}^{1(1,2,3)}+g_{00,1}^{3(1,2,3)}$ \\ \hline
$c_{T}^{(1,2,3)}$ & $-g_{00,1}^{4(1,2,3)}+2g_{00,1}^{5(1,2,3)}$ \\ \hline
$c_{5,BBBB}^{(1,2,3)}$ &$\frac{1}{4\Lambda^2}(\,g_{00,1}^{3(1,2,3)}+2g_{00,1}^{5(1,2,3)})$ \\ \hline
$c_{6,BBBB}^{(1,2,3)}$ &$\frac{1}{\Lambda^2}(\,g_{00,1}^{1(1,2,3)}+g_{00,1}^{3(1,2,3)})$ \\ \hline
$c_{7,BBBB}^{(1,2,3)}$ & $-\frac{1}{4\Lambda^2}(\,g_{00,1}^{3(1,2,3)}-2g_{00,1}^{5(1,2,3)})$ \\ \hline
$c_{8,BBBB}^{(1,2,3)}$ &$-\frac{1}{\Lambda^2}(\,g_{00,1}^{4(1,2,3)}-g_{00,1}^{5(1,2,3)})$ \\ \hline
$c_{9,BBBB}^{(1,2,3)}$ & $-\frac{1}{4\Lambda^2}(\,g_{00,1}^{1(1,2,3)}+g_{00,1}^{3(1,2,3)}-g_{00,1}^{4(1,2,3)}+2g_{00,1}^{5(1,2,3)})$ \\ \hline
$c_{10,BBBB}^{(1,2,3)}$ & $-\frac{1}{4\Lambda^2}(\,g_{00,1}^{2(1,2,3)}-g_{00,1}^{3(1,2,3)}-g_{00,1}^{4(1,2,3)}+2g_{00,1}^{5(1,2,3)})$ \\ \hline
$c_{11,BBBB}^{(1,2,3)}$ & $-\frac{1}{\Lambda^2}(\,2g_{00,1}^{4(1,2,3)}-g_{00,1}^{5(1,2,3)})$ \\ \hline
\end{tabular}

\caption{\color{black}The relations between the potential's low-energy constants, $c_{n,BBBB}^{(\alpha)}$ and the chiral Lagrangian coupling constants, $g_{00,1}^{i(\alpha)}$ for $BB$ to $BB$ transition. These LECs characterize the various components of the two-baryon potential in Eq.(\ref{4B}): $c_{S}^{(\alpha)} $ up to $c_{8,BBBB}^{(\alpha)}$ are the central force terms, $c_{9,BBBB}^{(\alpha)}$ constitute the spin-orbit force, and the last two LECs, i.e., $c_{10,BBBB}^{(\alpha)}$, and $c_{11,BBBB}^{(\alpha)}$ correspond to the tensor force. This transition have 27 LECs in total: 6 at LO and 21 at NLO. }
\label{BBBB LECs Table}
\end{table}
\vspace{-1em}
\begin{table}[H]
\color{black}
\centering
    \setlength{\tabcolsep}{6pt} 
    \renewcommand{\arraystretch}{2} 
\begin{tabular}{|c||c|}
\hline
\text{The Potential's LECs} & \text{The Lagrangian's coupling constants} \\ \hline
$c_{2,BBDB}^{(1,2)}$ & $g_{01,1}^{4(1,2)}$ \\ \hline
$c_{7,BBDB}^{(1,2)}$ & $\frac{1}{16\Lambda^2}g_{01,1}^{4(1,2)}$ \\ \hline
$c_{8,BBDB}^{(1,2)}$ & $\frac{1}{\Lambda^2}g_{01,1}^{4(1,2)}$ \\ \hline
$c_{9,BBDB}^{(1,2)}$ & $-\frac{1}{4\Lambda^2}g_{01,1}^{4(1,2)}$ \\ \hline
$c_{10,BBDB}^{(1,2)}$ & $-\frac{1}{2\Lambda^2}g_{01,1}^{4(1,2)}$ \\ \hline
$c_{11,BBDB}^{(1,2)}$ & $\frac{1}{8\Lambda^2}g_{01,1}^{4(1,2)}$ \\ \hline
\end{tabular}
\caption{\color{black}The relations between the LECs, $c_{n,BBDB}^{(\alpha)}$ and the chiral Lagrangian coupling constants, $g_{01,1}^{i(\alpha)}$ for $BB$ to $DB$ transition.  These LECs characterize the various components of the two-baryon potential in Eq.(\ref{3B1D}): $c_{2,BBDB}^{(\alpha)} $, $c_{7,BBDB}^{(\alpha)}$, $c_{8,BBDB}^{(\alpha)}$ are the central force, $c_{9,BBDB}^{(\alpha)}$ constitute the spin-orbit force, and $c_{10,BBDB}^{(\alpha)}$, $c_{11,BBDB}^{(\alpha)}$ correspond to the tensor force components. This transition have 12 LECs in total: 2 at LO and 10 at NLO.}
\label{BBDB LECs Table}
\end{table}
\newpage
\begin{table}[H]
\color{black}
\centering
    \setlength{\tabcolsep}{6pt} 
    \renewcommand{\arraystretch}{2} 
\begin{tabular}{|c||c|}
\hline
\text{The Potential's LECs} & \multicolumn{1}{l|}{The Lagrangian's coupling constants} \\ \hline
$c_{1,DBDB}^{(1,2,3,4)}$ & $g_{11,1}^{1(1,2,3,4)}+g_{11,1}^{3(1,2,3,4)}$ \\ \hline
$c_{2,DBDB}^{(1,2,3,4)}$ & $-g_{11,1}^{4(1,2,3,4)}+2g_{11,1}^{5(1,2,3,4)}$ \\ \hline
$c_{5,DBDB}^{(1,2,3,4)}$ & $\frac{1}{4\Lambda^2}(\,g_{11,1}^{3(1,2,3,4)}+2g_{11,1}^{5(1,2,3,4)})$ \\ \hline
$c_{6,DBDB}^{(1,2,3,4)}$ & $\frac{1}{\Lambda^2}(\,g_{11,1}^{1(1,2,3,4)}+g_{11,1}^{3(1,2,3,4)})$ \\ \hline
$ c_{7,DBDB}^{(1,2,3,4)}$ & $-\frac{1}{4\Lambda^2}(\,g_{11,1}^{3(1,2,3,4)}-2g_{11,1}^{5(1,2,3,4)})$ \\ \hline
$c_{8,DBDB}^{(1,2,3,4)}$ & $-\frac{1}{\Lambda^2}(\,g_{11,1}^{4(1,2,3,4)}-g_{11,1}^{5(1,2,3,4)})$ \\ \hline
$c_{9,DBDB}^{(1,2,3,4)}$ & $-\frac{1}{4\Lambda^2}(\,g_{11,1}^{1(1,2,3,4)}+g_{11,1}^{3(1,2,3,4)}-g_{11,1}^{4(1,2,3,4)}+2g_{11,1}^{5(1,2,3,4)})$ \\ \hline
$c_{10,DBDB}^{(1,2,3,4)}$ & $-\frac{1}{4\Lambda^2}(\,g_{11,1}^{2(1,2,3,4)}-g_{11,1}^{3(1,2,3,4)}-g_{11,1}^{4(1,2,3,4)}+2g_{11,1}^{5(1,2,3,4)})$ \\ \hline
$c_{11,DBDB}^{(1,2,3,4)}$ &$-\frac{1}{\Lambda^2}(\,2g_{11,1}^{4(1,2,3,4)}-g_{11,1}^{5(1,2,3,4)})$ \\ \hline
\end{tabular}
\caption{\color{black}The relations between the LECs, $c_{n,DBDB}^{(\alpha)}$ and the chiral Lagrangian coupling constants, $g_{11,1}^{i(\alpha)}$ for $DB$ to $DB$ transition are summarized. These LECs characterize the various components of the two-baryon potential in Eq.(\ref{2D2B}): $c_{1,DBDB}^{(\alpha)} $ $\rightarrow$ $c_{8,DBDB}^{(\alpha)}$ are the central force contributions, $c_{9,DBDB}^{(\alpha)}$ constitute the spin-orbit force, and the last two LECs, i.e., $c_{10,DBDB}^{(\alpha)}$, and $c_{11,DBDB}^{(\alpha)}$ are the tensor force terms. This transition have 36 LECs in total: 8 at LO and 28 at NLO.}
\label{DBDB LECs Table}
\end{table}

\begin{table}[H]
\color{black}
\centering
    \setlength{\tabcolsep}{6pt} 
    \renewcommand{\arraystretch}{2} 
\begin{tabular}{|c||c|}
\hline
\text{The Potential's LECs} & \multicolumn{1}{l|}{The Lagrangian's coupling constants}  \\ \hline
$ c_{2,BBDD}^{(1)}$ & $\frac{1}{4}(\,2g_{02,1}^{1(1)}+2g_{02,1}^{3(1)}+g_{02,1}^{4(1)}-g_{02,2}^{4(1)}+2g_{02,1}^{5(1)}+2g_{02,2}^{5(1)})$ \\ \hline
 $c_{3,BBDD}^{(1)}$ & $-\frac{3}{2}(\,g_{02,1}^{4(1)}+g_{02,2}^{4(1)}+2g_{02,1}^{5(1)}-2g_{02,2}^{5(1)})$ \\ \hline
 $c_{7,BBDD}^{(1)}$&  $\frac{1}{32\Lambda^2}(\,g_{02,1}^{2(1)}+5g_{02,1}^{3(1)}-4g_{02,2}^{3(1)}+g_{02,1}^{4(1)}+6g_{02,1}^{5(1)}+g_{02,2}^{5(1)})$ \\ \hline
$c_{8,BBDD}^{(1)}$ &  $\frac{1}{8\Lambda^2}(\,4g_{02,1}^{1(1)}+4g_{02,1}^{3(1)}+4g_{02,1}^{4(1)}-g_{02,2}^{4(1)}+g_{02,1}^{5(1)}) $ \\ \hline
 $c_{10,BBDD}^{(1)}$& $-\frac{1}{32\Lambda^2}(g_{02,1}^{2(1)}+g_{02,1}^{3(1)}+g_{02,1}^{4(1)}-2g_{02,2}^{4(1)}+2g_{02,1}^{5(1)}-4g_{02,2}^{5(1)}) $ \\ \hline
$c_{11,BBDD}^{(1)}$ &  $\frac{1}{8\Lambda^2}(\,2g_{02,1}^{4(1)}-2g_{02,2}^{4(1)}+g_{02,1}^{5(1)}-4g_{02,2}^{5(1)})$ \\ \hline
 $c_{12,BBDD}^{(1)}$& $-\frac{3}{16\Lambda^2}(\,g_{02,1}^{3(1)}+2g_{02,2}^{3(1)}-2g_{02,1}^{5(1)}-2g_{02,2}^{5(1)})$ \\ \hline
$c_{13,BBDD}^{(1)}$ & $ -\frac{3}{4\Lambda^2}(\,g_{02,1}^{4(1)}+g_{02,2}^{4(1)}+g_{02,1}^{5(1)})$ \\ \hline
$c_{14,BBDD}^{(1)}$ & $-\frac{3}{16\Lambda^2}(\,g_{02,1}^{2(1)}-g_{02,1}^{3(1)}+g_{02,1}^{4(1)}-g_{02,2}^{4(1)}+2g_{02,1}^{5(1)}-2g_{02,2}^{5(1)})$ \\ \hline
$c_{15,BBDD}^{(1)}$ & \hspace{.15cm} $-\frac{3}{4\Lambda^2}(\,2g_{02,1}^{4(1)}+g_{02,2}^{4(1)}-g_{02,1}^{5(1)}+2g_{02,2}^{5(1)})$ \\ \hline
\end{tabular}
\caption{\color{black}The relations between the LECs, $c_{n,BBDD}^{(1)}$ and the chiral Lagrangian coupling constants, $g_{02,z}^{i(1)}$ for $BB$ to $DD$ transition are summarized. These LECs characterize the various components of the two-baryon potential in Eq.(\ref{2B2D}): four LECs as $c_{10,BBDD}^{(1)}$, $c_{11,BBDD}^{(1)}$, $c_{14,BBDD}^{(1)}$, $c_{15,BBDD}^{(1)}$ are the tensor force terms, while remaining LECs constitute the  central force components, and there is no spin-orbit force contribution for this particular sector. This transition have 10 LECs in total: 2 at LO and 8 at NLO.}
\label{BBDD LECs Table}
\end{table}
\newpage
\begin{table}[H]
\color{black}
\centering
    \setlength{\tabcolsep}{6pt} 
    \renewcommand{\arraystretch}{2} 
\begin{tabular}{|c||c|}
\hline
\text{The Potential's LECs} & \text{The Lagrangian's coupling constants}\\ \hline
$ c_{2,DBDD}^{(1)}$ & $\frac{1}{12}(12g_{12,1}^{4(1)}+g_{12,2}^{4(1)}-g_{12,3}^{4(1)})$ \\ \hline
$c_{3,DBDD}^{(1)}$ & $\frac{1}{15}\sqrt{\frac{3}{2}}(4g_{12,2}^{4(1)}+6g_{12,3}^{4(1)})$ \\ \hline
$c_{7,DBDD}^{(1)}$ & $-\frac{1}{96\Lambda^2}(g_{12,2}^{4(1)}-g_{12,3}^{4(1)})$ \\ \hline
$c_{8,DBDD}^{(1)}$ & $\frac{1}{8\Lambda^2}(8g_{12,1}^{4(1)}+g_{12,2}^{4(1)}-g_{12,3}^{4(1)})$ \\ \hline
$c_{9,DBDD}^{(1)}$ & $\frac{1}{48\Lambda^2}(12g_{12,1}^{4(1)}+3g_{12,2}^{4(1)}+5g_{12,3}^{4(1)})$ \\ \hline
$c_{10,DBDD}^{(1)}$ & $-\frac{1}{96\Lambda^2}(12g_{12,1}^{4(1)}-g_{12,2}^{4(1)}-g_{12,3}^{4(1)})$ \\ \hline
$c_{11,DBDD}^{(1)}$ & $\frac{1}{24\Lambda^2}(12g_{12,1}^{4(1)}+g_{12,2}^{4(1)}-g_{12,3}^{4(1)})$ \\ \hline
$c_{12,DBDD}^{(1)}$& $-\frac{1}{60\Lambda^2}\sqrt{\frac{3}{2}}(g_{12,2}^{4(1)}+g_{12,3}^{4(1)})$ \\ \hline
$c_{13,DBDD}^{(1)}$ & $\frac{1}{15\Lambda^2}\sqrt{\frac{3}{2}}(4g_{12,2}^{4(1)}+6g_{12,3}^{4(1)})$ \\ \hline
$c_{14,DBDD}^{(1)}$ & $ -\frac{1}{40\Lambda^2}\sqrt{\frac{3}{2}}(g_{12,2}^{4(1)}+g_{12,3}^{4(1)})$ \\ \hline
 $c_{15,DBDD}^{(1)}$& $\frac{1}{5\Lambda^2}\sqrt{\frac{3}{2}}(g_{12,2}^{4(1)}+g_{12,3}^{4(1)})$ \\ \hline
\end{tabular}
\caption{\color{black}The relations between the LECs, $c_{n,DBDD}^{(1)}$ and the chiral Lagrangian coupling constants, $g_{12,z}^{i1}$ for $DB$ to $DD$ transition are summarized.  These LECs characterize the various components of the two-baryon potential in Eq.(\ref{3D1B}): one LEC,$c_{9,DBDD}^{(1)}$ is the spin-orbit force, Four LECs as $c_{10,DBDD}^{(1)}$, $c_{11,DBDD}^{(1)}$, $c_{14,DBDD}^{(1)}$, $c_{15,DBDD}^{(1)}$) are the tensor force, and remaining LECs contain the  central force components. This transition have 11 LECs in total: 2 at LO and 9 at NLO.}
\label{DBDD LECs Table}
\end{table}
\vspace{-1.5em}
\begin{table}[H]
\color{black}
\centering
    \setlength{\tabcolsep}{6pt} 
    \renewcommand{\arraystretch}{2} 
\begin{tabular}{|c||l|}
\hline
\text{The Potential's LECs} & \text{The Lagrangian's coupling constants}\\ \hline
$c_{1,DDDD}^{(1,2)}$ &$g_{22,1}^{1(1,2)}+g_{22,1}^{3(1,2)}\textcolor{black}{-\frac{1}{9}(\,4g_{22,2}^{1(1,2)}+4g_{22,2}^{3(1,2)}-g_{22,2}^{4(1,2)}+2g_{22,2}^{5(1,2)})}$ \\ 
 & $\textcolor{black}{-\frac{1}{36}(\,16g_{22,3}^{1(1,2)}+16g_{22,3}^{3(1,2)}-g_{22,3}^{4(1,2)}+2g_{22,3}^{5(1,2)})+\frac{1}{9}(\,g_{22,4}^{4(1,2)}-2g_{22,4}^{5(1,2)})}$ \\ 
 & $\textcolor{black}{-\frac{1}{72}(\,3g_{22,5}^{4(1,2)}-g_{22,5}^{5(1,2)})+\frac{1}{72}(\,3g_{22,6}^{4(1,2)}-8g_{22,6}^{5(1,2)})+\frac{1}{36}(\,g_{22,7}^{4(1,2)}-g_{22,7}^{5(1,2)})}$  \\ \hline
$c_{2,DDDD}^{(1,2)}$ & $-g_{22,1}^{4(1,2)}+2g_{22,1}^{5(1,2)}$ \\ \hline
$c_{3,DDDD}^{(1,2)}$ & $\frac{1}{36}(\,4g_{22,2}^{1(1,2)}+4g_{22,2}^{3(1,2)})+\frac{1}{36}(\,4g_{22,3}^{1(1,2)}+4g_{22,3}^{3(1,2)}+g_{22,3}^{4(1,2)}-2g_{22,3}^{5(1,2)})$ \\ 
 & $-\frac{1}{36}(\,g_{22,5}^{4(1,2)}-2g_{22,5}^{5(1,2)})+\frac{1}{36}g_{22,6}^{4(1,2)}+\frac{1}{36}(\,g_{22,7}^{4(1,2)}+g_{22,7}^{5(1,2)})$ \\ \hline
$c_{4,DDDD}^{(1,2)}$ &  $\frac{1}{648}(\,g_{22,2}^{1(1,2)}+g_{22,2}^{3(1,2)}-g_{22,3}^{1(1,2)}-g_{22,3}^{3(1,2)})$ \\ \hline
$c_{5,DDDD}^{(1,2)}$ &  $ \frac{1}{4\Lambda^2}(\,g_{22,1}^{3(1,2)}+2g_{22,1}^{5(1,2)})-\frac{1}{36\Lambda^2}(\,g_{22,2}^{2(1,2)}+2g_{22,2}^{3(1,2)}-g_{22,2}^{4(1,2)}+8g_{22,2}^{5(1,2)})$ \\ 
 & $\textcolor{black}{-\frac{1}{144\Lambda^2}(\,g_{22,3}^{2(1,2)}+14g_{22,3}^{3(1,2)}+g_{22,3}^{4(1,2)}+32g_{22,3}^{5(1,2)})}-\frac{1}{36\Lambda^2}(\,g_{22,4}^{4(1,2)}+4g_{22,4}^{5(1,2)})$ \\
 & $\textcolor{black}{+\frac{1}{288\Lambda^2}(\,3g_{22,5}^{4(1,2)}+2g_{22,5}^{5(1,2)})-\frac{1}{96\Lambda^2}(\,g_{22,6}^{4(1,2)}+\frac{16}{3}g_{22,6}^{5(1,2)})-\frac{1}{144\Lambda^2}(\,g_{22,7}^{4(1,2)}+2g_{22,7}^{5(1,2)})}$ \\ \hline
 
\end{tabular}
\caption{\color{black}The relations between the LECs, $c_{n,DDDD}^{(\alpha)}$ and the chiral Lagrangian coupling constants, $g_{22,z}^{i\alpha}$ for $DD$ to $DD$ transition are summarized. These LECs characterize the various components of the two-baryon potential in Eq.(\ref{4D}): the spin-orbit force represent as $c_{9,DDDD}^{(\alpha)}$, there are six the tensor force terms ($c_{10,DDDD}^{(\alpha)}$, $c_{11,DDDD}^{(\alpha)}$, $c_{14,DDDD}^{(\alpha)}$, $c_{15,DDDD}^{(\alpha)}$, $c_{18,DDDD}^{(\alpha)}$, $c_{19,DDDD}^{(\alpha)}$), and remaining LECs contain the  central force components. This transition have 38 LECs in total: 8 at LO and 30 at NLO.}
\label{DDDD LECs Table 1}
\end{table}

\begin{table}[H]
\color{black}
\centering
    \setlength{\tabcolsep}{6pt} 
    \renewcommand{\arraystretch}{2} 
\begin{tabular}{|c||l|}
\hline
\text{The Potential's LECs} & \text{The Lagrangian's coupling constants} \\ \hline
$c_{6,DDDD}^{(1,2)}$ & $\frac{1}{\Lambda^2}(\,g_{22,1}^{1(1,2)}+g_{22,1}^{3(1,2)})\textcolor{black}{-\frac{1}{9\Lambda^2}(\,4g_{22,2}^{1(1,2)}+4g_{22,2}^{3(1,2)}+g_{22,2}^{4(1,2)})}$ \\ 
 & $\textcolor{black}{-\frac{1}{36\Lambda^2}(\,8g_{22,3}^{1(1,2)}+8g_{22,3}^{3(1,2)}+g_{22,3}^{4(1,2)})+\frac{1}{9\Lambda^2}(\,g_{22,4}^{3(1,2)}+2g_{22,4}^{4(1,2)}+2g_{22,4}^{5(1,2)})}$ \\ 
 & $\textcolor{black}{+\frac{1}{72\Lambda^2}(\,8g_{22,5}^{3(1,2)}-6g_{22,5}^{4(1,2)}-g_{22,5}^{5(1,2)})+\frac{1}{9\Lambda^2}(\,g_{22,6}^{3(1,2)}+\frac{3}{4}g_{22,6}^{4(1,2)}+g_{22,6}^{5(1,2)})}$ \\ 
 & $\textcolor{black}{+\frac{1}{36\Lambda^2}(\,4g_{22,7}^{3(1,2)}+2g_{22,7}^{4(1,2)}+g_{22,7}^{5(1,2)})}$ \\ \hline
$c_{7,DDDD}^{(1,2)}$ & $-\frac{1}{4\Lambda^2}(\,g_{22,1}^{3(1,2)}+2g_{22,1}^{5(1,2)})\textcolor{black}{+\frac{1}{48\Lambda^2}(\,g_{22,2}^{2(1,2)}-g_{22,2}^{4(1,2)})+\frac{1}{48\Lambda^2}(\,g_{22,4}^{2(1,2)}+g_{22,4}^{4(1,2)}}$ \\
 & $\textcolor{black}{+2g_{22,4}^{5(1,2)})-\frac{1}{48\Lambda^2}g_{22,5}^{3(1,2)}+\frac{1}{48\Lambda^2}(\,g_{22,6}^{3(1,2)}+g_{22,6}^{5(1,2)})}$ \\ \hline
$c_{8,DDDD}^{(1,2)}$ & $-\frac{1}{\Lambda^2}(\,g_{22,1}^{4(1,2)}-g_{22,1}^{5(1,2)})\textcolor{black}{-\frac{1}{12\Lambda^2}g_{22,2}^{5(1,2)}}+\frac{1}{6\Lambda^2}(\,g_{22,3}^{4(1,2)}-g_{22,3}^{5(1,2)})\textcolor{black}{-\frac{1}{12\Lambda^2}g_{22,4}^{4(1,2)}}$ \\
 & $\textcolor{black}{-\frac{1}{12\Lambda^2}g_{22,5}^{5(1,2)}-\frac{1}{12\Lambda^2}g_{22,6}^{5(1,2)}}$ \\ \hline
$c_{9,DDDD}^{(1,2)}$ & $-\frac{1}{4\Lambda^2}(\,g_{22,1}^{1(1,2)}+g_{22,1}^{3(1,2)}-g_{22,1}^{4(1,2)}+2g_{22,1}^{5(1,2)})+\frac{1}{36\Lambda^2}(\,g_{22,2}^{1(1,2)}+g_{22,2}^{3(1,2)}-4g_{22,2}^{4(1,2)}$ \\ 
 &$+2g_{22,2}^{5(1,2)})\textcolor{black}{+\frac{1}{144\Lambda^2}(g_{22,3}^{1(1,2)}+g_{22,3}^{3(1,2)}-g_{22,3}^{4(1,2)}+2g_{22,3}^{5(1,2)})+\frac{1}{144\Lambda^2}(2g_{22,4}^{3(1,2)}}$ \\ 
 &  $\textcolor{black}{-g_{22,4}^{4(1,2)}+2g_{22,4}^{5(1,2)})+\frac{1}{144\Lambda^2}(8g_{22,5}^{3(1,2)}-4g_{22,5}^{4(1,2)}-5g_{22,5}^{5(1,2)})+\frac{1}{144\Lambda^2}(2g_{22,6}^{3(1,2)}}$ \\ 
 & $\textcolor{black}{-g_{22,6}^{4(1,2)}+g_{22,6}^{5(1,2)})-\frac{1}{144\Lambda^2}(2g_{22,7}^{3(1,2)}-g_{22,7}^{4(1,2)}+g_{22,7}^{5(1,2)})}$ \\ \hline
 $c_{10,DDDD}^{(1,2)}$& $-\frac{1}{4\Lambda^2}(g_{22,1}^{2(1,2)}-g_{22,1}^{3(1,2)}-g_{22,1}^{4(1,2)}+g_{22,1}^{5(1,2)})+\frac{1}{24\Lambda^2}(g_{22,2}^{2(1,2)}-g_{22,2}^{4(1,2)})$ \\ 
 & $+\frac{1}{24\Lambda^2}(g_{22,4}^{2(1,2)}+g_{22,4}^{4(1,2)}+2g_{22,4}^{5(1,2)})-\frac{1}{24\Lambda^2}g_{22,5}^{3(1,2)}+\frac{1}{24\Lambda^2}(g_{22,6}^{3(1,2)}+g_{22,6}^{5(1,2)})$ \\ \hline
$c_{11,DDDD}^{(1,2)}$ & $-\frac{1}{\Lambda^2}(2g_{22,1}^{4(1,2)}+g_{22,1}^{5(1,2)})-\frac{1}{6\Lambda^2}g_{22,2}^{5(1,2)}+\frac{1}{3\Lambda^2}(g_{22,3}^{4(1,2)}-g_{22,3}^{5(1,2)})\textcolor{black}{+\frac{1}{36\Lambda^2}(g_{22,4}^{3(1,2)}}$ \\ 
 &  $\textcolor{black}{-6g_{22,4}^{4(1,2)})-\frac{1}{36\Lambda^2}(g_{22,5}^{3(1,2)}-6g_{22,5}^{5(1,2)})+\frac{1}{36\Lambda^2}(g_{22,6}^{3(1,2)}-6g_{22,6}^{5(1,2)})}-\frac{1}{2\Lambda^2}g_{22,7}^{3(1,2)}$\\ \hline
 $c_{12,DDDD}^{(1,2)}$& $\frac{1}{144\Lambda^2}(3g_{22,2}^{3(1,2)}+10g_{22,2}^{5(1,2)})+\frac{1}{144\Lambda^2}(g_{22,3}^{2(1,2)}+4g_{22,3}^{3(1,2)}-g_{22,3}^{4(1,2)}+12g_{22,3}^{5(1,2)})$ \\ 
 & $+\frac{5}{144\Lambda^2}g_{22,5}^{5(1,2)}+\frac{1}{144\Lambda^2}(2g_{22,7}^{3(1,2)}+g_{22,7}^{4(1,2)}+5g_{22,7}^{5(1,2)})$ \\ \hline
$ c_{13,DDDD}^{(1,2)}$ & $\frac{1}{36\Lambda^2}(4g_{22,2}^{1(1,2)}+4g_{22,2}^{3(1,2)}+2g_{22,2}^{4(1,2)}-g_{22,2}^{5(1,2)})+\frac{1}{18\Lambda^2}(2g_{22,3}^{1(1,2)}+2g_{22,3}^{3(1,2)})$ \\ 
 & $+\frac{1}{18\Lambda^2}g_{22,4}^{5(1,2)}-\frac{1}{36\Lambda^2}g_{22,5}^{4(1,2)}+\frac{1}{36\Lambda^2}g_{22,6}^{4(1,2)}-\frac{1}{36\Lambda^2}g_{22,7}^{5(1,2)}$ \\ \hline
$c_{14,DDDD}^{(1,2)}$ & $\frac{4}{144\Lambda^2}g_{22,2}^{3(1,2)}-\frac{1}{72\Lambda^2}(g_{22,3}^{2(1,2)}-g_{22,3}^{3(1,2)}-g_{22,3}^{4(1,2)}+2g_{22,3}^{5(1,2)})+\frac{1}{144\Lambda^2}(g_{22,5}^{4(1,2)}$ \\ 
 & $-g_{22,5}^{5(1,2)})-\frac{1}{144\Lambda^2}g_{22,6}^{4(1,2)}-\frac{1}{72\Lambda^2}(g_{22,7}^{3(1,2)}+g_{22,7}^{4(1,2)}+g_{22,7}^{5(1,2)})$ \\ \hline
$c_{15,DDDD}^{(1,2)}$ & $\frac{1}{18\Lambda^2}g_{22,2}^{4(1,2)}+\frac{1}{36\Lambda^2}(4g_{22,4}^{3(1,2)}+2g_{22,4}^{5(1,2)})+\frac{1}{36\Lambda^2}(4g_{22,5}^{3(1,2)}-g_{22,5}^{4(1,2)})$ \\ 
 & $+\frac{1}{36\Lambda^2}(4g_{22,6}^{3(1,2)}+g_{22,6}^{4(1,2)})+\frac{1}{18\Lambda^2}(2g_{22,7}^{3(1,2)}+g_{22,7}^{4(1,2)}+g_{22,7}^{5(1,2)})$ \\ \hline
$c_{16,DDDD}^{(1,2)}$ & $\frac{1}{2592\Lambda^2}(g_{22,2}^{3(1,2)}-2g_{22,2}^{5(1,2)})-\frac{1}{2592\Lambda^2}(g_{22,3}^{3(1,2)}+2g_{22,3}^{5(1,2)})$ \\ \hline
$c_{17,DDDD}^{(1,2)}$ & $\frac{1}{648\Lambda^2}(g_{22,2}^{1(1,2)}+g_{22,2}^{3(1,2)})-\frac{1}{648\Lambda^2}(g_{22,3}^{1(1,2)}+g_{22,3}^{3(1,2)})+\frac{1}{648\Lambda^2}(g_{22,4}^{3(1,2)}-g_{22,5}^{3(1,2)}$ \\
 & $+g_{22,6}^{3(1,2)}-g_{22,7}^{3(1,2)})$ \\ \hline
 $ c_{18,DDDD}^{(1,2)}$& $-\frac{3}{160\Lambda^2}(g_{22,2}^{2(1,2)}-g_{22,2}^{4(1,2)})-\frac{3}{160\Lambda^2}(g_{22,4}^{2(1,2)}+g_{22,4}^{4(1,2)}+2g_{22,4}^{5(1,2)})+\frac{3}{160\Lambda^2}g_{22,5}^{3(1,2)}$ \\
 &  $-\frac{3}{160\Lambda^2}(g_{22,6}^{3(1,2)}+g_{22,6}^{5(1,2)})$ \\ \hline
$c_{19,DDDD}^{(1,2)}$& $\frac{3}{40\Lambda^2}g_{22,2}^{5(1,2)}-\frac{3}{20\Lambda^2}(g_{22,3}^{4(1,2)}-g_{22,3}^{5(1,2)})+\frac{3}{40\Lambda^2}g_{22,4}^{4(1,2)}+\frac{3}{40\Lambda^2}g_{22,5}^{5(1,2)}+\frac{3}{40\Lambda^2}g_{22,6}^{5(1,2)}$ \\ \hline
\end{tabular}
\caption{\color{black}The relations between the LECs, $c_{n,DDDD}^{(\alpha)}$ and the chiral Lagrangian coupling constants, $g_{22,z}^{i\alpha}$ for $DD$ to $DD$ transition are summarized. These LECs characterize the various components of the two-baryon potential in Eq.(\ref{4D}): the spin-orbit force represent as $c_{9,DDDD}^{(\alpha)}$, there are six the tensor force terms ($c_{10,DDDD}^{(\alpha)}$, $c_{11,DDDD}^{(\alpha)}$, $c_{14,DDDD}^{(\alpha)}$, $c_{15,DDDD}^{(\alpha)}$, $c_{18,DDDD}^{(\alpha)}$, $c_{19,DDDD}^{(\alpha)}$), and remaining LECs contain the  central force components. This transition have 38 LECs in total: 8 at LO and 30 at NLO. (cont.)}
\label{DDDD LECs Table 2}
\end{table}

{These linear combinations can be derived using the Fierz identities for the Gell-Mann matrices ($\lambda^a$), and Pauli matrices ($\sigma_i$), along with the corresponding expressions and relations provided in \cite{asemke2012,susanne2020,borodulin2022core32compendiumrelations}. In addition, the power counting scheme ($Q/M$), where $Q$ denotes a typical three-momentum, has also been discussed in \cite{Epelbaum_2005}. More importantly, in this work, we replaced $M\rightarrow \Lambda$ after performing the non-relativistic expansion. In the single-baryon sector, the baryon mass is treated on the same footing as the chiral symmetry breaking scale ($\Lambda \sim 1$ GeV).} For a few baryon sectors, on the other hand, the baryon mass should be larger than the chiral symmetry breaking scale. {If the baryon mass ($M$) is assumed to be proportional to the chiral symmetry breaking scale ($\Lambda$), then the power counting scheme adopted in this work follows $(Q/M) \sim (Q/\Lambda)^2$, as also discussed in \cite{Ordonez:1995rz}.}\\
\indent \textcolor{black}{The chiral Lagrangians given in Eqs.(\ref{LBBBB}--\ref{LDDDD}) yield a total of 104 independent coupling constants ($g_{xy,z}^{i(\alpha)}$). By performing the non-relativistic expansion up to NLO, we derive the minimal sets of chiral baryon-baryon potentials for the octet-octet,  Mixed octet-decuplet, and decuplet-decuplet states, incorporating Lorentz and flavor structures as shown in Eqs.(\ref{4B}--\ref{4D}). These chiral potentials yield a total of 134 LECs ($c_{n,\text{t}}^{(\alpha)}$): 28 at LO and 106 at NLO, under small-momentum scaling ($Q/\Lambda$). The LECs given in Table \ref{BBBB LECs Table}--\ref{DDDD LECs Table 2}, each representing a linear combination of coupling constants, $g_{xy,z}^{i(\alpha)}$, complicate conventional calculations of the chiral potentials at higher orders. Therefore, in the following section, we will introduce the $1/N_c$ operator analysis to reduce the number of LECs in the SU(3) two-baryon interactions within the ChEFT framework.}

\section{The $1/N_c$ operator analysis of octet and decuplet baryon-baryon interaction}\label{sec-3}
\subsection{The Hartree Hamiltonian and the effective operators in the $1/N_c$ expansion}
\indent In this section, we begin by analyzing $1/N_c$ expansion for the matrix element of the octet-octet baryon.The baryon-baryon scattering should have a scale of about $N_c$ \cite{EdWITTEN1979,Kaplan_1996,Kaplan_1997}, while single baryon bilinear matrix elements in the SU(3) flavor symmetry have a scale like $N_c^0$. We can systematically expand its matrix elements in terms of effective spin-flavor baryon states and quark operators within the $1/N_c$ expansion framework as \cite{Luty_1994}\\
\begin{equation}\label{1/Ncexpandbaryon}
    \langle B|\mathcal{O}^i|B \rangle=(B|\sum_mc_{m}^{(i)}\frac{\mathcal{O}^{m}}{N_c^{m}}|B),
\end{equation}
{where $\mathcal{O}^i$ denotes the $i$-th quark current operator, $c_{m}^{(i)}$ represents a function encapsulating certain dynamical properties of the system, and $|B)$ refers to effective baryon states characterized by specific Lorentz (spin) and flavor structures\cite{Dashen_1995}. $\mathcal{O}^m$ denotes an $m$-body operator that can be expanded in terms of the generators of the contracted SU(6) spin–flavor symmetry \cite{Vonk_2025,Kaplan_1996,Kaplan_1997}, as given by:}
\begin{equation}
    \frac{\mathcal{O}^{m}}{N_c^{m}}=\biggl(\frac{J}{N_{c}}\biggr)^n\biggl(\frac{T}{N_{c}}\biggr)^o\biggl(\frac{G}{N_{c}}\biggr)^p, \text{with} \hspace{.2cm}m=n+o+p,
\end{equation}
{where the generators $J$, $T$, and $G$ represent the spin, flavor, and spin–flavor operators, respectively, which in this work are expressed as follows:}
\begin{equation}\label{JTG}
    \begin{split}
        \mathbb{1}& = q^{\dagger}(\mathbb{1}\otimes\mathbb{1})q, \hspace{1cm} J_i = q^{\dagger}\biggl(\frac{\sigma_i}{2}\otimes\mathbb{1}\biggr)q,\\[1em]
        T^a& =  q^{\dagger}\biggl(\mathbb{1}\otimes\frac{\lambda_a}{2}\biggr)q, \hspace{1cm} G^{a}_{i}=q^{\dagger}\biggl(\frac{\sigma_i}{2}\otimes\frac{\lambda_a}{2}\biggr)q.
    \end{split}
\end{equation}
{Note that $q$ and $q^\dagger$ are the annihilation and creation operators for quarks, respectively. These are treated as three-flavor bosonic operators, since the spin-flavor wavefunctions of ground-state baryons in the large-$N_c$ limit are fully symmetric. However, this symmetry does not extend to the color degrees of freedom. Here, the commutation relation between the quark operators is given by $[q, q^\dagger] = 1$.}\\
\indent In order to decompose a $1/N_c$ baryon state, we start to consider the $N_c$ scaling of the matrix elements between initial and final baryon states. The m-body operator $\mathcal{O}^{m}$ and its coefficient $c^{(i)}_{m}$ in Eq.(\ref{1/Ncexpandbaryon}) are given the $N_c$ scale as,
\begin{equation}
    (B|\mathcal{O}^{m}|B)\lesssim N_c^{m}, \hspace{1cm} c^{(i)}_{m}\sim N_c^0.
\end{equation}
Furthermore, the generators of Eq.(\ref{JTG}) are as follows,
\begin{equation}\label{NcRule}
    \begin{split}
         &(B|\mathbb{1}|B)\sim N_c, \hspace{1cm} (B|J^i|B)\sim N_c^0,\\[1em]
         &(B|T^a|B)\sim N_c^0 , \hspace{.7cm} (B|G_i^a|B)\sim N_c, \hspace{.9cm} \text{for}\hspace{.1cm} a=1,2,3,\\[1em]
         &(B|T^a|B)\sim \sqrt{N_c} , \hspace{.4cm} (B|G_i^a|B)\sim \sqrt{N_c}, \hspace{.5cm} \text{for}\hspace{.1cm} a=4,5,6,7,\\[1em]
         &(B|T^a|B)\sim N_c , \hspace{.7cm} (B|G_i^a|B)\sim N_c^0, \hspace{.9cm} \text{for}\hspace{.1cm} a=8.
    \end{split}
\end{equation}
{The $N_c$ scaling behaviors of the matrix elements $(B|T^a|B)$ and $(B|G_i^a|B)$ differ in their orders within the $1/N_c$ expansion. These differences depend on the direction in flavor space, specified by the adjoint representation indices ($a = 1,2,\dots,8$), as the isospin and baryons with strangeness contribute at order $\mathcal{O}(N_c^0)$ \cite{Dashen_1995}.} The external momentum variables in the center of mass frames, as mentioned in the previous section, exhibit their own $N_c$ scaling as follows \cite{Kaplan_1997},
\begin{equation}
    \begin{split}
        &\vec{p}_+ \sim \frac{1}{N_c}, \hspace{1cm}  \vec{p}_- \sim N_{c}^{0},\\
        &p_+^2 \sim N_c^0, \hspace{1cm} p_-^2 \sim  N_c^0.
    \end{split}
\end{equation}
In the higher-order correction term, the $\vec{p}_+$ is always accompanied by the baryon mass factor $1/M$. We have mentioned that $\vec{p}_+ \sim 1/N_c$, so the baryon mass ($M$) is proportional to $N_c$.\\
\indent In this basis, the two-baryon potential in terms of $1/N_c$ expansion can be written in the following form of the Hartree Hamiltonian \cite{Luty_1994,Dashen_1995},
\begin{equation}\label{HartreeHamil}
    \begin{split}
        \mathcal{H}=N_c\sum_m\sum_{no}c_{m,no}\biggl(\frac{J}{N_{c}}\biggr)^n\biggl(\frac{T}{N_{c}}\biggr)^o\biggl(\frac{G}{N_{c}}\biggr)^{m-n-o},
    \end{split}
\end{equation}
where the coefficient functions $c_{m,no}$ still contain the scale of $N_c^0$. It is a well-known fact that the spin-$1/2$ and -$3/2$ baryon sectors manifest degeneracy states at the large-$N_c$ limit. The $1/N_c$ baryon-baryon potential can be derived from this Hamiltonian, which is given by \cite{Kaplan_1997,Liu_2019},\\
\begin{equation}\label{Hamiltonian}
    \begin{split}
        \mathcal{H}_{LO}& = U_1^{LO}(p_-^2)\mathbb{1}_1\cdot\mathbb{1}_2+U_2^{LO}(p_-^2)T_1\cdot T_2+U_3^{LO}(p_-^2)G_1\cdot G_2+U_{4}^{LO}(p_-^2)(p^i_-p^j_-)_{(2)}\cdot(G_1^{i,a}G_2^{j,a})_{(2)},\\[1em]
        \mathcal{H}_{NNLO} & = U_1^{NNLO}(p_-^2)p_+^2\mathbb{1}_1\cdot\mathbb{1}_2+U_2^{NNLO}(p_-^2)\vec{J}_1\cdot\vec{J}_2+U_3^{NNLO}(p_-^2)\vec{J}_1\cdot\vec{J}_2T_1\cdot T_2+U_4^{NNLO}(p_-^2)p_+^2T_1\cdot T_2\\[1em]
        &\hspace{.5cm}+U_5^{NNLO}(p_-^2)p_+^2G_1\cdot G_2+U_6^{NNLO}(p_-^2)i(\vec{p}_+\times\vec{p}_-)\cdot(\vec{J}_1+\vec{J}_2)\\[1em]
        &\hspace{.5cm}+U_7^{NNLO}(p_-^2)i(\vec{p}_+\times\vec{p}_-)\cdot(T^a_1\vec{G}^a_2+\vec{G}^a_1T^a_2)+U_8^{NNLO}(p_-^2)i(\vec{p}_+\times\vec{p}_-)\cdot(\vec{J}_1+\vec{J}_2)T_1\cdot T_2\\[1em]
        &\hspace{.5cm}+U_9^{NNLO}(p_-^2)(p^i_-p^j_-)_{(2)}\cdot(J_1^iJ_2^j)_{(2)}+U_{10}^{NNLO}(p_-^2)(p^i_-p^j_-)_{(2)}\cdot(J_1^iJ_2^j)_{(2)}T_1\cdot T_2\\[1em]
        &\hspace{.5cm}+U_{11}^{NNLO}(p_-^2)(p^i_+p^j_+)_{(2)}\cdot(G_1^{i,a}G_2^{j,a})_{(2)}.\\[1em]
    \end{split}
\end{equation}
The Hartree Hamiltonian shown above is decomposed, within the framework of the $1/N_c$ expansion, into 4 LO parameters and 11 NNLO parameters. Note that there is no NLO for the Hartree Hamiltonian in 1/$N_c$ expansion. We impose $T_1 \cdot T_2 = T_{1}^aT_{2}^a$, $G_1 \cdot G_2=G_1^{i,a}G_2^{i,a}$ and the notation,
\begin{equation}
    (\sigma_1^i\sigma_2^j)_{(2)}\cdot(p_{\pm}^ip_{\pm}^j)_{(2)}= (\Vec{\sigma}_1 \cdot \Vec{p}_{\pm})(\Vec{\sigma}_2^j \cdot \Vec{p}_{\pm}) - \frac{1}{3}(\Vec{\sigma}_1 \cdot \Vec{\sigma}_2)p^2_{\pm}.
\end{equation}
After applying the $1/N_c$ expansion, any term of order $1/N_c^2$ will be eliminated from our analysis. The scale factor $1/N_c$ is implicit within each effective operators ($\mathbb{1}, J, T, G$). The arbitrary functions $U_i^{LO}(p_-^2)$ and $U_i^{NNLO}(p_-^2)$ contain scale of $N_c^0$. Due to $N_c$ scale counting in Eq.(\ref{NcRule}), we find out $T^aG^{i,a}/N_c \sim N_c^0$, which means $T^aG^{i,a} \sim N_c$.\\
\indent Next step  involves evaluating the $1/N_c$ potential of the Hartree Hamiltonian for each interaction state, which is provided by
\begin{equation}
    \begin{split}
        V &=({\rm Out}_2 ,{\rm Out}_1|\hat{H} |{\rm In}_1, {\rm In}_2)\\[1em]
        &=(\chi'_2,a'_2;\chi'_1,a'_1|\hat{H}|a_1,\chi_1;a_2,\chi_2)+(\chi'_2,a'_2;\chi'_1,i'_1j'_1k'_1|\hat{H}|a_1,\chi_1;a_2,\chi_2)\\[1em]
        &\hspace{.5cm}+(\chi'_2,a'_2;\chi'_1,i'_1j'_1k'_1|\hat{H}|i_1j_1k_1,\chi_1;a_2,\chi_2)+(\chi'_2,i'_2j'_2k'_2;\chi'_1,i'_1j'_1k'_1|\hat{H}|a_1,\chi_1;a_2,\chi_2)\\[1em]
        &\hspace{.5cm}+(\chi'_2,i'_2j'_2k'_2;\chi'_1,i'_1j'_1k'_1|\hat{H}|i_1j_1k_1,\chi_1;a_2,\chi_2)+(\chi'_2,i'_2j'_2k'_2;\chi'_1,i'_1j'_1k'_1|\hat{H}|i_1j_1k_1,\chi_1;i_2j_2k_2,\chi_2),
    \end{split}
\end{equation}
where $|a_1\,(a_2),\chi_1\,(\chi_2)\bigr)$ represent incoming $1^{\rm st}$ $ (2^{\rm nd})$  octet baryon flavor that contains spin $\chi_i=\pm1/2$, and $|i_1j_1k_1\,(i_2j_2k_2),\chi_1\,(\chi_2)\bigr)$ are incoming $1^{\rm st}$ $(2^{\rm nd})$  decuplet baryon flavor which have spin $\chi_i=\pm1/2,\,\pm3/2$.\\
\indent First of all, we need to review the actions of the effective operators on the octet and decuplet baryon states, which are obtained by contacting each possible state with proper spin-flavor tensors. The contraction forms for both baryon states are summarized as \cite{asemke2012},
\begin{alignat}{2}
    J^i|a,\chi) &= \frac{1}{2}\sigma^{(i)}_{\chi'\chi}|a,\chi')\notag\\[1em]
        T^e|a,\chi) &= if^{a_1a'_1e}|a',\chi)\notag\\[1em]
        G^{ie}|a,\chi) &= \frac{1}{2}\sigma^{(i)}_{\chi'\chi}(d^{eaa'}+\frac{2}{3}if^{eaa'})|a',\chi')+\frac{1}{2\sqrt{2}}S^{(i),\dagger}_{\chi'\chi}(\epsilon_{ijk}\lambda^{(e)}_{li}\lambda^{(a)}_{mj})_{sym(klm)}|klm,\chi')\label{Operatorcontact}\\[1em]
        J^i|klm,\chi) &= \frac{3}{2}(\vec{S}^\dagger\sigma^{(i)}\vec{S})_{\chi'\chi}|klm,\chi')\notag\\[1em]
        T^e|klm,\chi) &= \frac{3}{2}(\lambda^{(e)}_{nk}\delta_{ol}\delta_{pm})_{sym(klm)}|nop,\chi)\notag\\[1em]
        G^{ie}|klm,\chi) &= \frac{3}{4}(\vec{S}^\dagger\sigma^{(i)}\vec{S})_{\chi'\chi}(\lambda^{(e)}_{nk}\delta_{ol}\delta_{pm})_{sym(klm)}|nop,\chi')+\frac{1}{2\sqrt{2}}S^{(i)}_{\chi'\chi}(\epsilon_{ijk}\lambda^{(e)}_{il}\lambda^{(a)}_{jm})_{sym(klm)}|a,\chi').\notag
\end{alignat}
Before deriving the potential by matching the operator, the arbitrary functions $U_i^{LO}(p_-^2)$ and $U_i^{NNLO}(p_-^2)$ will be defined by the ansatz as $q_i$ and $h_i$, respectively. Then, by substituting Eq.(\ref{Operatorcontact}) to (\ref{Hamiltonian}), the potential at the LO and NNLO of $1/N_c$ expansion is evaluated in Appendix \ref{A-2-HHPotential}, where the $N_c$ scale of the potential are $V_{LO} \sim N_c$ and $V_{NNLO} \sim 1/N_c$, respectively. By comparing the potential between the chiral Lagrangian and the $1/N_c$ Hartree Hamiltonian, we can extract the $N_c$ scale of these coupling constants as follows. 
\begin{itemize}
\item Octet-Octet (\scalebox{1}{$BBBB$})
\begin{equation}\label{4BNc}
\begin{split}
        &g_{00,1}^{1(1,2,3)}, g_{00,1}^{3(1,2,3)}, g_{00,1}^{5(1,2,3)}\sim N_c, \quad\quad g_{00,1}^{2(1,2,3)}, g_{00,1}^{4(1,2,3)}\sim 1/N_c,\\[.5em]
\end{split}
\end{equation}
\item  Mixed Octet-Decuplet (\scalebox{1}{$BBDB$})
    \begin{equation}
    \begin{split}
        g_{01,1}^{41}, g_{01,1}^{42} \sim 1/N_c
    \end{split}
\end{equation}
\item Mixed Octet-Decuplet (\scalebox{1}{$DBDB$})
    \begin{equation}
    \begin{split}
         &g_{11,1}^{1(1,2,3,4)}, g_{11,1}^{3(1,2,3,4)}, g_{11,1}^{5(1,2,3,4)}\sim N_c, \quad\quad g_{11,1}^{2(1,2,3,4)}, g_{11,1}^{4(1,2,3,4)}\sim 1/N_c,\\[.5em]
    \end{split}
\end{equation}
\item Mixed Octet-Decuplet (\scalebox{1}{$BBDD$})
    \begin{equation}
     \begin{split}
        g_{02,1}^{11},g_{02,(1,2)}^{31},g_{02,(1,2)}^{51}\sim N_c , \quad\quad g_{02,1}^{21},g_{02,(1,2)}^{41}\sim 1/N_c
    \end{split}
    \end{equation}
\item Mixed Octet-Decuplet (\scalebox{1}{$DBDD$})

    \begin{equation}
    \begin{split}
         g_{12,1}^{41}, g_{12,2}^{41},  g_{12,3}^{41}\sim 1/N_c
    \end{split}
\end{equation}
\item Decuplet-Decuplet (\scalebox{1}{$DDDD$})
    \begin{equation}
    \begin{split}
        g_{22,(1,2,3)}^{1(1,2)}, g_{22,(1,2,...,7)}^{3(1,2)}, g_{22,(1,2,...,7)}^{5(1,2)}\sim N_c, \quad\quad g_{22,(1,2,3)}^{2(1,2)}, g_{22,(1,2,...,7)}^{4(1,2)}\sim 1/N_c,\\[.5em]
    \end{split}
\end{equation}
\end{itemize}
Furthermore, we continue to estimate the $N_c$ order of the coupling constants by considering Eqs. (\ref{BBBBp}--\ref{DDDDp}) compared to Eqs. (\ref{4B}--\ref{4D}) one by one. For example, the case of the octet-octet transition, starting with $c_{S,BBBB}^{\alpha}$, \\
\begin{equation}\label{sscale}
    \begin{split}
        \frac{1}{3}c_{S,BBBB}^{1}-\frac{1}{3}c_{S,BBBB}^{2}+c_{S,BBBB}^{3} &=\,\,\,9q_{1}N_c+\mathcal{O}(1/N_c^2),\\
        \frac{1}{2}c_{S,BBBB}^{1}+c_{S,BBBB}^{2} &=\,\,\,\mathcal{O}(1/N_c^2),\\
        -\frac{1}{2}c_{S,BBBB}^{1}-\frac{1}{2}c_{S,BBBB}^{2} &=\,\,\,-q_{2}N_c+\mathcal{O}(1/N_c^2),
    \end{split}
\end{equation}
or
\begin{equation}
    c_{S,BBBB}^{1}\sim c_{S,BBBB}^{2} \sim c_{S,BBBB}^{3} \sim N_c,
\end{equation}
where the liner combination of  $c_{S}^{(1,2,3)}=\,\,\,g_{00,1}^{1(1,2,3)}+g_{00,1}^{3(1,2,3)}$ . Turning to $c_{T,BBBB}^{\alpha}$, the matching yields
\begin{equation}\label{tscale}
    \begin{split}
        \frac{1}{3}c_{T,BBBB}^{1}-\frac{1}{3}c_{T,BBBB}^{2}+c_{T,BBBB}^{3} &=\,\,\,-\frac{1}{4}\frac{h_{2}}{N_c}+\mathcal{O}(1/N_c^2),\\
        \frac{1}{2}c_{T,BBBB}^{1}+c_{T,BBBB}^{2} &=\,\,\,-\frac{1}{4}q_{3}N_c+\mathcal{O}(1/N_c^2),\\
        -\frac{1}{2}c_{T,BBBB}^{1}-\frac{1}{2}c_{T,BBBB}^{2} &=\,\,\,\frac{1}{9}q_{3}N_c+\frac{1}{4}\frac{h_{3}}{N_c}+\mathcal{O}(1/N_c^2),
    \end{split}
\end{equation}
or 
\begin{equation}
    \begin{split}
        c_{T,BBBB}^{1} \sim c_{T,BBBB}^{2} \sim c_{T,BBBB}^{3} \sim N_c + \mathcal{O}\big(1/N_c\big)\,,
    \end{split}
\end{equation}
where $c_{T}^{(1,2,3)}=\,\,\,-\big(g_{00,1}^{4(1,2,3)}-2g_{00,1}^{5(1,2,3)}\big)$. It is clear that at the next-leading order, $c_{S,BBBB}^{\alpha}$ is manifestly of order $N_c$, while $c_{T,BBBB}^{\alpha}$ consists of contributions at both $N_c$ and $1/N_c$. Therefore, it is necessary to examine the remaining coefficients $c_{n,BBBB}^{\alpha}$ to find the exact $N_c$ scaling behavior of each coefficient. The results obtained from these matching conditions lead to the conclusion that each coupling constant holds a specific order $N_c$, as shown in Eq.(\ref{4BNc}). \textcolor{black}{Our results in this work are consistent with the results in Ref.\cite{Vonk_2025} for the octet-octet sector.} \textcolor{black}{However, we note that the coefficients on the left-hand sides of Eq.(\ref{sscale}) and (\ref{tscale}) are different. Since the flavor structures of the potential depend on choice of the tensor identities $d$ and $f$.} Other transitions are treated through the same procedure as demonstrated above systematically. \textcolor{black}{As a result, in summary, we found the $N_c$ scalings of the coupling constants, $g_{xy,z}^{i(\alpha)}$ from the chiral Lagrangians that there are 61 couplings in order of $N_c$ and 43 terms in order of $1/N_c$. Next subsection, we will determine the $1/N_c$ sum rules in order to reduce the number of the LECs of the baryon-baryon potentials, $V_{BBBB},\,V_{BBDB},\,V_{DBDB},\,V_{BBDD},\,V_{DBDD},\,V_{DDDD}$ in Eqs.(\ref{4B}--\ref{4D}), respectively.}
\newpage
\subsection{$1/N_c$ sum rules of the LO Hartree potentials}
\indent Now, we have the LECs, $c_{n,t}^{(\alpha)}$ from the potential of the SU(3) chiral Lagrangian and couplings, $q_i$ and $h_i$ from $1/N_c$ expansion of the Hartree Hamiltonian. Matching the spin and flavor structures between them allows our $1/N_c$ operator analysis to establish the relations (sum rules) of LECs of the SU(3) baryon contact interaction up to NLO of the three-momentum expansion. \textcolor{black}{At the LO of the $1/N_c$ Hartree potential, the sum rules of LECs are given by Tables \ref{tab:4BLO}--\ref{tab:4DLO} 
for $BBBB$ $BBDB$, $DBDB$, $BBDD$, $DBDD$, and $DDDD$, respectively.}\\

\begin{table}[H]
\color{black}
\centering
    \setlength{\tabcolsep}{6pt} 
    \renewcommand{\arraystretch}{2} 
\begin{tabular}{|lll|}
\hline
\multicolumn{3}{|l|}{The 1/$N_c$ sum rules at LO for the Octet-Octet ($BBBB$) of LECs at LO $(Q/M)^0$ } \\ \hline
\multicolumn{1}{|l}{$c_{S,BBBB}^{(1)} = -2c_{S,BBBB}^{(2)}$,} & \multicolumn{1}{l}{} &  \\ 
\multicolumn{1}{|l}{$c_{T,BBBB}^{(1)} = -\frac{1}{2}c_{T,BBBB}^{(3)}$,} & \multicolumn{1}{l}{$c_{T,BBBB}^{(2)} = \frac{5}{2}c_{T,BBBB}^{(3)}$} &  \\ \hline
\multicolumn{3}{|l|}{The 1/$N_c$ sum rules at LO for the Octet-Octet ($BBBB$) of LECs at NLO $(Q/M)^2$} \\ \hline
\multicolumn{1}{|l}{$c_{5,BBBB}^{(1)} = 0$,} & \multicolumn{1}{l}{$c_{5,BBBB}^{(2)} =0$,} & $c_{5,BBBB}^{(3)} = 0$, \\
\multicolumn{1}{|l}{$c_{6,BBBB}^{(1)} = 0$,} & \multicolumn{1}{l}{$c_{6,BBBB}^{(2)} =0$,} & $c_{6,BBBB}^{(3)} = 0$, \\ 
\multicolumn{1}{|l}{$c_{7,BBBB}^{(1)} = \frac{17}{6}c_{10,BBBB}^{(3)}$,} & \multicolumn{1}{l}{$c_{7,BBBB}^{(2)} =- \frac{13}{6}c_{10,BBBB}^{(3)}$,} & $c_{7,BBBB}^{(3)} = -\frac{5}{3}c_{10,BBBB}^{(3)}$, \\ 
\multicolumn{1}{|l}{$c_{8,BBBB}^{(1)} = 0$,} & \multicolumn{1}{l}{$c_{8,BBBB}^{(2)} =0$,} & $c_{8,BBBB}^{(3)} = 0$, \\ 
\multicolumn{1}{|l}{$c_{9,BBBB}^{(1)} = 0$,} & \multicolumn{1}{l}{$c_{9,BBBB}^{(2)} =0$,} & $c_{9,BBBB}^{(3)} = 0$, \\ 
\multicolumn{1}{|l}{$c_{10,BBBB}^{(1)} = -\frac{1}{2}c_{10,BBBB}^{(3)}$,} & \multicolumn{1}{l}{$c_{10,BBBB}^{(2)} = \frac{5}{2}c_{10,BBBB}^{(3)}$} &  \\ 
\multicolumn{1}{|l}{$c_{11,BBBB}^{(1)} = 0$,} & \multicolumn{1}{l}{$c_{11,BBBB}^{(2)} =0$,} & $c_{11,BBBB}^{(3)} = 0$, \\ \hline
\end{tabular}
\caption{\color{black}The sum rules for the Octet-Octet ($BBBB$) sector under the 1/$N_c$ expansion at LO are summarized. These results demonstrate a significant reduction in independent parameters due to the LO   $1/N_c$  Hartree potential constraint, several NLO LECs vanish because the LO couplings, $q_i$ cannot account for their structures. Consequently, other LECs are represented in terms of 3 parameters as $c_{S,BBBB}^{(2)}$, $c_{T,BBBB}^{(3)}$ and, $c_{10,BBBB}^{(3)}$. A total of 23 sum rules are derived for this sector, consisting of 8  non-vanished sum rules and 15 vanished sum rules.
  }
\label{tab:4BLO}
\end{table}

\begin{table}[H]
\color{black}
\centering
    \setlength{\tabcolsep}{6pt} 
    \renewcommand{\arraystretch}{2} 
\begin{tabular}{|lll|}
\hline
\multicolumn{3}{|l|}{The 1/$N_c$ sum rules at LO for the Mixed Octet-Decuplet ($BBDB$) of LECs at LO $(Q/M)^0$ } \\ \hline
\multicolumn{1}{|l}{$c_{2,BBDB}^{(1)} = \frac{15}{4}c_{T,BBBB}^{(3)}$,} & \multicolumn{1}{l}{$c_{2,BBDB}^{(2)} = -\frac{3}{4}c_{T,BBBB}^{(3)}$} &  \\ \hline
\multicolumn{3}{|l|}{The 1/$N_c$ sum rules at LO for the Mixed Octet-Decuplet ($BBDB$) of LECs at NLO $(Q/M)^2$} \\ \hline
\multicolumn{1}{|l}{$c_{7,BBDB}^{(1)} = -\frac{5}{2}c_{10,BBBB}^{(3)}$,} & \multicolumn{1}{l}{$c_{7,BBDB}^{(2)} =\frac{3}{2}c_{10,BBBB}^{(3)}$,} &  \\ 
\multicolumn{1}{|l}{$c_{8,BBDB}^{(1)} = 0$,} & \multicolumn{1}{l}{$c_{8,BBDB}^{(2)} = 0$,} &  \\ 
\multicolumn{1}{|l}{$c_{9,BBDB}^{(1)} = 0$,} & \multicolumn{1}{l}{$c_{9,BBDB}^{(2)} = 0$,} &  \\ 
\multicolumn{1}{|l}{$c_{10,BBDB}^{(1)} = 0$,} & \multicolumn{1}{l}{$c_{10,BBDB}^{(2)} = 0$,} &  \\ 
\multicolumn{1}{|l}{$c_{11,BBDB}^{(1)}=\frac{15}{4}c_{10,BBBB}^{(3)}$,} & \multicolumn{1}{l}{$c_{11,BBDB}^{(2)}=-\frac{3}{4}c_{10,BBBB}^{(3)}$} &  \\ \hline
\end{tabular}
\caption{\color{black}The sum rules for the Mixed Octet-Decuplet ($BBDB$) sector under  the 1/$N_c$ expansion at LO are summarized. These results demonstrate a significant reduction in independent parameters due to the LO of the $1/N_c$ Hartree potential, several LECs at $\mathcal{O}\big((Q/M)^2\big)$ vanish because the LO couplings in the $1/N_c$, $q_i$ cannot account for their structures. Consequently, other LECs are constrained to be proportional to 2  parameters i.e., $c_{T,BBBB}^{(3)}$, $c_{10,BBBB}^{(3)}$. A total of 12 sum rules are derived for this transition, consisting of 6 non-vanished sum rules and 6 vanished sum rules.}
\label{tab:3b1dLO}
\end{table}

\begin{table}[H]
\color{black}
\centering
    \setlength{\tabcolsep}{6pt} 
    \renewcommand{\arraystretch}{2} 
\begin{tabular}{|lll|}
\hline
\multicolumn{3}{|l|}{The 1/$N_c$ sum rules at LO for the Mixed Octet-Decuplet ($DBDB$) of LECs at LO $(Q/M)^0$} \\ \hline
\multicolumn{1}{|l}{$c_{1,DBDB}^{(1)} = -c_{S,BBBB}^{(2)}+c_{S,BBBB}^{(3)}$,} & \multicolumn{1}{l}{$c_{1,DBDB}^{(+)} = \frac{3}{4}c_{S,BBBB}^{(2)}$,} & $c_{1,DBDB}^{(3)} =-\frac{3}{4}c_{S,BBBB}^{(2)}$ \\
\multicolumn{1}{|l}{$c_{2,DBDB}^{(1)} = -\frac{27}{8}c_{T,BBBB}^{(3)}$,} & \multicolumn{1}{l}{$c_{2,DBDB}^{(+)} = \frac{45}{16}c_{T,BBBB}^{(3)}$,} & $c_{2,DBDB}^{(3)} =\frac{117}{16}c_{T,BBBB}^{(3)}$ \\ \hline
\multicolumn{3}{|l|}{The 1/$N_c$ sum rules at LO for the Mixed Octet-Decuplet ($DBDB$) of LECs at NLO $(Q/M)^2$} \\ \hline
\multicolumn{1}{|l}{$c_{5,DBDB}^{(1)} = 0$,} & \multicolumn{1}{l}{$c_{5,DBDB}^{(+)} = 0 $,} & $c_{5,DBDB}^{(3)} = 0 $, \\ 
\multicolumn{1}{|l}{$c_{6,DBDB}^{(1)} = 0$,} & \multicolumn{1}{l}{$c_{6,DBDB}^{(+)} = 0 $,} & $c_{6,DBDB}^{(3)} = 0 $, \\
\multicolumn{1}{|l}{$c_{7,DBDB}^{(1)} = \frac{9}{8}c_{10,BBBB}^{(3)}$,} & \multicolumn{1}{l}{$c_{7,DBDB}^{(+)} =- \frac{15}{16}c_{10,BBBB}^{(3)} $,} & $c_{7,DBDB}^{(3)} = - \frac{39}{16}c_{10,BBBB}^{(3)} $ \\
\multicolumn{1}{|l}{$c_{8,DBDB}^{(1)} = 0$,} & \multicolumn{1}{l}{$c_{8,DBDB}^{(+)} =0 $,} & $c_{8,DBDB}^{(3)} = 0 $, \\ 
\multicolumn{1}{|l}{$c_{9,DBDB}^{(1)} = 0$,} & \multicolumn{1}{l}{$c_{9,DBDB}^{(+)} = 0 $,} & $c_{9,DBDB}^{(3)} = 0 $, \\
\multicolumn{1}{|l}{$c_{10,DBDB}^{(1)} = 0$,} & \multicolumn{1}{l}{$c_{10,DBDB}^{(+)} =-\frac{9}{4}c_{10,BBBB}^{(3)} $,} & $c_{10,DBDB}^{(3)} = \frac{9}{4}c_{10,BBBB}^{(3)} $, \\
\multicolumn{1}{|l}{$c_{11,DBDB}^{(1)} = 0$,} & \multicolumn{1}{l}{$c_{11,DBDB}^{(+)} =0 $,} & $c_{11,DBDB}^{(3)} = 0 $ \\ \hline
\end{tabular}
\caption{\color{black}The sum rules for the Mixed Octet-Decuplet ($DBDB$) sector under the 1/$N_c$ expansion at LO  are summarized. These relations  demonstrate a significant reduction in independent parameters due to the LO of the $1/N_c$  Hartree potential, some LECs at $\mathcal{O}\big((Q/M)^2\big)$ vanish because the LO couplings, $q_i$ cannot account for their structures. In contrast, remaining LECs are proportional to 4 octet-octet parameters as $c_{S,BBBB}^{(2)}$, $c_{S,BBBB}^{(3)}$, $c_{T,BBBB}^{(3)}$, $c_{10,BBBB}^{(3)}$, and 9 free parameters from $DBDB$ sector as $c^{(2)}_{i,DBDB}$, $i=1,2,5,6,7,8,9,10,11$. A total of 27 sum rules are derived for this sector, consisting of 11 non-vanished sum rules and 16 vanished sum rules. In addition, we defined $c^{(+)}_{i,DBDB} = c^{(2)}_{i,DBDB} + c^{(4)}_{i,DBDB}$, $i=1,2,5,6,7,8,9,10,11$. 
}
\label{tab:dbdbLO}
\end{table}

\begin{table}[H]
\color{black}
\centering
    \setlength{\tabcolsep}{6pt} 
    \renewcommand{\arraystretch}{2} 
\begin{tabular}{|lll|}
\hline
\multicolumn{3}{|l|}{The 1/$N_c$ sum rules at LO for the Mixed Octet-Decuplet ($BBDD$) of LECs at LO $(Q/M)^0$} \\ \hline
\multicolumn{1}{|l}{$c_{2,BBDD}^{(1)} = 0$,} & \multicolumn{1}{l}{$c_{3,BBDD}^{(1)} = 0$} & \\ \hline
\multicolumn{3}{|l|}{The 1/$N_c$ sum rules at LO for the Mixed Octet-Decuplet ($BBDD$) of LECs at NLO $(Q/M)^2$} \\ \hline
\multicolumn{1}{|l}{$c_{7,BBDD}^{(1)} = \frac{3}{4}c_{10,BBBB}^{(3)}$,} & \multicolumn{1}{l}{$c_{8,BBDD}^{(1)} =0$,} &  $c_{10,BBDD}^{(1)}= \frac{9}{4}c_{10,BBBB}^{(3)}$,\\ 
\multicolumn{1}{|l}{$c_{11,BBDD}^{(1)} = 0$,} & \multicolumn{1}{l}{$c_{12,BBDD}^{(1)} =0 $,} & $c_{13,BBDD}^{(1)} = 0 $, \\
\multicolumn{1}{|l}{$c_{14,DBDB}^{(1)} = 0$,} & \multicolumn{1}{l}{$c_{15,DBDB}^{(1)} =0 $,} & \\ \hline
\end{tabular}
\caption{\color{black}The sum rules for the Mixed Octet-Decuplet ($BBDD$) sector under the 1/$N_c$ expansion at LO are summarized. These relations demonstrate a significant reduction in independent parameters due to the LO of the $1/N_c$ Hartree potential, several LECs vanish because the LO couplings, $q_i$ cannot account for their structures. However, remaining LECs are proportional to 1 octet-octet parameter, $c_{10,BBBB}^{(3)}$. A total of 10 sum rules are derived for this transition, consisting of 2 non-vanished sum rules and 8 vanished sum rules .}
\label{tab:BBDDLO}
\end{table}

\begin{table}[H]
\color{black}
\centering
    \setlength{\tabcolsep}{6pt} 
    \renewcommand{\arraystretch}{2} 
\begin{tabular}{|lll|}
\hline
\multicolumn{3}{|l|}{The 1/$N_c$ sum rules at LO for the Mixed Octet-Decuplet ($DBDD$) of LECs at LO $(Q/M)^0$} \\ \hline
\multicolumn{1}{|l}{$c_{2,DBDD}^{(1)} = \frac{27}{4}c_{T,BBBB}^{(3)}$,} & \multicolumn{1}{l}{$c_{3,DBDD}^{(1)} = 0$} & \\ \hline
\multicolumn{3}{|l|}{The 1/$N_c$ sum rules at LO for the Mixed Octet-Decuplet ($DBDD$) of LECs at NLO $(Q/M)^2$} \\ \hline
\multicolumn{1}{|l}{$c_{7,DBDD}^{(1)} = -\frac{9}{4}c_{10,BBBB}^{(3)}$,} & \multicolumn{1}{l}{$c_{8,DBDD}^{(1)} =0$,} &  $c_{9,DBDD}^{(1)}= 0$,\\ 
\multicolumn{1}{|l}{$c_{10,DBDD}^{(1)} = \frac{27}{4}c_{10,BBBB}^{(3)}$,} & \multicolumn{1}{l}{$c_{11,DBDD}^{(1)} =0 $,} & $c_{12,BBDD}^{(1)} = 0 $, \\
\multicolumn{1}{|l}{$c_{13,DBDB}^{(1)} = 0$,} & \multicolumn{1}{l}{$c_{14,DBDB}^{(1)} =0 $,} & $c_{15,DBDB}^{(1)} =0 $ \\ \hline
\end{tabular}
\caption{\color{black}The sum rules for the Mixed Octet-Decuplet ($DBDD$) sector under the 1/$N_c$ expansion at LO are summarized. These relations  demonstrate a significant reduction in independent parameters due to the LO  $1/N_c$  Hartree potential constraint, several LECs vanish because the LO couplings ($q_i$) cannot account for their structures. However, remaining LECs are proportional to 2  octet-octet parameters, $c_{T,BBBB}^{(3)}$ and $c_{10,BBBB}^{(3)}$. A total of 11 sum rules are derived for this transition, consisting of 3 non-vanished sum rules and 8 vanished sum rules.}
\label{tab:DBDDLO}
\end{table}

\begin{table}[H]
\color{black}
\centering
    \setlength{\tabcolsep}{6pt} 
    \renewcommand{\arraystretch}{2} 
\begin{tabular}{|llll|}
\hline
\multicolumn{4}{|l|}{The 1/$N_c$ sum rules at LO for the Decuplet-Decuplet ($DDDD$) of LECs at LO $(Q/M)^0$} \\ \hline
\multicolumn{1}{|l}{$c_{1,DDDD}^{(1)} = -\frac{13}{16}c_{S,BBBB}^{(2)}+c_{S,BBBB}^{(3)}$,} & \multicolumn{3}{l|}{$c_{1,DDDD}^{(2)} = -\frac{9}{16}c_{S,BBBB}^{(2)}$,}   \\
\multicolumn{1}{|l}{$c_{2,DDDD}^{(1)} = -\frac{27}{16}c_{T,BBBB}^{(3)}$,} & \multicolumn{3}{l|}{$c_{2,DDDD}^{(2)} = \frac{81}{16}c_{T,BBBB}^{(3)}$,}  \\
\multicolumn{1}{|l}{$c_{3,DDDD}^{(1)} = 0$,} & \multicolumn{3}{l|}{$c_{3,DDDD}^{(2)} = 0$,}   \\
\multicolumn{1}{|l}{$c_{4,DDDD}^{(1)} = 0$,} & \multicolumn{3}{l|}{$c_{4,DDDD}^{(2)} = 0$,}   \\ \hline
\multicolumn{4}{|l|}{The 1/$N_c$ sum rules at LO for the Decuplet-Decuplet ($DDDD$) of LECs at NLO $(Q/M)^2$} \\ \hline
\multicolumn{1}{|l}{$c_{5,DDDD}^{(1)} = 0$,} & \multicolumn{1}{l}{$c_{5,DDDD}^{(2)} = 0 $,} & \multicolumn{1}{l}{$c_{13,DDDD}^{(1)} = 0$,} & \multicolumn{1}{l|}{$c_{13,DDDD}^{(2)} =0 $,}  \\ 
\multicolumn{1}{|l}{$c_{6,DDDD}^{(1)} = 0$,} & \multicolumn{1}{l}{$c_{6,DDDD}^{(2)} = 0 $,} & \multicolumn{1}{l}{$c_{14,DDDD}^{(1)} = 0$,} & \multicolumn{1}{l|}{$c_{14,DDDD}^{(2)} =0 $,} \\
\multicolumn{1}{|l}{$c_{7,DDDD}^{(1)} = \frac{9}{8}c_{10,BBBB}^{(3)}$,} & \multicolumn{1}{l}{$c_{7,DDDD}^{(2)} =-\frac{27}{8}c_{10,BBBB}^{(3)} $,} & \multicolumn{1}{l}{$c_{15,DDDD}^{(1)} = 0$,} & \multicolumn{1}{l|}{$c_{15,DDDD}^{(2)} =0 $,} \\
\multicolumn{1}{|l}{$c_{8,DDDD}^{(1)} = 0$,} & \multicolumn{1}{l}{$c_{8,DDDD}^{(2)} =0 $,} & \multicolumn{1}{l}{$c_{16,DDDD}^{(1)} = 0$,} & \multicolumn{1}{l|}{$c_{16,DDDD}^{(2)} =0 $,} \\ 
\multicolumn{1}{|l}{$c_{9,DDDD}^{(1)} = 0$,} & \multicolumn{1}{l}{$c_{9,DDDD}^{(2)} =0$,} & \multicolumn{1}{l}{$c_{17,DDDD}^{(1)} = 0$,} & \multicolumn{1}{l|}{$c_{17,DDDD}^{(2)} =0 $,} \\
\multicolumn{1}{|l}{$c_{10,DDDD}^{(1)} = -\frac{27}{8}c_{10,BBBB}^{(3)}$,} & \multicolumn{1}{l}{$c_{10,DDDD}^{(2)} =\frac{81}{8}c_{10,BBBB}^{(3)}$,} & \multicolumn{1}{l}{$c_{18,DDDD}^{(1)} = 0$,} & \multicolumn{1}{l|}{$c_{18,DDDD}^{(2)} =0 $,} \\
\multicolumn{1}{|l}{$c_{11,DDDD}^{(1)} = 0$,} & \multicolumn{1}{l}{$c_{11,DDDD}^{(2)} =0 $,} & \multicolumn{1}{l}{$c_{19,DDDD}^{(1)} = 0$,} & \multicolumn{1}{l|}{$c_{19,DDDD}^{(2)} =0 $,} \\ 
\multicolumn{1}{|l}{$c_{12,DDDD}^{(1)} = 0$,} & \multicolumn{1}{l}{$c_{12,DDDD}^{(2)} =0 $,} & & \\  \hline
\end{tabular}
\caption{\color{black}The sum rules for the Decuplet-Decuplet ($DDDD$) sector under the 1/$N_c$ expansion at LO are summarized. These results demonstrate a significant reduction in independent parameters due to the LO  $1/N_c$  Hartree potential constraint, several  LECs vanish because the LO couplings ($q_i$) cannot account for their structure. Consequently, other LECs are proportional to 4 free parameters, $c_{S,BBBB}^{(2)}$, $c_{S,BBBB}^{(3)}$, $c_{T,BBBB}^{(3)}$, $c_{10,BBBB}^{(3)}$. A total of 38 sum rules are derived for this sector, consisting of 8  non-vanished sum rules and 30 vanished sum rules.}
\label{tab:4DLO}
\end{table}
\newpage
Here, by matching the spin and Gell-Mann matrices with their algebraic properties between the chiral potentials and the matrix elements of the $1/N_c$ Hartree potentials. \textcolor{black}{This leads to systematic reduction of the number of independent coupling constants across various sectors. In this matching conditions at the LO of the $1/N_c$ Hartree Hamiltonian, a total of 121 sum rules are derived, given as follows: 23 ($BB \rightarrow BB$), 12 ($BB \rightarrow DB$), 27 ($DB \rightarrow DB$), 10 ($BB \rightarrow DD$), 11 ($DB \rightarrow DD$), and 38 ($DD \rightarrow DD$). Subsequently, the number of LECs is reduced from 134 to 13, all non-vanished LECs are expressed in terms of 4 octet-octet LECs i.e., $c_{S,BBBB}^{(2)},\,c_{S,BBBB}^{(3)},\,c_{T,BBBB}^{(3)},\,c_{10,BBBB}^{(3)}$, see details of the sum rules in Tables \ref{tab:4BLO} to \ref{tab:4DLO}, respectively. Except the LECs for $DBDB$ sector, we have additional 9 free parameter as $c_{i,DBDB}^{(2)}$, $i=1,2,5,6,7,8,9,10,11$.} These LECs are free parameters for describing two-body baryon scattering. Note that the LECs of the $BBBB$ sector can describe across all baryon-baryon interaction transitions, since the couplings, $q_i$ in the $1/N_c$ Hartree potential assign the same coefficients to all sectors as shown in Appendix \ref{A-2-HHPotential}. Moreover, even though several LECs, as indicated in Table \ref{tab:4BLO}--\ref{tab:4DLO}, are found to be zero, these values do not represent an actual null. Instead, they suggest the possibility that a sum rule could be formed upon them, which the LO condition alone are insufficient. This leads to further investigation by including the NNLO $1/N_c$ Hartree potentials, we anticipate that the new sum rules will provide insight to understand the behavior of LECs.

\subsection{$1/N_c$ sum rules up to the NNLO Hartree potentials}
Subsequently, we further include NNLO terms with the couplings $h_i$, $i=1,\cdots 11$ from the $1/N_c$  Hartree potential, then derive the sum rules of LECs again by matching the spin and flavor structures of both potentials. \textcolor{black}{In the case of the $1/N_c$ Hartree potential up to NNLO, The sum rules are presented in Tables, \ref{tab:4BNNLO}--\ref{tab:4DNNLO} as follows,}

\begin{table}[H]
\color{black}
\centering
    \setlength{\tabcolsep}{6pt} 
    \renewcommand{\arraystretch}{2} 
\begin{tabular}{|ll|}
\hline
\multicolumn{2}{|l|}{The 1/$N_c$ sum rules at NNLO for the Octet-Octet ($BBBB$) of LECs at LO $(Q/M)^0$} \\ \hline
\multicolumn{1}{|l}{$c_{S,BBBB}^{(1)} = -2c_{S,BBBB}^{(2)}$,} &   \\ \hline
\multicolumn{2}{|l|}{The 1/$N_c$ sum rules at NNLO for the Octet-Octet ($BBBB$) of LECs at NLO $(Q/M)^2$} \\ \hline
\multicolumn{1}{|l}{$c_{5,BBBB}^{(1)} = 0$, \hspace{2.3cm} $c_{5,BBBB}^{(2)} =0$} & \multicolumn{1}{l|}{$c_{5,BBBB}^{(3)} = 0$,} \\
\multicolumn{1}{|l}{$c_{6,BBBB}^{(1)} = -2c_{6,BBBB}^{(2)}$,} & \\ 
\multicolumn{1}{|l}{$c_{7,BBBB}^{(1)} = \frac{1}{27}(7c_{10,BBBB}^{(1)}+32c_{10,BBBB}^{(2)})$,} & \multicolumn{1}{l|}{$c_{7,BBBB}^{(2)} = -\frac{1}{27}(8c_{10,BBBB}^{(1)}+25c_{10,BBBB}^{(2)})$,}   \\ 
\multicolumn{1}{|l}{$c_{7,BBBB}^{(3)} = -\frac{1}{27}(8c_{10,BBBB}^{(1)}+16c_{10,BBBB}^{(2)}+9c_{10,BBBB}^{(3)})$,} & \multicolumn{1}{l|}{}   \\ 
\multicolumn{1}{|l}{$c_{8,BBBB}^{(1)} = -\frac{1}{2}c_{8,BBBB}^{(3)}$,} & \multicolumn{1}{l|}{$c_{8,BBBB}^{(2)} =\frac{5}{2}c_{8,BBBB}^{(3)}$,}  \\ 
\multicolumn{1}{|l}{$c_{9,BBBB}^{(1)} = -2c_{9,BBBB}^{(2)}$,} & \multicolumn{1}{l|}{}   \\ 

\multicolumn{1}{|l}{$c_{11,BBBB}^{(1)} = -\frac{1}{2}c_{11,BBBB}^{(3)}$,} & \multicolumn{1}{l|}{$c_{11,BBBB}^{(2)} =\frac{5}{2}c_{11,BBBB}^{(3)}$}   \\ \hline
\end{tabular}
\caption{\color{black}The sum rules for the Octet-Octet ($BBBB$) sector under $1/N_c$ expansion at NNLO are summarized. These relations show a significant reduction in independent parameters due to the LO and NNLO  $1/N_c$  Hartree potential constraints. Only $c_{5,BBBB}^{(\alpha)}$ vanish because the couplings up to NNLO ($q_i$ and $h_i$) are insufficient to account for their structure. However, remaining LECs are proportional to 8  parameters, $c_{S,BBBB}^{(2)}$, $c_{6,BBBB}^{(2)}$, $c_{8,BBBB}^{(3)}$, $c_{9,BBBB}^{(2)}$, $c_{10,BBBB}^{(1,2,3)}$,  $c_{11,BBBB}^{(3)}$. A total of 13 sum rules are derived for this sector, consisting of 10 non-vanished sum rules and 3 vanished sum rules.}
\label{tab:4BNNLO}
\end{table}

\begin{table}[H]
\color{black}
\centering
    \setlength{\tabcolsep}{6pt} 
    \renewcommand{\arraystretch}{2} 
\begin{tabular}{|lll|}
\hline
\multicolumn{3}{|l|}{The 1/$N_c$ sum rules at NNLO for the Mixed Octet-Decuplet ($BBDB$) of LECs at LO $(Q/M)^0$} \\ \hline
\multicolumn{1}{|l}{$c_{2,BBDB}^{(1)} = \frac{5}{6}(c_{T,BBBB}^{(1)}+2c_{T,BBBB}^{(2)})$,} & \multicolumn{1}{l}{$c_{2,BBDB}^{(2)} = -\frac{1}{6}(c_{T,BBBB}^{(1)}+2c_{T,BBBB}^{(2)})$} &  \\ \hline
\multicolumn{3}{|l|}{The 1/$N_c$ sum rules at NNLO for the Mixed Octet-Decuplet ($BBDB$) of LECs at NLO $(Q/M)^2$} \\ \hline
\multicolumn{1}{|l}{$c_{7,BBDB}^{(1)} = \frac{5}{3}c_{11,BBDB}^{(2)}$,} & \multicolumn{1}{l}{$c_{7,BBDB}^{(2)} =- \frac{1}{3}c_{11,BBDB}^{(2)}$,} &  \\ 
\multicolumn{1}{|l}{$c_{8,BBDB}^{(1)} = \frac{15}{4}c_{8,BBBB}^{(3)}$,} & \multicolumn{1}{l}{$c_{8,BBDB}^{(2)} = -\frac{3}{4}c_{8,BBBB}^{(3)}$,} &  \\ 
\multicolumn{1}{|l}{$c_{9,BBDB}^{(1)} = -\frac{3}{4}c_{9,BBBB}^{(2)}$,} & \multicolumn{1}{l}{$c_{9,BBDB}^{(2)}=-\frac{3}{4}c_{9,BBBB}^{(2)}$} &  \\ 
\multicolumn{1}{|l}{$c_{10,BBDB}^{(1)} = \frac{15}{4}c_{11,BBBB}^{(3)}$,} & \multicolumn{1}{l}{$c_{10,BBDB}^{(2)} = -\frac{3}{4}c_{11,BBBB}^{(3)}$,} &  \\ 
\multicolumn{1}{|l}{$c_{11,BBDB}^{(1)}=-5c_{11,BBDB}^{(2)}$} & \multicolumn{1}{l}{} &  \\ \hline
\end{tabular}
\caption{\color{black}The sum rules for the Mixed Octet-Decuplet ($BBDB$) sector under $1/N_c$ expansion at NNLO are summarized. These relations show a significant reduction in independent parameters due to the LO and NNLO   $1/N_c$  Hartree potential constraint. Specifically, all LECs are proportional to 5 octet-octet parameters, $c_{T,BBBB}^{(1,2)}$, $c_{8,BBBB}^{(3)}$, $c_{9,BBBB}^{(2)}$, $c_{11,BBBB}^{(3)}$, and 1 mixed octet-decuplet parameter, $c_{11,BBDB}^{(2)}$. A total of 11 sum rules are derived for this transition}
\label{tab:BBDBNNLO}
\end{table}
\vspace{-1.5em}
\begin{table}[H]
\color{black}
\centering
    \setlength{\tabcolsep}{3pt} 
    \renewcommand{\arraystretch}{2} 
\begin{tabular}{|lll|}
\hline
\multicolumn{3}{|l|}{The 1/$N_c$ sum rules at NNLO for the Mixed Octet-Decuplet ($DBDB$) of LECs at LO $(Q/M)^0$} \\ \hline
\multicolumn{1}{|l}{$c_{1,DBDB}^{(1)} = -c_{S,BBBB}^{(2)}+c_{S,BBBB}^{(3)}$,} & \multicolumn{1}{l}{$c_{1,DBDB}^{(3)} = -\frac{3}{4}c_{S,BBBB}^{(2)}$,} & $c_{1,DBDB}^{(+)} =+\frac{3}{4}c_{S,BBBB}^{(2)}$ \\
\multicolumn{1}{|l}{$c_{2,DBDB}^{(1)} = \frac{1}{4}c_{T,BBBB}^{(1)}-\frac{5}{2}c_{T,BBBB}^{(2)}+3c_{T,BBBB}^{(3)}$,} & \multicolumn{1}{l}{$c_{2,DBDB}^{(3)} = \frac{23}{8}c_{T,BBBB}^{(1)}+\frac{7}{2}c_{T,BBBB}^{(2)}$,} &  \\ 
\multicolumn{1}{|l}{$c_{2,DBDB}^{(+)} =-\frac{5}{8}c_{T,BBBB}^{(1)}-c_{T,BBBB}^{(2)}$,} & \multicolumn{1}{l}{} &  \\\hline
\multicolumn{3}{|l|}{The 1/$N_c$ sum rules at NNLO for the Mixed Octet-Decuplet ($DBDB$) of LECs at NLO $(Q/M)^2$} \\ \hline
\multicolumn{1}{|l}{$c_{5,DBDB}^{(1)} = 0$,} & \multicolumn{1}{l}{$c_{5,DBDB}^{(+)} =0 $,} & $c_{5,DBDB}^{(3)} = 0 $, \\ 
\multicolumn{1}{|l}{$c_{6,DBDB}^{(1)} = -c_{6,BBBB}^{(2)}-c_{6,BBBB}^{(3)}$,} & \multicolumn{1}{l}{$c_{6,DBDB}^{(3)} =-\frac{3}{4}c_{6,BBBB}^{(2)} $,} & $c_{6,DBDB}^{(+)} = \frac{3}{4}c_{6,BBBB}^{(2)} $, \\
\multicolumn{1}{|l}{$c_{7,DBDB}^{(1)} = -\frac{1}{12}c_{10,BBBB}^{(1)}+\frac{5}{6}c_{10,BBBB}^{(2)}-c_{10,BBBB}^{(3)}$,} & \multicolumn{2}{l|}{$c_{7,DBDB}^{(3)} =-\frac{23}{24}c_{10,BBBB}^{(1)}-\frac{7}{6}c_{10,BBBB}^{(2)}$,} \\
\multicolumn{1}{|l}{$c_{7,DBDB}^{(+)} = \frac{5}{24}c_{10,BBBB}^{(1)}-\frac{1}{3}c_{10,BBBB}^{(2)}$,} & \multicolumn{2}{l|}{} \\
\multicolumn{1}{|l}{$c_{8,DBDB}^{(1)} = -\frac{27}{8}c_{8,BBBB}^{(3)}$,} & \multicolumn{1}{l}{$c_{8,DBDB}^{(3)} =\frac{117}{16}c_{8,BBBB}^{(3)} $,} & $c_{8,DBDB}^{(+)} = \frac{45}{16}c_{8,BBBB}^{(3)} $, \\ 
\multicolumn{1}{|l}{$c_{9,DBDB}^{(1)} =\frac{1}{8}c_{9,BBBB}^{(2)}+c_{9,BBBB}^{(3)}$,} & \multicolumn{1}{l}{$c_{9,DBDB}^{(3)} = -\frac{57}{16}c_{9,BBBB}^{(2)}$,} & $c_{9,DBDB}^{(+)} =\frac{3}{16}c_{9,BBBB}^{(2)}$ \\
\multicolumn{1}{|l}{$c_{10,DBDB}^{(1)} = \frac{1}{2}c_{10,BBBB}^{(1)}-\frac{1}{2}c_{10,BBBB}^{(2)}+\frac{3}{2}c_{10,BBBB}^{(3)}$,} & \multicolumn{2}{l|}{$c_{10,DBDB}^{(3)} =\frac{5}{2}c_{10,BBBB}^{(1)}+\frac{1}{2}c_{10,BBBB}^{(2)}+\frac{9}{4}c_{10,BBBB}^{(3)}$,} \\
\multicolumn{2}{|l}{$c_{10,DBDB}^{(+)} =-c_{10,BBBB}^{(1)}-2c_{10,BBBB}^{(2)}+\frac{9}{4}c_{10,BBBB}^{(3)}$,} &  \\
\multicolumn{1}{|l}{$c_{11,DBDB}^{(1)} = -9c_{11,BBBB}^{(3)}$,} & \multicolumn{1}{l}{$c_{11,DBDB}^{(3)} =\frac{63}{4}c_{11,BBBB}^{(3)} $,} & $c_{11,DBDB}^{(+)} = \frac{45}{4}c_{11,BBBB}^{(3)} $ \\ \hline
\end{tabular}
\caption{\color{black}The sum rules for the Mixed Octet-Decuplet ($DBDB$) sector under under $1/N_c$ expansion at NNLO are summarized. These relations show a significant reduction in independent parameters due to the LO and NNLO  $1/N_c$  Hartree potential constraint. The non-vanished LECs are expressed as proportional to 14 octet-octet parameters, $c_{S,BBBB}^{(2,3)}$, $c_{T,BBBB}^{(1,2,3)}$, $c_{6,BBBB}^{(2,3)}$, $c_{8,BBBB}^{(3)}$, $c_{9,BBBB}^{(2,3)}$, $c_{10,BBBB}^{(1,2,3)}$, $c_{11,BBBB}^{(3)}$ and 9 free parameters from $DBDB$ sector as $c^{(2)}_{i,DBDB}$, $i=1,2,5,6,7,8,9,10,11$. A total of 27 sum rules are derived for this transition, consisting of 24 non-vanished sum rules and 3 vanished sum rules. In addition, we defined $c^{(+)}_{i,DBDB} = c^{(2)}_{i,DBDB} + c^{(4)}_{i,DBDB}$, $i=1,2,5,6,7,8,9,10,11$.}
\label{tab:DBDBNNLO}
\end{table}
\newpage
\begin{table}[H]
\color{black}
\centering
    \setlength{\tabcolsep}{6pt} 
    \renewcommand{\arraystretch}{2} 
\begin{tabular}{|lll|}
\hline
\multicolumn{3}{|l|}{The 1/$N_c$ sum rules at NNLO for the Mixed Octet-Decuplet ($BBDD$) of LECs at LO $(Q/M)^0$} \\ \hline
\multicolumn{1}{|l}{$c_{2,BBDD}^{(1)} = -\frac{5}{18}c_{T,BBBB}^{(1)}-\frac{1}{18}c_{T,BBBB}^{(2)}$,} & \multicolumn{1}{l}{$c_{3,BBDD}^{(1)} = 0$} & \\ \hline
\multicolumn{3}{|l|}{The 1/$N_c$ sum rules at NNLO for the Mixed Octet-Decuplet ($BBDD$) of LECs at NLO $(Q/M)^2$} \\ \hline
\multicolumn{1}{|l}{$c_{7,BBDD}^{(1)} = \frac{1}{6}c_{10,BBBB}^{(1)}+\frac{1}{3}c_{10,BBBB}^{(2)}$,} & \multicolumn{1}{l}{$c_{8,BBDD}^{(1)} =-\frac{9}{4}c_{8,BBBB}^{(3)}$,} &  $c_{10,BBDD}^{(1)}= \frac{1}{2}c_{10,BBBB}^{(1)}+c_{10,BBBB}^{(2)}$,\\ 
\multicolumn{1}{|l}{$c_{11,BBDD}^{(1)} = \frac{9}{4}c_{11,BBBB}^{(3)}$,} & \multicolumn{1}{l}{$c_{12,BBDD}^{(1)} =0 $,} & $c_{13,BBDD}^{(1)} = 0 $, \\
\multicolumn{1}{|l}{$c_{14,DBDB}^{(1)} = 0$,} & \multicolumn{1}{l}{$c_{15,DBDB}^{(1)} =0 $,} & \\ \hline
\end{tabular}
\caption{\color{black}The sum rules for the Mixed Octet-Decuplet ($BBDD$) sector under $1/N_c$ expansion at NNLO are summarized. These relations provide a significant reduction in independent parameters due to the LO and NNLO  $1/N_c$  Hartree potential constraint. The LECs are proportional to 6 octet-octet parameters, $c_{T,BBBB}^{(1,2)}$, $c_{8,BBBB}^{(3)}$, $c_{10,BBBB}^{(1,2)}$, $c_{11,BBBB}^{(3)}$. Several LECs vanish since the couplings up to NNLO ($q_i$ and $h_i$) are insufficient to account for their structure. A total of 10 sum rules are derived for this transition, consisting of non-vanished 5 sum rules and 5 vanished sum rules.}
\label{tab:BBDDNNLO}
\end{table}

\begin{table}[H]
\color{black}
\centering
    \setlength{\tabcolsep}{6pt} 
    \renewcommand{\arraystretch}{2} 
\begin{tabular}{|lll|}
\hline
\multicolumn{3}{|l|}{The 1/$N_c$ sum rules at NNLO for the Mixed Octet-Decuplet ($DBDD$) of LECs at LO $(Q/M)^0$} \\ \hline
\multicolumn{1}{|l}{$c_{2,DBDD}^{(1)} = \frac{3}{2}c_{T,BBBB}^{(1)}+3c_{T,BBBB}^{(2)}$,} & \multicolumn{1}{l}{$c_{3,DBDD}^{(1)} = 0$} & \\ \hline
\multicolumn{3}{|l|}{The 1/$N_c$ sum rules at NNLO for the Mixed Octet-Decuplet ($DBDD$) of LECs at NLO $(Q/M)^2$} \\ \hline
\multicolumn{1}{|l}{$c_{7,DBDD}^{(1)} = -\frac{1}{2}c_{10,BBBB}^{(1)}-c_{10,BBBB}^{(2)}$,} & \multicolumn{1}{l}{$c_{8,DBDD}^{(1)} =\frac{9}{4}c_{8,BBBB}^{(3)}-\frac{3}{2}c_{11,BBBB}^{(3)}$,} &  $c_{9,DBDD}^{(1)}= -\frac{9}{4}c_{9,BBBB}^{(2)}$,\\ 
\multicolumn{1}{|l}{$c_{10,DBDD}^{(1)} = \frac{3}{2}c_{10,BBBB}^{(1)}+3c_{10,BBBB}^{(2)}$,} & \multicolumn{1}{l}{$c_{11,DBDD}^{(1)} =\frac{27}{4}c_{11,BBBB}^{(3)} $,} & $c_{12,BBDD}^{(1)} = 0 $, \\
\multicolumn{1}{|l}{$c_{13,DBDB}^{(1)} = 0$,} & \multicolumn{1}{l}{$c_{14,DBDB}^{(1)} =0 $,} & $c_{15,DBDB}^{(1)} =0 $ \\ \hline
\end{tabular}
\caption{\color{black}The sum rules for the Mixed Octet-Decuplet ($DBDD$) sector under $1/N_c$ expansion at NNLO are summarized. These relations provide a significant reduction in independent parameters due to the LO and NNLO  $1/N_c$  Hartree potential constraint. The LECs are proportional to 7 octet-octet parameters, $c_{T,BBBB}^{(1,2)}$, $c_{8,BBBB}^{(3)}$, $c_{9,BBBB}^{(2)}$, $c_{10,BBBB}^{(1,2)}$, $c_{11,BBBB}^{(3)}$. Several LECs vanish since the couplings up to NNLO ($q_i$ and $h_i$) are insufficient to account for their structure. A total of 11 sum rules are derived for this transition, consisting of 6 non-vanished sum rules and 5 vanished sum rules that lead to zero.}
\label{tab:DBDDNNLO}
\end{table}

\begin{table}[H]
\color{black}
\centering
    \setlength{\tabcolsep}{6pt} 
    \renewcommand{\arraystretch}{2} 
\begin{tabular}{|ll|}
\hline
\multicolumn{2}{|l|}{The 1/$N_c$ sum rules at NNLO for the Decuplet-Decuplet ($DDDD$) of LECs at LO $(Q/M)^0$} \\ \hline
\multicolumn{1}{|l}{$c_{1,DDDD}^{(1)} = -\frac{13}{16}c_{S,BBBB}^{(2)}+c_{S,BBBB}^{(3)}-\frac{15}{4}c_{T,BBBB}^{(1)}-\frac{3}{4}c_{T,BBBB}^{(2)}$,} & \multicolumn{1}{l|}{$c_{1,DDDD}^{(2)} = -\frac{9}{16}c_{S,BBBB}^{(2)}+\frac{45}{4}c_{T,BBBB}^{(1)}+\frac{9}{4}c_{T,BBBB}^{(2)}$,}  \\
\multicolumn{1}{|l}{$c_{2,DDDD}^{(1)} =-\frac{3}{8}c_{T,BBBB}^{(1)}-\frac{3}{4}c_{T,BBBB}^{(2)} +\frac{1}{4}c_{6,BBBB}^{(2)}- \frac{1}{4}c_{6,BBBB}^{(3)}$,} & \multicolumn{1}{l|}{$c_{2,DDDD}^{(2)} = \frac{9}{8}c_{T,BBBB}^{(1)}+\frac{9}{4}c_{T,BBBB}^{(2)}$,}\\
\multicolumn{1}{|l}{$c_{3,DDDD}^{(1)} = 0$, \hspace{2cm}$c_{3,DDDD}^{(2)} = 0$,} & \multicolumn{1}{l|}{$c_{4,DDDD}^{(1)} = 0$, \hspace{2cm} $c_{4,DDDD}^{(2)} = 0$}\\ \hline
\multicolumn{2}{|l|}{The 1/$N_c$ sum rules at NNLO for the Decuplet-Decuplet ($DDDD$) of LECs at NLO $(Q/M)^2$} \\ \hline
$c_{5,DDDD}^{(1)} = 0$, & $c_{5,DDDD}^{(2)} = 0 $,  \\ 
$c_{6,DDDD}^{(1)} = -\frac{1}{4}c_{6,BBBB}^{(2)}+c_{6,BBBB}^{(3)}$, & $c_{6,DDDD}^{(2)} = -\frac{9}{4}c_{6,BBBB}^{(2)}$,   \\
$c_{7,DDDD}^{(1)} = \frac{13}{36}c_{10,BBBB}^{(1)}+\frac{31}{18}c_{10,BBBB}^{(2)}-3c_{10,BBBB}^{(3)}$, & $c_{7,DDDD}^{(2)} =-\frac{49}{12}c_{10,BBBB}^{(1)}-\frac{26}{12}c_{10,BBBB}^{(2)} $,   \\
$c_{8,DDDD}^{(1)} = -\frac{27}{8}c_{8,BBBB}^{(3)}$, & $c_{8,DDDD}^{(2)} =\frac{81}{8}c_{8,BBBB}^{(3)} $,  \\ 
$c_{9,DDDD}^{(1)} = \frac{9}{24}c_{9,BBBB}^{(2)}+3c_{9,BBBB}^{(3)}$, & $c_{9,DDDD}^{(2)}=-\frac{81}{8}c_{9,BBBB}^{(2)}$   \\
$c_{10,DDDD}^{(1)} = -\frac{3}{2}c_{10,BBBB}^{(1)}-\frac{21}{4}c_{10,BBBB}^{(2)}+9c_{10,BBBB}^{(3)}$, & $c_{10,DDDD}^{(2)} =\frac{27}{2}c_{10,BBBB}^{(1)}+\frac{27}{4}c_{10,BBBB}^{(2)}$,   \\
$c_{11,DDDD}^{(1)} = -\frac{27}{8}c_{11,BBBB}^{(3)}$, & $c_{11,DDDD}^{(2)} =\frac{81}{8}c_{11,BBBB}^{(3)} $,   \\ 
\multicolumn{1}{|l}{$c_{12,DDDD}^{(1)} = 0$, \hspace{2cm} $c_{12,DDDD}^{(2)} =0 $,} & \multicolumn{1}{l|}{$c_{13,DDDD}^{(1)} = 0$, \hspace{2cm}$c_{13,DDDD}^{(2)} =0 $,}  \\  
\multicolumn{1}{|l}{$c_{14,DDDD}^{(1)} = 0$, \hspace{2cm} $c_{14,DDDD}^{(2)} =0 $,} &  \multicolumn{1}{l|}{$c_{15,DDDD}^{(1)} = 0$, \hspace{2cm} $c_{15,DDDD}^{(2)} =0 $,} \\ 
\multicolumn{1}{|l}{$c_{16,DDDD}^{(1)} = 0$, \hspace{2cm} $c_{16,DDDD}^{(2)} =0 $,} & \multicolumn{1}{l|}{$c_{17,DDDD}^{(1)} = 0$, \hspace{2cm} $c_{17,DDDD}^{(2)} =0 $,}  \\ 
\multicolumn{1}{|l}{$c_{18,DDDD}^{(1)} = 0$, \hspace{2cm} $c_{18,DDDD}^{(2)} =0 $,}  & \multicolumn{1}{l|}{$c_{19,DDDD}^{(1)} = 0$, \hspace{2cm} $c_{19,DDDD}^{(2)} =0 $,} \\  \hline
\end{tabular}
\caption{\color{black}The sum rules for the Decuplet-Decuplet ($DDDD$) sector under $1/N_c$ expansion at NNLO  are summarized. These relations provide a significant reduction in independent parameters due to the LO and NNLO  $1/N_c$  Hartree potential constraint. The LECs are proportional to 12 octet-octet parameters, $c_{S,BBBB}^{(1,2)}$, $c_{T,BBBB}^{(1,2)}$, $c_{6,BBBB}^{(2,3)}$, $c_{8,BBBB}^{(3)}$, $c_{9,BBBB}^{(2)}$, $c_{10,BBBB}^{(1,2,3)}$, $c_{11,BBBB}^{(3)}$. Several LECs vanish since the couplings up to NNLO ($q_i$ and $h_i$) are insufficient to account for their structure. A total of 38 sum rules are derived for this transition, consisting of 16 non-vanished sum rules and 22 vanished sum rules.}
\label{tab:4DNNLO}
\end{table}
\indent \textcolor{black}{According to the matching condition up to NNLO of the 1/$N_c$ expansion of the Hartree Hamiltonian, a total of 110 sum rules are derived, described as follows: 13 ($BB \rightarrow BB$), 11 ($BB \rightarrow DB$), 27 ($DB \rightarrow DB$), 10 ($BB \rightarrow DD$), 11 ($DB \rightarrow DD$), and 38 ($DD \rightarrow DD$). Consequently, the number of LECs is reduced from 134 to 24.  All non-vanished LECs are appeared in terms of 14 LECs for the $BBBB$ sector as, $c_{S,BBBB}^{(2,3)}$, $c_{T,BBBB}^{(1,2,3)}$, $c_{6,BBBB}^{(2,3)}$, $c_{8,BBBB}^{(3)}$, $c_{9,BBBB}^{(2,3)}$, $c_{10,BBBB}^{(1,2,3)}$, $c_{11,BBBB}^{(3)}$, 1 LECs from Mixed Octet-Decuplet ($BBDB$) sector, $c_{11,BBDB}^{(2)}$. Except the $DBDB$ sector, there are additional 9 free parameters as $c_{i,DBDB}^{(2)}$, $i=1,2,5,6,7,8,9,10,11$.} 
Interestingly, the number of vanishing LECs in the NNLO of the 1/$N_c$ expansion has decreased significantly compared to the LO and this provides more insight into LECs' behaviors. These relations hold up the correction of $1/N_c^2 \approx 10\%$ at $N_c=3$.\\ 

\section{THE IMPLICATION OF SUM RULES ON $\Omega N$ AND $\Omega\Omega$ SCATTERING}\label{sec-4}
\indent To examine the phenomenological implications of the $1/N_c$ sum rules, we apply them to the LO of the hyperon–nucleon ($YN$) and hyperon–hyperon ($YY$) contact interactions presented in \cite{haiden2017scattdecupletbary}. The $ \Omega N $ and  $ \Omega\Omega $ systems serve as primary candidates for this application, as they allow us to implement the sum rules at the level of partial-wave projected LECs and to construct an additional sum rule in the partial-wave basis. \textcolor{black}{Let us consider the octet-octet partial-wave LECs ($\mathcal{C}_{xy}^{{\rm rep.}}$), which are the linear combination of the coefficients $c_{n,t}^{(\alpha)}$ as shown in Appendix \ref{A-5-PartialWave}, by utilizing the $1/N_c$ octet-octet coupling constant's sum rules in Table \ref{tab:4BLO} on these coefficients.} The new sum rules for octet-octet partial-wave LECs at LO of the 1/$N_c$ expansion are given by,\\
\textcolor{black}{
\begin{equation}\label{4Bpatialsumrule}
    \begin{split}
        &\mathcal{C}_{00}^{10}=\frac{1}{173}\biggl(607\mathcal{C}_{00}^{1}+201\mathcal{C}_{00}^{8a}-635\mathcal{C}_{00}^{8s}\biggr),\\[1em]
        &\mathcal{C}_{00}^{\overline{10}}=\frac{1}{173}\biggl(-115\mathcal{C}_{00}^{1}+160\mathcal{C}_{00}^{8a}+128\mathcal{C}_{00}^{8s}\biggr),\\[1em]
        &\mathcal{C}_{00}^{27}=\frac{1}{173}\biggl(-335\mathcal{C}_{00}^{1}+90\mathcal{C}_{00}^{8a}+418\mathcal{C}_{00}^{8s}\biggr),
    \end{split}
\end{equation}}
where these $1/N_c$ sum rules give us an obvious relation between partial-wave LECs, causing coefficient suppression, so the number of partial-wave LECs is dropped from 6 to 3. In the same manner, the octet-decuplet mixing ($DBDB$) and decuplet-decuplet ($DDDD$) sectors are treated with their own $1/N_c$ sum rules in Table \ref{tab:dbdbLO} and \ref{tab:4DLO} on a partial-wave basis, respectively. Their new sum rules can be written down, the octet-decuplet mixing ($DBDB$),\\
\textcolor{black}{
\begin{alignat}{2}
    &\mathcal{C}_{11}^{10,^5S_2}=-\frac{287}{2768}\mathcal{C}_{00}^{1}+\frac{2007}{5536}\mathcal{C}_{00}^{8a}+\frac{187}{5536}\mathcal{C}_{00}^{8s},\notag\\[1em]
        &\mathcal{C}_{11}^{27,^5S_2}=\frac{1193}{2768}\mathcal{C}_{00}^{1}+\frac{2961}{5536}\mathcal{C}_{00}^{8a}-\frac{2579}{5536}\mathcal{C}_{00}^{8s},\notag\\[1em]
        &\mathcal{C}_{11}^{35,^5S_2}=-\frac{1747}{2768}\mathcal{C}_{00}^{1}+\frac{1845}{5536}\mathcal{C}_{00}^{8a}+\frac{4417}{5536}\mathcal{C}_{00}^{8s},\label{DBpartialSumRule}\\[1em]
        &\mathcal{C}_{11}^{10,^3S_1}= \frac{1607}{2768}\mathcal{C}_{00}^{1}+\frac{2847}{5536}\mathcal{C}_{00}^{8a}-\frac{3293}{5536}\mathcal{C}_{00}^{8s},\notag\\[1em]
        &\mathcal{C}_{11}^{27,^3S_1}= -\frac{1535}{2768}\mathcal{C}_{00}^{1}+\frac{1225}{5536}\mathcal{C}_{00}^{8a}+\frac{4613}{5536}\mathcal{C}_{00}^{8s},\notag\\[1em]
        &\mathcal{C}_{11}^{35,^3S_1}=\frac{101}{2768}\mathcal{C}_{00}^{1}+\frac{3021}{5536}\mathcal{C}_{00}^{8a}-\frac{455}{5536}\mathcal{C}_{00}^{8s},\notag
\end{alignat}}
and decuplet-decuplet sectors ($DDDD$),
\textcolor{black}{
\begin{equation}\label{DDPartialSumRule}
    \begin{split}
        &\mathcal{C}_{22}^{\overline{10},^3S_1}= -\frac{1855}{1384}\mathcal{C}_{00}^{1}+\frac{927}{2768}\mathcal{C}_{00}^{8a}+\frac{5551}{2768}\mathcal{C}_{00}^{8s},\\[1em]
        &\mathcal{C}_{22}^{\overline{10},^7S_3}=  \frac{2105}{1384}\mathcal{C}_{00}^{1}+\frac{3447}{2768}\mathcal{C}_{00}^{8a}-\frac{4889}{2768}\mathcal{C}_{00}^{8s},\\[1em]
        &\mathcal{C}_{22}^{27,^1S_0}=  -\frac{905}{692}\mathcal{C}_{00}^{1}+\frac{525}{1384}\mathcal{C}_{00}^{8a}+\frac{2669}{1384}\mathcal{C}_{00}^{8s},\\[1em]
        &\mathcal{C}_{22}^{27,^5S_2}=  -\frac{113}{692}\mathcal{C}_{00}^{1}+\frac{1029}{1384}\mathcal{C}_{00}^{8a}+\frac{581}{1384}\mathcal{C}_{00}^{8s},\\[1em]
        &\mathcal{C}_{22}^{28,^1S0}=   \frac{2375}{1384}\mathcal{C}_{00}^{1}+\frac{4185}{2768}\mathcal{C}_{00}^{8a}-\frac{6167}{2768}\mathcal{C}_{00}^{8s},\\[1em]
        &\mathcal{C}_{22}^{28,^5S2}=  -\frac{1}{1384}\mathcal{C}_{00}^{1}+\frac{2673}{2768}\mathcal{C}_{00}^{8a}+\frac{97}{2768}\mathcal{C}_{00}^{8s},\\[1em]  
        &\mathcal{C}_{22}^{35,^3S_1}=  -\frac{17}{173}\mathcal{C}_{00}^{1}+\frac{144}{173}\mathcal{C}_{00}^{8a}+\frac{46}{173}\mathcal{C}_{00}^{8s},\\[1em]
        &\mathcal{C}_{22}^{35,^7S_3}=  -\frac{17}{173}\mathcal{C}_{00}^{1}+\frac{144}{173}\mathcal{C}_{00}^{8a}+\frac{46}{173}\mathcal{C}_{00}^{8s}.
    \end{split}
\end{equation}}
In the octet-decuplet ($DBDB$) scattering sector, 6 sum rules are obtained, which reduce the number of partial-wave LECs from 8 to 2. In the decuplet-decuplet ($DDDD$) sector,  8 sum rules are found, completely lowering the partial-wave LECs from 8 to 0.  Consequently, all partial-wave LECs in the decuplet-decuplet sector can be expressed in terms of the octet–octet partial-wave LECs. \textcolor{black}{We can summarize that these 3 partial-wave sum rule sectors are governed by $\mathcal{C}_{00}^{1}$, $\mathcal{C}_{00}^{8a}$, and $\mathcal{C}_{00}^{8s}$ as free parameters according to the $1/N_c$ operator analysis at LO. In addition, the $DBDB$ sector, there are free parameters as, $\mathcal{C}_{11}^{8,^3S_1}$, and $\mathcal{C}_{11}^{8,^5S_2}$.}

\subsection{The $1/N_c$ sum rules in the $\Omega N$ scattering}
\indent The scattering of $\Omega N$ is particularly interesting due to its stability against hadronic decay and its ability to couple to various other channels, i.e., $BB$ or $BD$ sectors \cite{HALQCD:2015qmg}. We consider the potential of $\Omega N$ interaction in the $^5S_2$ partial-wave regime \cite{haiden2017scattdecupletbary}. Using the partial-wave projection of the contact interaction, they are
\begin{equation}
    V_{\Omega N\Omega N}^{^5S_2}=\frac{1}{40}\biggl(10\mathcal{C}_{11}^{10,^5S_2}+9\mathcal{C}_{11}^{27,^5S_2}+5\mathcal{C}_{11}^{35,^5S_2}+16\mathcal{C}_{11}^{8,^5S_2}\biggr).
\end{equation}
Then, applying our $1/N_c$ at LO constraints given by Eq.(\ref{DBpartialSumRule}) and choosing the relevant potentials provided in \cite{haiden2017scattdecupletbary}, we found that
\textcolor{black}{
\begin{equation}\label{ONPoten}
\begin{split}
      V_{\Omega N\Omega N}^{^5S_2} &= \frac{1}{2}V_{\Delta\Sigma\Delta\Sigma}^{^5S_2}+\frac{27}{64}V_{NNNN}^{^1S_0}-\frac{23}{32}V_{NNNN}^{^3S_1}-\frac{15}{32}V_{\Lambda N\Lambda N}^{^1S_0}+\frac{65}{64}V_{\Xi N \Xi N}^{^3S_1}.
\end{split}
\end{equation}}

\indent \textcolor{black}{The result of the derived expression indicates that the potential governing the $\Omega N$ scattering process, initially parameterized by four partial-wave LECs can, in fact, be effectively reduced to a formulation involving only two distinct types of partial-wave LECs. These include the LECs within the $^5S_2$  regime on same transition, along with partial-wave LECs inherited from the octet–octet sector in $^3S_1$ and $^1S_0$ channels. In addition, the $S$-wave potentials for $NN$, $\Lambda N$, and, $\Xi N$ scatterings, $V^{^3S_1}$ and $V^{^1S_0}$ in above equation are represented by 3 LECs, $c_{S,BBBB}^{(2)}$, $c_{S,BBBB}^{(3)}$, and $c_{T,BBBB}^{(3)}$ only. One can see the relations between $c_{n,{\rm t}}^{(i)}$ and the Lagragian couplings, $g_{xy,z}^{i(\alpha)}$ in Table \ref{BBBB LECs Table}. While the $S$-wave potential $V_{\Delta\Sigma\Delta\Sigma}^{^5S_2}$ can be considered as a free parameter in this case and will be determined in the latter.} Notably, these original LECs are replaced or absorbed by possible contributions from lower-strangeness channels, which are provided in \cite{haiden2017scattdecupletbary}. When reformulating the partial-wave LECs in terms of the potential of the scattering process, it becomes clear that the $\Omega N$ to $\Omega N$ can be further constrained through related scattering processes involving more accessible states. In particular, these cross-channel connections demonstrate a relation that significantly enhances predictive capability. They offer a valuable alternative approach to estimate the potential of $\Omega N$ to $\Omega N$ through more accessible and experimentally reliable channels, in contrast to the traditional method, which still suffers from a lack of empirical data.\\
\subsection{The $1/N_c$ sum rules in the $\Omega \Omega$ scattering}
\indent According to the Pauli Exclusion Principle, only $^1S_0$ and $^5S_2$ $S$-wave states are allowed for $ \Omega\Omega $. The potential depends only on a single individual partial-wave coupling constant, corresponding to the irreducible representation 28 of SU (3). The scattering of $\Omega\Omega$ is studied using lattice QCD simulations by two research groups: one led by Buchoff et al. and the other by the HAL QCD collaboration \cite{Buchoff_2012,HALQCD:2015qmg}. They indicate that a strongly attractive interaction occurs in the $^1S_0$ channel, so we decide to consider a partial-wave LECs in the $^1S_0$ sector at LO \cite{haiden2017scattdecupletbary}. Using the partial-wave projection of the contact interaction, they are
\begin{equation}
    \begin{split}
        V_{\Omega\Omega\Omega\Omega}^{^1S_0}=\mathcal{C}_{22}^{28,^1S_0}
    \end{split}
\end{equation}
Then, applying our $1/N_c$ constraints given by Eq.(\ref{DDPartialSumRule}), we find
\textcolor{black}{
\begin{equation}
    \begin{split}
        V_{\Omega\Omega\Omega\Omega}^{^1S_0}&= -\frac{17825}{12}V_{N NN N}^{^1S_0}-\frac{6871}{12}V_{\Lambda\Lambda\Lambda\Lambda}^{^1S_0}+2058V_{\Lambda N\Lambda N}^{^1S_0}
    \end{split}
\label{V1S0OOOO}
\end{equation}}
\indent The result indicates that the potential for the $\Omega\Omega$ scattering process, which was initially governed by a single partial-wave LECs, can now be described by using three octet-octet partial-wave LECs. The LECs are then replaced or absorbed by possible contributions from the octet-octet channels $^1S_0$, which are provided in \cite{haiden2017scattdecupletbary}, to construct the potential. This extension of $ V_{\Omega\Omega\Omega\Omega}^{^1S_0}$ provides an enhancement of predictive power. \textcolor{black}{Remarkably, our analysis reveals that a compelling possibility of the potential $\Omega \Omega$, which is experimentally challenging to access directly, can be studied indirectly through the $NN$, $\Lambda N$, and $\Lambda\Lambda$ scattering channels. These $S$-wave potential scatterings, $V^{^1S_0}$ can be expressed by 3 LECs, $c_{S,BBBB}^{(2)}$, $c_{S,BBBB}^{(3)}$, and $c_{T,BBBB}^{(3)}$. Thus, the $\Omega\Omega$ $S$-wave interaction is not controlled by a large number of unrelated couplings. It is predicted in terms of a small, explicitly identifiable number of octet–octet parameters.} {In addition, a related study on $S$-wave $\Omega\Omega$ interaction has been done by using the $1/N_c$ analysis in \cite{richardson2024}. However, their approach focuses on the partial-wave coupling constants of the $\Delta\Delta$ channel within the flavor symmetry of SU(2). Then, they subsequently generalized to the SU(3) flavor symmetry. Based on this method, they derived 4 expression relations, allowing the number of partial-wave couplings in \cite{haiden2017scattdecupletbary} to be reduced, and their coupling constant relations are only within the $\Delta\Delta$ channel.}
\subsection{Applications of the $1/N_c$ sum rules to baryon-baryon scattering in KSW formalism}
\indent Furthermore, the scattering length of $a_0^{(\Omega N)}$ and  $a_0^{(\Omega\Omega)}$, which are provided by the Lattice QCD calculations in Refs.\cite{Iritani:2018lvw,Gongyo:2017fjb,Aoki:2023qih}, can be used to \textcolor{black}{estimate $\Omega N$ and $\Omega\Omega$ potential by following the KSW scheme formalism} in \cite{Kaplan:1998we, Meng:2022ozq}. The momentum-space ($S$-wave) EFT potential is written down as,
\begin{equation}
    V(p',p)= \mathcal{C}_0(\mu)+\frac{1}{2}\mathcal{C}_2(\mu)(p'^2+p^2)+ \dots
\end{equation}
where $\mathcal{C}_0(\mu)$ and $\mathcal{C}_2(\mu)$ are couplings to the low-energy scattering data (scattering length or effective range). Nevertheless, $\mathcal{C}_2(\mu)$ will be neglected because these scatterings are considered at LO. The coupling $\mathcal{C}_0(\mu)$ is given by
\begin{equation}
    \mathcal{C}_0(\mu)=\frac{4\pi}{M}\biggl(\frac{1}{-\mu+1/a_0}\biggr),
\end{equation}
and we identify $V_{\Omega N\Omega N}^{^5S_2}, V_{\Omega\Omega\Omega\Omega}^{^1S_0} \equiv \mathcal{C}_0(\mu)$ where $M$ is the mass of the baryon, $a_0$ represents the scattering length, and  $\mu$ is the dimensional regularization scale, which is related to the cutoff ($\Lambda$) as $\mu\simeq1.2\,\Lambda$ \cite{Oller:1998hw}. We define the following parameters: $M_{N} = 0.938\,\, {\rm GeV}$, $M_{\Omega} = 1.672\,\, {\rm GeV} $, $a_0^{(\Omega N)}= 5.30\,\, {\rm fm} = 26.8604\,\, {\rm GeV}^{-1}$, $a_0^{(\Omega\Omega)}= 4.6\,\, {\rm fm} = 23.3082\,\, {\rm GeV}^{-1}$, $\Lambda= 0.6\,\, {\rm GeV}$, where the scattering length $a_0^{(\Omega N)}$ and $a_0^{(\Omega\Omega)}$ are taken from the lattice QCD calculation in \cite{Iritani:2018lvw,Gongyo:2017fjb,Aoki:2023qih}. The potentials of $\Omega N $ at $^5S_2$ channel and $\Omega\Omega$ at $^1S_0$ channel are read,
\textcolor{black}{
\begin{equation}\label{Vnumer}
   V_{\Omega N\Omega N}^{^5S_2}= -14.103 \,\, {\rm GeV}^{-2}, \quad\quad\quad  V_{\Omega\Omega\Omega\Omega}^{^1S_0}= -11.100 \,\, {\rm GeV}^{-2}.
\end{equation}}
According to the potential value for the octet-octet sector in Eq.(\ref{ONPoten}) and (\ref{V1S0OOOO}), these potentials can be obtained by utilizing the same procedure. The scattering lengths of $NN$, $\Lambda N$, $\Lambda\Lambda$ and $\Xi N$ are given in \cite{NPLQCD:2013bqy,Rijken:2010zzb,2016EPJA...52...21R,Gasparyan:2011kg,Haidenbauer:2007ra}, so their potentials in KSW formalism read,   \\
\textcolor{black}{
\begin{equation}\label{ococpoten}
    \begin{split}
         &V_{NNNN}^{^1S_0}= -21.042 \,\, {\rm GeV}^{-2},  \quad V_{NNNN}^{^3S_1}= -21.495 \,\, {\rm GeV}^{-2},\quad V_{\Lambda N \Lambda N}^{^1S_0}= -18.924 \,\, {\rm GeV}^{-2}, \\[1em]
        &\hspace{2.5cm} V_{\Lambda\Lambda\Lambda\Lambda}^{^1S_0}= -13.41\,\, {\rm GeV}^{-2}, \quad   V_{\Xi N \Xi N}^{^3S_1}= -14.908 \,\, {\rm GeV}^{-2}.
    \end{split}
\end{equation}}
\\
Subsequently, these numerical potentials are employed to determine $V_{\Delta\Sigma\Delta\Sigma}^{^5S_2}$ since there is no theoretical evaluation in the literature. \textcolor{black}{By substituting these values into Eq.(\ref{ONPoten}), we obtain the prediction of the  $V_{\Delta\Sigma\Delta\Sigma}^{^5S_2}$ potential from the 1/$N_c$ expansion at LO as,}
\textcolor{black}{
\begin{equation}
    V_{\Delta\Sigma\Delta\Sigma}^{^5S_2}= -25.009 \,\, {\rm GeV}^{-2}.
\end{equation}}
Although direct theoretical results for this potential are currently unavailable, we anticipate that our predictions will provide a valuable benchmark for $\Delta\Sigma$ $\to$ $\Delta\Sigma$ scattering in future experimental studies or lattice QCD simulations. \textcolor{black}{Next, by substituting the numerical potentials into Eq.(\ref{V1S0OOOO}), we obtain the $V_{\Omega\Omega\Omega\Omega}^{^1S_0}$ potential as well,}
\textcolor{black}{
\begin{equation}
    V_{\Omega\Omega\Omega\Omega}^{^1S_0}= -10.539\,\, {\rm GeV}^{-2}.
\end{equation}}
\textcolor{black}{We find that the $V_{\Omega\Omega\Omega\Omega}^{^1S_0}$ potential value in Eq. (\ref{Vnumer}) from the KSW approach is consistent with the predicted value from the 1/$N_c$ expansion at LO, which is obtained using only the octet-octet ($BBBB$) sector potentials.} This suggests that the underlying octet-octet sector constrains the decuplet-decuplet interaction via the 1/$N_c$ sum rules.
\\

\section{Discussion and conclusion}\label{sec-5}
Our study uses a $1/N_c$ operator analysis to investigate two-body interactions between octet and decuplet baryons within SU(3) chiral effective field theory. This approach could significantly advance our understanding of hadron dynamics. We construct the minimal set of four-point contact terms of SU(3) chiral Lagrangians with non-derivative terms, as detailed in \cite{haiden2017scattdecupletbary}. \textcolor{black}{There are 104 chiral coupling constants, $g_{xy,z}^{i(\alpha)}$ corresponding to all possible configurations of the baryon–baryon system. Utilizing a non-relativistic expansion of chiral Lagrangians up to NLO, we have derived a total of 134 LECs, $c_{n,\text{t}}^{(\alpha)}$ under SU(3) flavor symmetry and the $Q/\Lambda$ expansion scale. This includes 28 terms in the LO and the 106 terms at NLO.}\\
\indent The Hartree baryon potential, when expanded in powers of $1/N_c$, contains 4 operator coefficients, $q_i$, $i=1,2,3,4$ at LO, 11 operator coefficients, $h_i$, $i=1,\cdots,11$ at NNLO. The LECs exhibit two scales: $N_c$ and $1/N_c$, as discussed in section \ref{sec-3}. Through the matching process between the chiral potentials and the Hartree potentials in section \ref{sec-4}, \textcolor{black}{at LO of the 1/$N_c$ expansion, the sum rules presented in Table \ref{tab:4BLO}--\ref{tab:4DLO} establish relations among the LECs. reducing their number from 134 to 13. These remaining LECs are expressed in terms of 4 free parameters as $c_{S,BBBB}^{(2)},\,c_{S,BBBB}^{(3)},\,c_{T,BBBB}^{(3)},\,c_{10,BBBB}^{(3)}$, and the $DBDB$ sector, there are additional 9 free parameters as $c_{i,DBDB}^{(2)}$, $i=1,2,5,6,7,8,9,10,11$.} Moreover, \textcolor{black}{the inclusion of NNLO terms provides additional sum rules in Table \ref{tab:4BNNLO}--\ref{tab:4DNNLO}, which further constrain the system, reducing the number of LECs to 24 and expressing them in terms of 14 free parameters as, $c_{S,BBBB}^{(2,3)}$, $c_{T,BBBB}^{(1,2,3)}$, $c_{6,BBBB}^{(2,3)}$, $c_{8,BBBB}^{(3)}$, $c_{9,BBBB}^{(2,3)}$, $c_{10,BBBB}^{(1,2,3)}$, $c_{11,BBBB}^{(3)}$,  1 LECs from the $BBDB$ sector, $c_{11,BBDB}^{(2)}$, and additional 9 free parameters in the $DBDB$ sector as $c_{i,DBDB}^{(2)}$, $i=1,2,5,6,7,8,9,10,11$.} These relations are under corrections of the order of $1/N_c^2$, which is approximately 10\%. \\
\indent The implications of the $1/N_c$ sum rules are further explored in phenomenological applications, particularly in the context of $\Omega N$ and $\Omega\Omega$ scattering. Although these channels are experimentally difficult to access directly, relations derived from the partial-wave sum rules allow their scattering potentials to be estimated via well-established scattering processes with more abundant and reliable data. \textcolor{black}{The potential of $\Omega N \rightarrow \Omega N $, initially governed by four partial-wave LECs within the $^5S_2$ regime, is deformed into contributions of the $^3S_1$ and $^1S_0$ regimes in the octet-octet sector. Similarly, the $\Omega \Omega$ $\to$ $\Omega \Omega$ scattering potential, originally described by a single partial-wave LEC, is extended to 3 parameters in the $^1S_0$ regime of the octet-octet sector.} These applications of $1/N_c$ analysis not only improve theoretical predictability, but also offer alternative pathways for investigations through other channels, such as $NN$, $\Lambda N$, or $\Xi N$, which are more accessible in lattice QCD information and experiment. \textcolor{black}{We can use the KSW approach to predict the potential of $V_{\Delta\Sigma\Delta\Sigma}^{^5S_2}$, which is around $ -25.009 \,\, {\rm GeV}^{-2}$, and to calculate the potential of $V_{\Omega\Omega\Omega\Omega}^{^1S_0} = -10.539\,\, {\rm GeV}^{-2}$.} However, our application of the $1/N_c$ partial-wave sum rules is still limited because of information lacking. So we anticipate future lattice QCD simulations or experiments may provide more reliable data to validate our $1/N_c$ operator analysis.\\

\acknowledgments
\indent The authors thank Tetsuo Hyodo for helpful discussions. C.B. gratefully acknowledges the financial support provided by the DPST (Development and Promotion of Science and Technology Talents Project) Scholarship, Royal Government of Thailand, and research collaboration with D.F.V., Jr. in connection with his Ph.D. dissertation. This collaboration was made possible through the Research Enhancement Program (Sandwich) of the Department of Physics, Mindanao State University–Iligan Institute of Technology (MSU-IIT), and the scholarship support provided through the Department of Science and Technology–Accelerated Science and Technology Human Resource Development Program (DOST-ASTHRDP), Philippines. DS is supported by the Fundamental Fund of Khon Kaen University and has received funding support from the National Science, Research, and Innovation Fund and supported by Thailand NSRF through PMU-B [grant number B39G680009].
\newpage
\section{Appendices}
\subsection{Non-Relativistic Expansion of Baryon Bilinear}\label{A-1-NR-Expansion}
\indent The two-baryon contact terms can be translated into the matrix element of two bilinears $(\bar{u}_3\Gamma_iu_1)(\bar{u}_4\Gamma_ju_2)$ with momentum factors from derivatives, which act on the baryon field, via the Feynman rules \cite{BERNARD_1995}. The free Dirac spinors have been read,
\begin{equation}
    \begin{split}
        &\bar{u}_i = \sqrt{\frac{E_i+M}{2M}}\begin{pmatrix} \mathbb{1} & -\frac{\Vec{\sigma}\cdot\Vec{p'}}{E_i+M} \end{pmatrix}, \hspace{1cm} u_j = \sqrt{\frac{E_j+M}{2M}} \begin{pmatrix}\mathbb{1} \\ \frac{\Vec{\sigma}\cdot\Vec{p}}{E_j+M}\end{pmatrix},\\
        &\text{where},\\
        &E_{i} = \sqrt{M^2+\Vec{p'}^2}, \hspace{.5cm} E_{i} = \sqrt{M^2+\Vec{p}^2},
    \end{split}
\end{equation}
$M$ represents baryon mass in the chiral limit for this work. The non-relativistic expansion of a single baryon bilinear in terms of inverse large mass up to order $\mathcal{O}(q^2)$, one received from \cite{PETSCHAUER20131}, we use the properties of Pauli spin ($\sigma$) to simplify those expansion forms. They yield the expression given in Table \ref{TB1}
\begin{table}[h]
    \centering
    \setlength{\tabcolsep}{10pt} 
    \renewcommand{\arraystretch}{2} 
    \caption{The non-relativistic expansion up to $q^2$ for bilinears of baryon,  where $\Vec{p}$ and $\Vec{p'}$ represent incoming and outgoing three-momentum in the c.m. frame.}
    \begin{tabular}{|c|c|} \hline 
        $\Gamma_i $ & Non-relativistic expansion up to $q^2$ \\ \hline 
         $\mathbb{1}$& $1+\frac{p^2+p'^2}{8M^2}+\frac{\vec{p'}\cdot\vec{p}}{4M^2}+\frac{i\Vec{\sigma}\cdot(\Vec{p'}\times\Vec{p})}{4M^2}$ \\ \hline 
         $\gamma^0$& $1+\frac{p^2+p'^2}{8M^2}-\frac{\vec{p'}\cdot\vec{p}}{4M^2}-\frac{i\Vec{\sigma}\cdot(\Vec{p'}\times\Vec{p})}{4M^2}$\\ \hline 
         $\Vec{\gamma}$&$\Vec{0}+\frac{(\Vec{p}+\Vec{p'})}{2M}+\frac{i(\Vec{p}-\Vec{p'})\times\Vec{\sigma}}{2M}$ \\ \hline 
         $\gamma_5$ & $0+\frac{\Vec{\sigma}\cdot(\Vec{p}-\Vec{p'})}{2M}$\\ \hline 
         $\gamma^0\gamma_5$& $0+\frac{\Vec{\sigma}\cdot(\Vec{p}+\Vec{p'})}{2M}$ \\ \hline 
         $\Vec{\gamma}\gamma_5$& $\Vec{\sigma}+\frac{p^2+p'^2}{8M^2}\Vec{\sigma}+\frac{(\Vec{p'}\cdot\Vec{p})}{4M^2}\Vec{\sigma}+\frac{i(\Vec{p'}\times\Vec{p})}{4M^2}-\frac{\Vec{p'}(\Vec{\sigma}\cdot\Vec{p})}{4M^2}-\frac{\Vec{p}(\Vec{\sigma}\cdot\Vec{p}')}{4M^2}$ \\ \hline 
         $\sigma^{0l}$& $0+\frac{i(p^l-p'^l)}{2M}-\frac{\varepsilon^{lmn}(p^m+p'^m)\sigma^n}{2M}$ \\ \hline 
         $\sigma^{kl}$&$\varepsilon^{klm}[\sigma_m+\frac{p^2+p'^2}{8m^2}\sigma_m-\frac{(\Vec{p'}\cdot\Vec{p})}{4M^2}\sigma_m-\frac{i(\Vec{p'}\times\Vec{p})_m}{4M^2}+\frac{p'_m(\Vec{\sigma}\cdot\Vec{p})}{4M^2}+\frac{(\Vec{\sigma}\cdot\Vec{p'})p_m}{4m^2}]$ \\ \hline
    \end{tabular}
    
    \label{TB1}
\end{table}
\newpage
\subsection{The Potential at the LO and NNLO of $1/N_c$ Expansion of Hartree Hamiltonian}\label{A-2-HHPotential}
Here, the potentials of the Hartree Hamiltonian corresponding to the 6 configurations of octet and decuplet baryon interactions up to NNLO in the $1/N_c$ expansion approach are presented,\\
\\
\indent $\bullet$ Octet-Octet (\scalebox{1}{$BBBB$})
\\
\small{\begin{equation}\label{BBBBp}
    \begin{split}
        V_{BBBB}=&\quad 9q_1(\delta^{\chi_1\chi'_1}\delta^{\chi_2\chi'_2})\delta^{a_1a'_1}\delta^{a_2a'_2}\,\,\,\,\,-q_2(\delta^{\chi_1\chi'_1}\delta^{\chi_2\chi'_2})f^{a_1a'_1e}f^{a_2a'_2e}\quad+\frac{1}{4}q_3(\vec{\sigma}_1\cdot\vec{\sigma}_2)d^{ea_1a'_1}d^{ea_2a'_2}\\[.5em]
        &+\frac{1}{6}q_3(\vec{\sigma}_1\cdot\vec{\sigma}_2)id^{ea_1a'_1}f^{ea_2a'_2}\quad+\frac{1}{6}q_3(\vec{\sigma}_1\cdot\vec{\sigma}_2)if^{ea_1a'_1}d^{ea_2a'_2}\quad-\frac{1}{9}q_3(\vec{\sigma}_1\cdot\vec{\sigma}_2)f^{ea_1a'_1}f^{ea_2a'_2}\\[.5em]
        &+\frac{1}{4}q_4(\sigma_1^i\sigma_2^j)(p_-^ip_-^j)d^{ea_1a'_1}d^{ea_2a'_2}\quad+\frac{1}{6}q_4(\sigma_1^i\sigma_2^j)(p_-^ip_-^j)id^{ea_1a'_1}f^{ea_2a'_2}\\[.5em]
            &+\frac{1}{6}q_4(\sigma_1^i\sigma_2^j)(p_-^ip_-^j)if^{ea_1a'_1}d^{ea_2a'_2}\quad-\frac{1}{9}q_4(\sigma_1^i\sigma_2^j)(p_-^ip_-^j)f^{ea_1a'_1}f^{ea_2a'_2}\\[.5em]
            &-\frac{1}{12}q_4(\vec{\sigma}_1\cdot\vec{\sigma}_2)(p_-^2)d^{ea_1a'_1}d^{ea_2a'_2}\quad-\frac{1}{18}q_4(\vec{\sigma}_1\cdot\vec{\sigma}_2)(p_-^2)id^{ea_1a'_1}f^{ea_2a'_2}\\[.5em]
            &-\frac{1}{18}q_4(\vec{\sigma}_1\cdot\vec{\sigma}_2)(p_-^2)if^{ea_1a'_1}d^{ea_2a'_2}\quad
            +\frac{1}{27}q_4(\vec{\sigma}_1\cdot\vec{\sigma}_2)(p_-^2)f^{ea_1a'_1}f^{ea_2a'_2}\\[.5em]
        &+9h_1(\delta^{\chi_1\chi'_1}\delta^{\chi_2\chi'_2})(p_+^2)\delta^{a_1a'_1}\delta^{a_2a'_2}\quad+\frac{1}{4}h_2(\vec{\sigma}_1\cdot\vec{\sigma}_2)\delta^{a_1a'_1}\delta^{a_2a'_2}\\[.5em]
        &-\frac{1}{4}h_3(\vec{\sigma}_1\cdot\vec{\sigma}_2)f^{a_1a'_1e}f^{a_2a'_2e}\quad-h_4(\delta^{\chi_1\chi'_1}\delta^{\chi_2\chi'_2})(p_+^2)f^{a_1a'_1e}f^{a_2a'_2e}\\[.5em]
        &+\frac{1}{4}h_5(\vec{\sigma}_1\cdot\vec{\sigma}_2)(p_+^2)d^{ea_1a'_1}d^{ea_2a'_2}\quad+\frac{1}{6}h_5(\vec{\sigma}_1\cdot\vec{\sigma}_2)(p_+^2)id^{ea_1a'_1}f^{ea_2a'_2}\\[.5em]
        &+\frac{1}{6}h_5(\vec{\sigma}_1\cdot\vec{\sigma}_2)(p_+^2)if^{ea_1a'_1}d^{ea_2a'_2}\quad-\frac{1}{9}h_5(\vec{\sigma}_1\cdot\vec{\sigma}_2)(p_+^2)f^{ea_1a'_1}f^{ea_2a'_2}\\[.5em]
        &+\frac{1}{2}h_6(\Vec{\sigma}_1+\Vec{\sigma}_2)\cdot i(\Vec{p}_+\times\Vec{p}_-)\delta^{a_1a'_1}\delta^{a_2a'_2}\quad+\frac{1}{2}h_7(\Vec{\sigma}_1)i(\Vec{p}_+\times\Vec{p}_-)id^{ea_1a'_1}f^{a_2a'_2e}\\[.5em]
        &+\frac{1}{2}h_7(\Vec{\sigma}_2)i(\Vec{p}_+\times\Vec{p}_-)if^{a_1a'_1e}d^{ea_2a'_2}\quad-\frac{1}{3}h_7(\Vec{\sigma}_1+\Vec{\sigma}_2)i(\Vec{p}_+\times\Vec{p}_-)f^{a_1a'_1e}f^{a_2a'_2e}\\[.5em]
        &-\frac{1}{2}h_8(\Vec{\sigma}_1+\Vec{\sigma}_2)i(\Vec{p}_+\times\Vec{p}_-)f^{a_1a'_1e}f^{a_2a'_2e}\quad+\frac{1}{4}h_9(\sigma^i_1\sigma^j_2)(p^i_-p^j_-)\delta^{a_1a'_1}\delta^{a_2a'_2}\\[.5em]
        &-\frac{1}{12}h_9(\vec{\sigma}_1\cdot\vec{\sigma}_2)(p^2_-)\delta^{a_1a'_1}\delta^{a_2a'_2}\quad-\frac{1}{4}h_{10}(\sigma^i_1\sigma^j_2)(p^i_-p^j_-)f^{a_1a'_1e}f^{a_2a'_2e}\\[.5em]
        &+\frac{1}{12}h_{10}(\vec{\sigma}_1\cdot\vec{\sigma}_2)(p_-^2)f^{a_1a'_1e}f^{a_2a'_2e}\quad+\frac{1}{4}h_{11}(\sigma^i_1\sigma^j_2)(p^i_+p^j_+)d^{ea_1a'_1}d^{ea_2a'_2}\\[.5em]
        &+\frac{1}{6}h_{11}(\sigma^i_1\sigma^j_2)(p^i_+p^j_+)id^{ea_1a'_1}f^{ea_2a'_2}\quad+\frac{1}{6}h_{11}(\sigma^i_1\sigma^j_2)(p^i_+p^j_+)if^{ea_1a'_1}d^{ea_2a'_2}\\[.5em]
        &-\frac{1}{9}h_{11}(\sigma^i_1\sigma^j_2)(p^i_+p^j_+)f^{ea_1a'_1}f^{ea_2a'_2}\quad-\frac{1}{12}h_{11}(\vec{\sigma}_1\cdot\vec{\sigma}_2)(p^2_+)d^{ea_1a'_1}d^{ea_2a'_2}\\[.5em]
        &-\frac{1}{18}h_{11}(\vec{\sigma}_1\cdot\vec{\sigma}_2)(p^2_+)id^{ea_1a'_1}f^{ea_2a'_2}\quad-\frac{1}{18}h_{11}(\vec{\sigma}_1\cdot\vec{\sigma}_2)(p^2_+)if^{ea_1a'_1}d^{ea_2a'_2}\\[.5em]
        &+\frac{1}{27}h_{11}(\vec{\sigma}_1\cdot\vec{\sigma}_2)(p^2_+)f^{ea_1a'_1}f^{ea_2a'_2}
    \end{split}
\end{equation}}
\newpage
\indent $\bullet$ Mixed Octet-Decuplet (\scalebox{1}{$BBDB$})
\\
\small{\begin{equation}
    \begin{split}
        V_{BBDB}=&\quad\frac{1}{4\sqrt{2}}q_3(\Vec{S}_1^{\dagger}\cdot\Vec{\sigma}_2)\Lambda_{a_1}^{e,  i_1j_1k_1}d^{ea_2a'_2}\delta^{i_1j_1k_1}_{i'_1j'_1k'_1}\quad+\frac{1}{6\sqrt{2}}q_3(\Vec{S}_1^{\dagger}\cdot\Vec{\sigma}_2)\Lambda_{a_1}^{e,  i_1j_1k_1}d^{ea_2a'_2}if^{ea_2a'_2}\delta^{i_1j_1k_1}_{i'_1j'_1k'_1}\\[.5em]
        &+\frac{1}{4\sqrt{2}}q_4(S_1^{\dagger,i}\sigma^j_2)(p^i_-p^j_-)\Lambda_{a_1}^{e, i_1j_1k_1}d^{ea_2a'_2}\delta^{i_1j_1k_1}_{i'_1j'_1k'_1}\quad+\frac{1}{6\sqrt{2}}q_4(S_1^{\dagger,i}\sigma^j_2)(p^i_-p^j_-)\Lambda_{a_1}^{e, i_1j_1k_1}if^{ea_2a'_2}\delta^{i_1j_1k_1}_{i'_1j'_1k'_1}\\[.5em]
        &-\frac{1}{12\sqrt{2}}q_4(\vec{S}_1^{\dagger}\cdot\vec{\sigma}_2)(p^2_-)\Lambda_{a_1}^{e, i_1j_1k_1}d^{ea_2a'_2}\delta^{i_1j_1k_1}_{i'_1j'_1k'_1}\quad-\frac{1}{18\sqrt{2}}q_4(\vec{S}_1^{\dagger}\cdot\vec{\sigma}_2)(p^2_-)\Lambda_{a_1}^{e, i_1j_1k_1}if^{ea_2a'_2}\delta^{i_1j_1k_1}_{i'_1j'_1k'_1}\\[.5em]
        &+\frac{1}{4\sqrt{2}}h_5(\Vec{S}_1^{\dagger}\cdot\Vec{\sigma}_2)(p^2_+)\Lambda_{a_1}^{e, i_1j_1k_1}d^{ea_2a'_2}\delta^{i_1j_1k_1}_{i'_1j'_1k'_1}\quad+\frac{1}{6\sqrt{2}}h_5(\Vec{S}_1^{\dagger}\cdot\Vec{\sigma}_2)(p^2_+)\Lambda_{a_1}^{e,  i_1j_1k_1}d^{ea_2a'_2}if^{ea_2a'_2}\delta^{i_1j_1k_1}_{i'_1j'_1k'_1}\\[.5em]
        &+\frac{1}{2\sqrt{2}}h_7(\Vec{S}_1^\dagger\delta^{\chi_2\chi'_2})\cdot i(\Vec{p}_+\times\Vec{p}_-) if^{ea_2a'_2}\delta^{i_1j_1k_1}_{i'_1j'_1k'_1}\quad+\frac{1}{4\sqrt{2}}h_{11}(S_1^{\dagger,i}\sigma^j_2)(p^i_+p^j_+)\Lambda_{a_1}^{e, i_1j_1k_1}d^{ea_2a'_2}\delta^{i_1j_1k_1}_{i'_1j'_1k'_1}\\[.5em]
        &+\frac{1}{6\sqrt{2}}h_{11}(S_1^{\dagger,i}\sigma^j_2)(p^i_+p^j_+)\Lambda_{a_1}^{e, i_1j_1k_1}if^{ea_2a'_2}\delta^{i_1j_1k_1}_{i'_1j'_1k'_1}\quad-\frac{1}{12\sqrt{2}}h_{11}(\vec{S}_1^{\dagger}\cdot\vec{\sigma}_2)(p^2_+)\Lambda_{a_1}^{e, i_1j_1k_1}d^{ea_2a'_2}\delta^{i_1j_1k_1}_{i'_1j'_1k'_1}\\[.5em]
        &-\frac{1}{18\sqrt{2}}h_{11}(\vec{S}_1^{\dagger}\cdot\vec{\sigma}_2)(p^2_+)\Lambda_{a_1}^{e, i_1j_1k_1}if^{ea_2a'_2}\delta^{i_1j_1k_1}_{i'_1j'_1k'_1}
    \end{split}
\end{equation}}
\\[1em]
\indent $\bullet$ Mixed Octet-Decuplet (\scalebox{1}{$DBDB$})
\\
\small{\begin{equation}
    \begin{split}
        V_{DBDB}=&\quad9q_1\delta^{\chi_1\chi'_1}\delta^{\chi_2\chi'_2}\delta^{i_1j_1k_1}_{i'_1j'_1k'_1}\delta^{a_2a'_2}\quad+\frac{3}{2}q_2\delta^{\chi_1\chi'_1}\delta^{\chi_2\chi'_2}\Lambda^{e, i'_1j'_1k'_1}_{i_1j_1k_1}if^{a_2a'_2e}\quad+\frac{3}{8}q_3(S^{r,\dagger}\vec{\sigma}_1S^r\cdot\vec{\sigma}_2)\Lambda^{e, i'_1j'_1k'_1}_{i_1j_1k_1}d^{ea_2a'_2}\\[.5em]
        &+\frac{1}{4}q_3(S^{r,\dagger}\vec{\sigma}_1S^r\cdot\vec{\sigma}_2)\Lambda^{e, i'_1j'_1k'_1}_{i_1j_1k_1}if^{ea_2a'_2}\quad+\frac{3}{8}q_4(p^i_-p^j_-)(S^{r,\dagger}\sigma_1^iS^r\cdot\sigma_2^j)\Lambda^{e, i'_1j'_1k'_1}_{i_1j_1k_1}d^{ea_2a'_2}\\[.5em]
        &+\frac{1}{4}q_4(p^i_-p^j_-)(S^{r,\dagger}\sigma_1^iS^r\cdot\sigma_2^j)\Lambda^{e, i'_1j'_1k'_1}_{i_1j_1k_1}if^{ea_2a'_2}\quad-\frac{1}{8}q_4(p^2_-)(S^{r,\dagger}\vec{\sigma}_1S^r\cdot\vec{\sigma}_2)\Lambda^{e, i'_1j'_1k'_1}_{i_1j_1k_1}d^{ea_2a'_2}\\[.5em]
        &-\frac{1}{12}q_4(p^2_-)(S^{r,\dagger}\vec{\sigma}_1S^r\cdot\vec{\sigma}_2)\Lambda^{e, i'_1j'_1k'_1}_{i_1j_1k_1}if^{ea_2a'_2}\quad+9h_1(p^2_+)\delta^{\chi_1\chi'_1}\delta^{\chi_2\chi'_2}\delta^{i_1j_1k_1}_{i'_1j'_1k'_1}\delta^{a_2a'_2}\\[.5em]
        &+\frac{3}{4}h_2(S^{r,\dagger}\vec{\sigma}_1S^r\cdot\vec{\sigma}_2)\delta^{i_1j_1k_1}_{i'_1j'_1k'_1}\delta^{a_2a'_2}\quad+\frac{9}{8}h_3(S^{r,\dagger}\vec{\sigma}_1S^r\cdot\vec{\sigma}_2)\Lambda^{e, i'_1j'_1k'_1}_{i_1j_1k_1}if^{a_2a'_2e}\\[.5em]
        &+\frac{3}{2}h_4(p^2_+)\delta^{\chi_1\chi'_1}\delta^{\chi_2\chi'_2}\Lambda^{e, i'_1j'_1k'_1}_{i_1j_1k_1}if^{a_2a'_2e}\quad+\frac{3}{8}h_5(p^2_+)(S^{r,\dagger}\vec{\sigma}_1S^r\cdot\vec{\sigma}_2)\Lambda^{e, i'_1j'_1k'_1}_{i_1j_1k_1}d^{ea_2a'_2}\\[.5em]
        &+\frac{1}{4}h_5(p^2_+)(S^{r,\dagger}\vec{\sigma}_1S^r\cdot\vec{\sigma}_2)\Lambda^{e, i'_1j'_1k'_1}_{i_1j_1k_1}if^{ea_2a'_2}\quad+\frac{3}{2}h_6i(\Vec{p}_+\times\Vec{p}_-)(S^{r,\dagger}\Vec{\sigma}_1S^r)\delta^{i_1j_1k_1}_{i'_1j'_1k'_1}\delta^{a_2a'_2}\\[.5em]
        &+\frac{1}{2}h_{6}i(\Vec{p}_+\times\Vec{p}_-)(\Vec{\sigma}_2)\delta^{i_1j_1k_1}_{i'_1j'_1k'_1}\delta^{a_2a'_2}\quad+\frac{3}{4}h_7i(\Vec{p}_+\times\Vec{p}_-)(S^{r,\dagger}\Vec{\sigma}_1S^r)\Lambda^{e, i'_1j'_1k'_1}_{i_1j_1k_1}if^{a_2a'_2e}\\[.5em]
        &+\frac{3}{4}h_7i(\Vec{p}_+\times\Vec{p}_-)(\Vec{\sigma}_2)\Lambda^{e, i'_1j'_1k'_1}_{i_1j_1k_1}d^{ea_2a'_2}\quad+\frac{1}{2}h_7i(\Vec{p}_+\times\Vec{p}_-)(\Vec{\sigma}_2)\Lambda^{e, i'_1j'_1k'_1}_{i_1j_1k_1}if^{a_2a'_2e}\\[.5em]
        &+\frac{9}{4}h_8i(\Vec{p}_+\times\Vec{p}_-)(S^{r,\dagger}\Vec{\sigma}_1S^r)\Lambda^{e, i'_1j'_1k'_1}_{i_1j_1k_1}if^{a_2a'_2e}\quad+\frac{3}{4}h_8i(\Vec{p}_+\times\Vec{p}_-)(\Vec{\sigma}_2)\Lambda^{e, i'_1j'_1k'_1}_{i_1j_1k_1}if^{a_2a'_2e}\\[.5em]
        &+\frac{3}{4}h_9(p^i_-p^j_-)(S^{r,\dagger}\sigma_1^iS^r\cdot\sigma_2^j)\delta^{i_1j_1k_1}_{i'_1j'_1k'_1}\delta^{a_2a'_2}\quad-\frac{1}{4}h_9(p^2_-)(S^{r,\dagger}\vec{\sigma}_1S^r\cdot\vec{\sigma}_2)\delta^{i_1j_1k_1}_{i'_1j'_1k'_1}\delta^{a_2a'_2}\\[.5em]
        &+\frac{9}{8}h_{10}(p^i_-p^j_-)(S^{r,\dagger}\sigma_1^iS^r\cdot\sigma_2^j)\Lambda^{e, i'_1j'_1k'_1}_{i_1j_1k_1}if^{a_2a'_2e}\quad-\frac{3}{8}h_{10}(p^2_-)(S^{r,\dagger}\vec{\sigma}_1S^r\cdot\vec{\sigma}_2)\Lambda^{e, i'_1j'_1k'_1}_{i_1j_1k_1}if^{a_2a'_2e}\\[.5em]
        &+\frac{3}{8}h_{11}(p^i_+p^j_+)(S^{r,\dagger}\sigma_1^iS^r\cdot\sigma_2^j)\Lambda^{e, i'_1j'_1k'_1}_{i_1j_1k_1}d^{ea_2a'_2}\quad+\frac{1}{4}h_{11}(p^i_+p^j_+)(S^{r,\dagger}\sigma_1^iS^r\cdot\sigma_2^j)\Lambda^{e, i'_1j'_1k'_1}_{i_1j_1k_1}if^{ea_2a'_2}\\[.5em]
        &-\frac{1}{8}h_{11}(p^2_+)(S^{r,\dagger}\vec{\sigma}_1S^r\cdot\vec{\sigma}_2)\Lambda^{e, i'_1j'_1k'_1}_{i_1j_1k_1}d^{ea_2a'_2}\quad-\frac{1}{12}h_{11}(p^2_+)(S^{r,\dagger}\vec{\sigma}_1S^r\cdot\vec{\sigma}_2)\Lambda^{e, i'_1j'_1k'_1}_{i_1j_1k_1}if^{ea_2a'_2}
    \end{split}
\end{equation}}
\newpage
\indent $\bullet$ Mixed Octet-Decuplet (\scalebox{1}{$BBDD$})
\\
\small{\begin{equation}
    \begin{split}
        V_{BBDD}=&\quad\frac{1}{8}q_3(\vec{S}_1^{\dagger}\cdot\vec{S}_2^{\dagger})\Lambda_{a_1}^{e, i_1j_1k_1}\Lambda_{a_2}^{e, i_2j_2k_2}\delta^{i_1j_1k_1}_{i'_1j'_1k'_1}\delta^{i_2j_2k_2}_{i'_2j'_2k'_2}\quad+\frac{1}{8}q_4(p^i_-p^j_-)(S_1^{i,\dagger}S_2^{j,\dagger})\Lambda_{a_1}^{e, i_1j_1k_1}\Lambda_{a_2}^{e, i_2j_2k_2}\delta^{i_1j_1k_1}_{i'_1j'_1k'_1}\delta^{i_2j_2k_2}_{i'_2j'_2k'_2}\\[.5em]
            &-\frac{1}{24}q_4(p^2_-)(\vec{S}_1^{\dagger}\cdot\vec{S}_2^{\dagger})\Lambda_{a_1}^{e, i_1j_1k_1}\Lambda_{a_2}^{e, i_2j_2k_2}\delta^{i_1j_1k_1}_{i'_1j'_1k'_1}\delta^{i_2j_2k_2}_{i'_2j'_2k'_2}\quad+\frac{1}{8}h_5(p^2_+)(\vec{S}_1^{\dagger}\cdot\vec{S}_2^{\dagger})\Lambda_{a_1}^{e, i_1j_1k_1}\Lambda_{a_2}^{e, i_2j_2k_2}\delta^{i_1j_1k_1}_{i'_1j'_1k'_1}\delta^{i_2j_2k_2}_{i'_2j'_2k'_2}\\[.5em]
        &+\frac{1}{8}h_{11}(p^i_+p^j_+)(S_1^{i,\dagger}S_2^{j,\dagger})\Lambda_{a_1}^{e, i_1j_1k_1}\Lambda_{a_2}^{e, i_2j_2k_2}\delta^{i_1j_1k_1}_{i'_1j'_1k'_1}\delta^{i_2j_2k_2}_{i'_2j'_2k'_2}\quad-\frac{1}{24}h_{11}(p^2_+)(\vec{S}_1^{\dagger}\cdot\vec{S}_2^{\dagger})\Lambda_{a_1}^{e, i_1j_1k_1}\Lambda_{a_2}^{e, i_2j_2k_2}\delta^{i_1j_1k_1}_{i'_1j'_1k'_1}\delta^{i_2j_2k_2}_{i'_2j'_2k'_2}
    \end{split}
\end{equation}}
\\[1em]
\indent $\bullet$ Mixed Octet-Decuplet (\scalebox{1}{$DBDD$})
\\
\small{\begin{equation}
    \begin{split}
        V_{DBDD}=&\quad\frac{3}{8\sqrt{2}}q_3(S^{r,\dagger}\vec{\sigma}_1S^r\cdot \vec{S}_2^{\dagger})\Lambda_{i_1j_1k_1}^{e, i'_1j'_1k'_1}\Lambda_{a_2}^{e, i_2j_2k_2}\delta^{i_2j_2k_2}_{i'_2j'_2k'_2}\quad+\frac{3}{8\sqrt{2}}q_4(p^i_-p^j_-)(S^{r,\dagger}\sigma_1^iS^r\cdot S_2^{j\dagger})\Lambda_{i_1j_1k_1}^{e, i'_1j'_1k'_1}\Lambda_{a_2}^{e, i_2j_2k_2}\delta^{i_2j_2k_2}_{i'_2j'_2k'_2}\\[.5em]
            &-\frac{1}{8\sqrt{2}}q_4(p^2_-)(S^{r,\dagger}\vec{\sigma}_1S^r\cdot \vec{S}_2)\Lambda_{i_1j_1k_1}^{e, i'_1j'_1k'_1}\Lambda_{a_2}^{e, i_2j_2k_2}\delta^{i_2j_2k_2}_{i'_2j'_2k'_2}\quad+\frac{3}{8\sqrt{2}}h_5(p^2_+)(S^{r,\dagger}\vec{\sigma}_1S^r\cdot \vec{S}_2^{\dagger})\Lambda_{i_1j_1k_1}^{e, i'_1j'_1k'_1}\Lambda_{a_2}^{e, i_2j_2k_2}\delta^{i_2j_2k_2}_{i'_2j'_2k'_2}\\[.5em]
        &+\frac{3}{4\sqrt{2}}h_7i(\Vec{p}_+\times\Vec{p}_-)(\Vec{S}_2^{\dagger})\Lambda_{i_1j_1k_1}^{e, i'_1j'_1k'_1}\Lambda_{a_2}^{e, i_2j_2k_2}\delta^{i_2j_2k_2}_{i'_2j'_2k'_2}\quad+\frac{3}{8\sqrt{2}}h_{11}(p^i_+p^j_+)(S^{r,\dagger}\sigma_1^iS^r\cdot S_2^{j\dagger})\Lambda_{i_1j_1k_1}^{e, i'_1j'_1k'_1}\Lambda_{a_2}^{e, i_2j_2k_2}\delta^{i_2j_2k_2}_{i'_2j'_2k'_2}\\[.5em]
            &-\frac{1}{8\sqrt{2}}h_{11}(p^2_+)(S^{r,\dagger}\vec{\sigma}_1S^r\cdot \vec{S}_2)\Lambda_{i_1j_1k_1}^{e, i'_1j'_1k'_1}\Lambda_{a_2}^{e, i_2j_2k_2}\delta^{i_2j_2k_2}_{i'_2j'_2k'_2}
    \end{split}
\end{equation}}
\\[1em]
\indent $\bullet$ Decuplet-Decuplet (\scalebox{1}{$DDDD$})
\small{\begin{alignat}{2}
    V_{DDDD}=& \quad 9q_1\delta^{\chi_1\chi'_1}\delta^{\chi_2\chi'_2}\delta^{i_1j_1k_1}_{i'_1j'_1k'_1}\delta^{i_2j_2k_2}_{i'_2j'_2k'_2}\quad+\frac{9}{4}q_2\delta^{\chi_1\chi'_1}\delta^{\chi_2\chi'_2}\Lambda_{i_1j_1k_1}^{e, i'_1j'_1k'_1}\Lambda_{i_2j_2k_2}^{e, i'_2j'_2k'_1}\notag\\[.5em]
    &+\frac{9}{16}q_3(S^{r,\dagger}\vec{\sigma}_1S^r\cdot S^{q,\dagger}\vec{\sigma}_2S^q)\Lambda_{i_1j_1k_1}^{e, i'_1j'_1k'_1}\Lambda_{i_2j_2k_2}^{e, i'_2j'_2k'_1}\quad+\frac{9}{16}q_4(p^i_-p^j_-)(S^{r,\dagger}\sigma_1^iS^r\cdot S^{q,\dagger}\sigma_2^jS^q)\Lambda_{i_1j_1k_1}^{e, i'_1j'_1k'_1}\Lambda_{i_2j_2k_2}^{e, i'_2j'_2k'_1}\notag\\[.5em]
        &-\frac{3}{16}q_4(p^2_-)(S^{r,\dagger}\vec{\sigma}_1S^r\cdot S^{q,\dagger}\vec{\sigma}_2S^q)\Lambda_{i_1j_1k_1}^{e, i'_1j'_1k'_1}\Lambda_{i_2j_2k_2}^{e, i'_2j'_2k'_1}\quad+9h_1(p^2_+)\delta^{\chi_1\chi'_1}\delta^{\chi_2\chi'_2}\delta^{i_1j_1k_1}_{i'_1j'_1k'_1}\delta^{i_2j_2k_2}_{i'_2j'_2k'_2}\notag\\[.5em]
        &+\frac{9}{4}h_2(S^{r,\dagger}\vec{\sigma}_1S^r\cdot S^{q,\dagger}\vec{\sigma}_2S^q)\delta^{i_1j_1k_1}_{i'_1j'_1k'_1}\delta^{i_2j_2k_2}_{i'_2j'_2k'_2}\quad+\frac{81}{16}h_3(S^{r,\dagger}\vec{\sigma}_1S^r\cdot S^{q,\dagger}\vec{\sigma}_2S^q)\Lambda_{i_1j_1k_1}^{e, i'_1j'_1k'_1}\Lambda_{i_2j_2k_2}^{e, i'_2j'_2k'_1}\notag\\[.5em]
        &+\frac{9}{4}h_4(p^2_+)\delta^{\chi_1\chi'_1}\delta^{\chi_2\chi'_2}\Lambda_{i_1j_1k_1}^{e, i'_1j'_1k'_1}\Lambda_{i_2j_2k_2}^{e, i'_2j'_2k'_1}\quad+\frac{9}{16}h_5(p^2_+)(S^{r,\dagger}\vec{\sigma}_1S^r\cdot S^{q,\dagger}\vec{\sigma}_2S^q)\Lambda_{i_1j_1k_1}^{e, i'_1j'_1k'_1}\Lambda_{i_2j_2k_2}^{e, i'_2j'_2k'_1}\notag\\[.5em]
        &+\frac{3}{2}h_6i(\Vec{p}_+\times\Vec{p}_-)(S^{r,\dagger}\Vec{\sigma}_1S^r + S^{q,\dagger}\Vec{\sigma}_2S^q)\delta^{i_1j_1k_1}_{i'_1j'_1k'_1}\delta^{i_2j_2k_2}_{i'_2j'_2k'_2}\notag\\[.5em]
        &+\frac{9}{8}h_7i(\Vec{p}_+\times\Vec{p}_-)(S^{r,\dagger}\Vec{\sigma}_1S^r + S^{q,\dagger}\Vec{\sigma}_2S^q)\Lambda_{i_1j_1k_1}^{e, i'_1j'_1k'_1}\Lambda_{i_2j_2k_2}^{e, i'_2j'_2k'_1}\label{DDDDp}\\[.5em]
        &+\frac{27}{8}h_8i(\Vec{p}_+\times\Vec{p}_-)(S^{r,\dagger}\Vec{\sigma}_1S^r + S^{q,\dagger}\Vec{\sigma}_2S^q)\Lambda_{i_1j_1k_1}^{e, i'_1j'_1k'_1}\Lambda_{i_2j_2k_2}^{e, i'_2j'_2k'_1}\notag\\[.5em]
        &+\frac{9}{4}h_9(p^i_-p^j_-)(S^{r,\dagger}\sigma_1^iS^r\cdot S^{q,\dagger}\sigma_2^jS^q)\delta^{i_1j_1k_1}_{i'_1j'_1k'_1}\delta^{i_2j_2k_2}_{i'_2j'_2k'_2}\notag\\[.5em]
        &-\frac{3}{4}h_9(p^2_-)(S^{r,\dagger}\vec{\sigma}_1S^r\cdot S^{q,\dagger}\vec{\sigma}_2S^q)\delta^{i_1j_1k_1}_{i'_1j'_1k'_1}\delta^{i_2j_2k_2}_{i'_2j'_2k'_2}\notag\\[.5em]
        &+\frac{81}{16}h_{10}(p^i_-p^j_-)(S^{r,\dagger}\sigma_1^iS^r\cdot S^{q,\dagger}\sigma_2^jS^q)\Lambda_{i_1j_1k_1}^{e, i'_1j'_1k'_1}\Lambda_{i_2j_2k_2}^{e, i'_2j'_2k'_1}\notag\\[.5em]
        &-\frac{27}{16}h_{10}(p^2_-)(S^{r,\dagger}\vec{\sigma}_1S^r\cdot S^{q,\dagger}\vec{\sigma}_2S^q)\Lambda_{i_1j_1k_1}^{e, i'_1j'_1k'_1}\Lambda_{i_2j_2k_2}^{e, i'_2j'_2k'_1}\notag\\[.5em]
        &+\frac{9}{16}h_{11}(p^i_+p^j_+)(S^{r,\dagger}\sigma_1^iS^r\cdot S^{q,\dagger}\sigma_2^jS^q)\Lambda_{i_1j_1k_1}^{e, i'_1j'_1k'_1}\Lambda_{i_2j_2k_2}^{e, i'_2j'_2k'_1}\notag\\[.5em]
        &-\frac{3}{16}h_{11}(p^2_+)(S^{r,\dagger}\vec{\sigma}_1S^r\cdot S^{q,\dagger}\vec{\sigma}_2S^q)\Lambda_{i_1j_1k_1}^{e, i'_1j'_1k'_1}\Lambda_{i_2j_2k_2}^{e, i'_2j'_2k'_1}.\notag
\end{alignat}}
\subsection{Pauli and Spin Transition Matrices, and Their Relation}\label{A-3-Pauli}
\indent This section has a collection of Pauli and spin transition matrices for spin $1/2$ and $3/2$. These definitions and expressions are frequently used to evaluate matrix elements \cite{susanne2020,asemke2012}. The Significant general properties of the Pauli matrices are as follows,
\begin{equation}
    \sigma_i^{\dagger}=\sigma_i, \hspace{.4cm} \text{tr}(\sigma_i) = 0, \hspace{.4cm} [\sigma_i, \sigma_j] = 2i\epsilon_{ijk}\sigma_k, \hspace{.4cm} \{\sigma_i, \sigma_j\} = 2\delta_{ij}, \hspace{.4cm} \text{det}\,\sigma_i = -1.
\end{equation}
By using the above identities, several expressions have emerged as
\begin{equation}
    \begin{split}
        &\sigma_i\sigma_i = \mathbb{1}, \hspace{.5cm} \sigma_i\sigma_j = \delta_{ij}+i\epsilon_{ijk}\sigma^k, \hspace{.5cm}  \sigma_i\sigma_j\sigma_k= i\epsilon_{ijk}+\delta_{ij}\sigma_k-\delta_{ik}\sigma_{j}+\delta_{jk}\sigma_i,\\[1em]
        & \sigma_i\sigma_j\sigma_i = -\sigma_j, \hspace{.5cm} \epsilon_{ijk} \sigma_i\sigma_j = 2i\sigma_k, \hspace{.5cm}  \sigma_j\sigma_i\sigma_k \sigma_j\sigma_i = 5\sigma_k,\\[1em]
        &\text{tr}(\sigma_i\sigma_j) = 2\delta_{ij}, \hspace{.5cm} \text{tr}(\sigma_i\sigma_j\sigma_k)= 2i\epsilon_{ijk}.
    \end{split}
\end{equation}
The Fiertz identities (Fiertz transformation) for the Pauli matrices are given as \cite{borodulin2022core32compendiumrelations},
\begin{equation}
    \begin{split}
        \sigma_{ab}^{i}\sigma_{cd}^{i} = 2\delta_{ad}\delta_{cb} - \delta_{ab}\delta_{cd} = \frac{3}{2}\delta_{ad}\delta_{cb} -\frac{1}{2}\sigma_{ad}^{i}\sigma_{cb}^{i}.
    \end{split}
\end{equation}
The spin-transition matrices for spin-$1/2$ and -$3/2$ are written down,
\begin{equation}
    \begin{split}
        &S_{1} = \frac{1}{\sqrt{6}}\begin{pmatrix} -\sqrt{3} & 0 & 1 & 0\\ 0 & - 1 & 0 &  \sqrt{3} \end{pmatrix}, \hspace{.2cm} S_{2} = \frac{-i}{\sqrt{6}}\begin{pmatrix}\sqrt{3} & 0 & 1 & 0\\ 0 & 1 & 0 &  \sqrt{3} \end{pmatrix},\\[1em]
        &S_{3} = \frac{1}{\sqrt{6}}\begin{pmatrix} 0 & 2 & 0 & 0\\ 0 & 0 & 2 &  0 \end{pmatrix},
    \end{split}
\end{equation}
with their hermitian conjugates ($S_{i}^{\dagger}$) are the inverse of spin-transition. There are the main properties as,
\begin{equation}
    \begin{split}
        &S_{i}S_{j}^{\dagger}=\delta_{ij}-\frac{1}{3}\sigma_i\sigma_j=\frac{2}{3}(\delta_{ij}-\frac{1}{2}i\epsilon_{ijk}\sigma_k), \hspace{.5cm} S^{\dagger}_i\sigma_j-S^{\dagger}_j\sigma_i= - i\epsilon_{ijk}S^{\dagger}_{k},\\[1em]
        &\Vec{S}\cdot\Vec{S}^{\dagger} = 2 \cdot\mathbb{1}_{2\times2}, \hspace{.4cm} \Vec{S}^{\dagger}\cdot\Vec{S} = \mathbb{1}_{4\times4}, \hspace{.4cm} \Vec{\sigma}\cdot\Vec{S} = 0, \hspace{.4cm} \Vec{S}^{\dagger}\cdot\Vec{\sigma} = 0,\\[1em]
        &S^{\dagger}_i\sigma_k\sigma_i= 2S_k, \hspace{.4cm} i\epsilon_{ijk}S^{\dagger}_i\sigma_j = \sigma_k.
    \end{split}
\end{equation}
Additional spin matrices of rank-2 and -3 are able to be represented through scalar and vector spin matrices, which are symmetric,
\begin{equation}
    \begin{split}
        &S_{ij} = -\frac{1}{\sqrt{6}}(\sigma_iS_j + \sigma_jS_i), \hspace{1cm} S_{ij}^{\dagger} = -\frac{1}{\sqrt{6}}(S^{\dagger}_i\sigma_j + S^{\dagger}_{j}\sigma_i),\\[1em]
        &\Sigma_{ij} = \frac{1}{8}(\Sigma_i\Sigma_{j}+\Sigma_j\Sigma_{i}-10\delta_{ij}\cdot\mathbb{1}) = \delta_{ij}\cdot\mathbb{1} - \frac{3}{2}(S^{\dagger}_iS_j+(S^{\dagger}_jS_i)\\[1em]
        &\Sigma_{ijk} = \frac{1}{36\sqrt{3}}\biggl(5(\Sigma_i\Sigma_j\Sigma_k+\Sigma_k\Sigma_i\Sigma_j+\Sigma_j\Sigma_k\Sigma_i+\Sigma_i\Sigma_k\Sigma_j+\Sigma_j\Sigma_i\Sigma_k+\Sigma_k\Sigma_j\Sigma_i)-82(\Sigma_i\delta_{jk}+\Sigma_j\delta_{ik}+\Sigma_{k}\delta_{ij})\biggr)
    \end{split}
\end{equation}
Here, the useful product of two matrices provides the  following expressions,
\begin{equation}
    \begin{split}
        &S^{\dagger}_i\sigma_j = -\sqrt{\frac{3}{2}}S_{ij}^{\dagger}-\frac{1}{2}i\epsilon_{ijk}S_{k}^{\dagger}, \hspace{1cm} \sigma_iS_j = -\sqrt{\frac{3}{2}}S_{ij}-\frac{1}{2}i\epsilon_{ijk}S_{k},\\[1em]
        &S^{\dagger}_iS_j = \frac{1}{3}\delta_{ij}\cdot\mathbb{1}-\frac{1}{3}\Sigma_{ij}+\frac{1}{6}i\epsilon_{ijk}\Sigma_{k}, \hspace{.6cm} S_iS_j^{\dagger} = \frac{1}{3}(2\delta_{ij}\cdot\mathbb{1}-i\epsilon_{ijk}\Sigma_{k}),\\[1em]
        &\Sigma_iS_j^{\dagger}= -\sqrt{\frac{3}{2}}S_{ij}^{\dagger}+\frac{5}{2}i\epsilon_{ijk}S_{k}^{\dagger}, \hspace{1cm} S_i\Sigma_j = -\sqrt{\frac{3}{2}}S_{ij}+\frac{5}{2}i\epsilon_{ijk}S_{k},\\[1em]
        &\Sigma_i\Sigma_j = 5\delta_{ij}\cdot\mathbb{1}+4\Sigma_{ij}+i\epsilon_{ijk}\Sigma_k, \hspace{1cm} \Sigma_i\Sigma_j\Sigma_i = 11 \Sigma_j\\[1em]
        &\Sigma_i\Sigma_i = 15\cdot\mathbb{1}.
    \end{split}
\end{equation}
\subsection{Gell-mann Matrix Properties in SU(3) and Tensor Relations}\label{A-4-Gell-mann}
\indent The Gell-Mann matrices ($\lambda_a$) are a set of $3\times3$ Hermitian and traceless matrices that serve as the generators of the SU(3) group. We summarize the general properties of the Gell-Mann matrix and relations used for our calculation \cite{borodulin2022core32compendiumrelations,asemke2012},
\begin{equation}
    \begin{split}
        &\lambda_a = \lambda^{\dagger}_a, \hspace{1cm} [\lambda^{a}, \lambda^{b}] = 2if^{abc}\lambda^{c}, \hspace{1cm} \{\lambda^{a}, \lambda^{b}\} = \frac{4}{3}\delta^{ij}+2d^{abc}\lambda^{c},\\[1em]
        &\lambda^{a}\lambda^{b} = \frac{2}{3}\delta^{ab}+(d^{abc}+if^{abc})\lambda^{c},\\[1em]
        &\text{tr}(\lambda^{a}) = 0, \hspace{.5cm} \text{tr}(\lambda^{a}\lambda^{b}) = 2\delta^{ab},\\[1em]
        &\text{tr}(\lambda^{a}\lambda^{b}\lambda^{c}) = 2(d^{abc}+if^{abc}),\\[1em]
        &\text{tr}(\lambda^{a}\lambda^{b}\lambda^{c}\lambda^{d}) = \frac{4}{3}\delta^{ab}\delta^{cd}+2(d^{abn}+if^{abn})(d^{ncd}+if^{ncd})
    \end{split}
\end{equation}
The tensors $f$ and $d$ are the anti-symmetry and symmetry tensors, respectively. the non-zero values of the SU(3) structure constants $f$ and $d$ are equal to,
\begin{equation}
    \begin{split}
        &f^{123} = 1,\,\,\, f^{147}=-f^{156}=f^{246}=f^{257} =f^{345}=-f^{367}=\frac{1}{2},\\[1em]
        &f^{458}=f^{678} = \frac{\sqrt{3}}{2},\\[1em]
        &d^{146}=d^{157}=-d^{247}=d^{256}=d^{344}=d^{355}=-d^{366}=-d^{377}=\frac{1}{2}\\[1em]
        &d^{118}=d^{228}=d^{338}=-d^{888}=\frac{1}{\sqrt{3}},\,\,\, d^{448}=d^{558}=d^{668}=d^{778}=-\frac{1}{2\sqrt{3}}.
    \end{split}
\end{equation}
The Fiertz identities (Fiertz transformation) for $\lambda$ matrices have the form as,
\begin{equation}
    \begin{split}
        &\lambda^{a}_{ij}\lambda^{a}_{kl} = 2\delta_{il}\delta_{jk}-\frac{2}{3}\delta_{ij}\delta_{kl},\\[1em]
        &\lambda^{a}_{ij}\lambda^{b}_{kl} = 2\delta^{ab}_{il}\delta_{jk}-\frac{2}{3}\delta_{ij}\delta^{ab}_{kl}. 
    \end{split}
\end{equation}
The coefficient $f$ and $d$ have the Jacobi identities equal ,
\begin{equation}
    \begin{split}
        &d^{abe}f^{ecl}+d^{bce}f^{eal}+d^{cae}f^{ebl}= 0,\\
        &f^{abe}f^{ecl}+f^{bce}f^{eal}+f^{cae}f^{ebl}= 0.
    \end{split}
\end{equation}
Here are some beneficial variations relevant to our work,
\begin{equation}
    \begin{split}
        d^{abe}d^{ecl}+d^{bce}d^{eal}+d^{cae}d^{ebl} &= \frac{1}{3}(\delta^{ab}\delta^{cl}+\delta^{ac}\delta^{bl}+\delta^{al}\delta^{bc}),\\[.1em]
        d^{ace}d^{ble}-d^{ale}d^{bce} &= f^{abe}f^{ecl}-\frac{2}{3}(\delta^{ac}\delta^{bl}-\delta^{al}\delta^{bc}),\\
        f^{ace}f^{ble}+f^{ale}f^{bce} &= \delta^{ac}\delta^{bl}+\delta^{al}\delta^{bc}-\delta^{ab}\delta^{cl}-3d^{abe}d^{ecl},\\
        f^{abe}f^{ecl}&=f^{ace}f^{ebl}-f^{ale}f^{ebc},\\
        f^{abe}d^{ecl}&=f^{ace}d^{ebl}+f^{ale}d^{ebc},\\
        d^{aac}&=f^{aac}=d^{abc}f^{abm}=0.
    \end{split}
\end{equation}
\newpage
\subsection{ The $1/N_c$ sum rule in terms of coupling constant}\label{A-4.5-Sumrule}
In this section, we want to investigate further on our $1/N_c$ operator analysis for couplings, $g_{xy,z}^{i(\alpha)}$. According the sum rules, which are given in \ref{sec-3}, we can construct the new sets of sum rule that represent in terms of the couplings. \textcolor{black}{At the LO of the $1/N_c$ Hartree potential, the sum rules of couplings are given as follows:}\\
\textcolor{black}{
\indent $\bullet$ Octet-Octet (\scalebox{1}{$BBBB$})
    \begin{equation}
    \begin{split}
        g_{00,1}^{1(1)}=&\frac{17}{10}g_{00,1}^{3(3)}, \quad\quad g_{00,1}^{1(2)}= -\frac{13}{10}g_{00,1}^{3(3)}, \quad\quad g_{00,1}^{1(3)}=-g_{00,1}^{3(3)},\\[.5em]
        g_{00,1}^{2(1)}=&-\frac{7}{5}g_{00,1}^{3(3)}, \quad\quad g_{00,1}^{2(2)}= -\frac{1}{5}g_{00,1}^{3(3)},  \quad\quad g_{00,1}^{2(3)}=\frac{2}{5}g_{00,1}^{3(3)}, \\[.5em]
        g_{00,1}^{3(1)}=&-\frac{17}{10}g_{00,1}^{3(3)}, \quad\quad g_{00,1}^{3(2)}= -\frac{13}{10}g_{00,1}^{3(3)},\\[.5em]
         g_{00,1}^{4(1)}=&0, \quad\quad\quad\quad\quad\quad g_{00,1}^{4(2)}= 0,  \quad\quad\quad\quad\quad\quad g_{00,1}^{4(3)}=0\\[.5em]
         g_{00,1}^{5(1)}=&0, \quad\quad\quad\quad\quad\quad g_{00,1}^{5(2)}= 0,  \quad\quad\quad\quad\quad\quad g_{00,1}^{5(3)}=0.
    \end{split}
\end{equation}
For the Lagrangian of octet-octet sector, by using  $1/N_c$ operator analysis method, we can establish 14 the couplings sum rules with single free parameter, $g_{00,1}^{3(3)}$.}\\[.5em]

\textcolor{black}{
\indent $\bullet$ Mixed Octet-Decuplet (\scalebox{1}{$BBDB$})\\
    \begin{equation}
        g_{01,1}^{4(1)}=-5g_{01,1}^{4(2)}.
    \end{equation}
Similarly, we also use the $1/N_c$ operator analysis on the $BB$ to $DB$ to form the sum rule with free parameters, $g_{01,1}^{4(2)}$.}\\[.5em]

\textcolor{black}{
\indent $\bullet$ Mixed Octet-Decuplet (\scalebox{1}{$DBDB$})
    \begin{equation}
    \begin{split}
        &g_{11,1}^{1(1)} = 0, \quad\quad\quad g_{11,1}^{1(2)} = -g_{11,1}^{1(4)}, \quad\quad g_{11,1}^{1(3)} = 0, \\[.5em]
            &g_{11,1}^{2(1)} = 0, \quad\quad\quad g_{11,1}^{2(2)} =-g_{11,1}^{2(4)}, \quad\quad\quad  g_{11,1}^{2(3)} = 0,\\[.5em]
            &g_{11,1}^{3(1)} = 0, \quad\quad\quad g_{11,1}^{3(2)} = -g_{11,1}^{3(4)}, \quad\quad\quad g_{11,1}^{3(3)} = 0,\\[.5em]
            &g_{11,1}^{4(1)} = 0, \quad\quad\quad g_{11,1}^{4(2)} = -g_{11,1}^{4(4)},  \quad\quad\quad g_{11,1}^{4(3)} = 0,\\[.5em]
            &g_{11,1}^{5(1)} = 0, \quad\quad\quad g_{11,1}^{5(2)} = - g_{11,1}^{5(4)},\quad\quad\quad g_{11,1}^{5(3)} = 0.
    \end{split}
\end{equation}
For the sector of $DB$ to $DB$, our $1/N_c$ operator analysis constructs 15 sum rules with these free parameters , $g_{11,1}^{1(4)}$, $g_{11,1}^{2(4)}$, $g_{11,1}^{3(4)}$, $g_{11,1}^{4(4)}$, and $g_{11,1}^{5(4)}$.} \\[.5em]

\textcolor{black}{
\indent $\bullet$ Mixed Octet-Decuplet (\scalebox{1}{$BBDD$})
    \begin{equation}
    \begin{split}
        &g_{02,1}^{1(1)} = -\frac{66}{5}g_{00,1}^{3(3)}, \quad\quad g_{02,1}^{2(1)} =  -\frac{12}{5}g_{00,1}^{3(3)},\\[.5em]
        &g_{02,1}^{3(1)} = \frac{12}{5}g_{00,1}^{3(3)}, \hspace{2.7em} g_{02,2}^{3(1)} = -\frac{6}{5}g_{00,1}^{3(3)},\\[.5em]
        &g_{02,1}^{4(1)}= 0, \hspace{6em} g_{02,2}^{4(1)}=0, \\[.5em]
         &g_{02,1}^{5(1)} =0, \hspace{6em} g_{02,2}^{5(1)}= 0.\\[.5em]
    \end{split}
\end{equation}
In $BB$ to $DD$ sector, we can summarize that 8  Lagrangian couplings are described by single octet-octet free parameters, $g_{00,1}^{3(3)}$.}\\[.5em]

\textcolor{black}{
\indent $\bullet$  Mixed Octet-Decuplet (\scalebox{1}{$DBDD$})
    \begin{equation}
    g_{12,1}^{4(1)} = -\frac{243}{85}g_{00,1}^{3(3)}, \quad\quad  g_{12,2}^{4(1)} = \frac{1782}{85}g_{00,1}^{3(3)}, \quad\quad  g_{12,3}^{4(1)} = \frac{162}{85}g_{00,1}^{3(3)}.
\end{equation}
The 3  couplings sum rules in sector $DB$ to $DD$ are described by single octet-octet free parameters, $g_{00,1}^{3(3)}$.} \\[.5em]

\textcolor{black}{
\indent $\bullet$ Decuplet-Decuplet (\scalebox{1}{$DDDD$})
    \begin{alignat}{2}
         g_{22,1}^{1(1)}=&-\frac{1}{2}g_{22,7}^{4(1)}-\frac{4}{3}g_{22,2}^{5(1)}-\frac{2}{3}g_{22,3}^{5(1)}-\frac{3}{4}g_{22,5}^{5(1)}-\frac{1}{3}g_{22,7}^{5(1)},\notag\\[.5em]
         g_{22,2}^{1(1)}=&-\frac{1}{9}g_{22,6}^{4(1)}-\frac{1}{3}g_{22,7}^{4(1)}-\frac{5}{2}g_{22,2}^{5(1)}+\frac{1}{5}g_{22,3}^{5(1)}+\frac{1}{5}g_{22,4}^{5(1)}-\frac{1}{2}g_{22,5}^{5(1)}-\frac{1}{4}g_{22,7}^{5(1)},\notag\\[.5em]
         g_{22,3}^{1(1)}=&-\frac{1}{9}g_{22,6}^{4(1)}-\frac{1}{3}g_{22,7}^{4(1)}-\frac{1}{2}g_{22,2}^{5(1)}+\frac{11}{5}g_{22,3}^{5(1)}+\frac{1}{5}g_{22,4}^{5(1)}-\frac{1}{2}g_{22,5}^{5(1)}-\frac{1}{4}g_{22,7}^{5(1)},\notag\\[.5em]
         g_{22,1}^{1(2)}=&\frac{1}{6}g_{22,3}^{3(2)}+\frac{1}{9}g_{22,5}^{4(1)}+\frac{1}{2}g_{22,7}^{4(1)}-\frac{1}{2}g_{22,2}^{4(2)}+\frac{1}{3}g_{22,3}^{4(2)}+\frac{13}{3}g_{22,2}^{5(1)}+\frac{5}{2}g_{22,3}^{5(1)}+\frac{1}{2}g_{22,4}^{5(1)}+\frac{5}{3}g_{22,5}^{5(1)}+\frac{1}{6}g_{22,6}^{5(1)}+\frac{1}{2}g_{22,7}^{5(1)},\notag\\[.5em]
         g_{22,2}^{1(2)}=&-\frac{3}{2}g_{22,2}^{3(2)}+\frac{1}{3}g_{22,3}^{3(2)}+\frac{1}{2}g_{22,5}^{3(2)}-\frac{2}{3}g_{22,7}^{3(2)}+\frac{1}{3}g_{22,5}^{4(1)}-\frac{2}{3}g_{22,6}^{4(1)}-\frac{15}{2}g_{22,7}^{4(1)}-\frac{6}{5}g_{22,2}^{4(2)}-\frac{1}{2}g_{22,3}^{4(2)}+\frac{189}{4}g_{22,2}^{5(1)}+\frac{1}{3}g_{22,3}^{5(1)}\notag\\[.5em]
                    &+\frac{26}{3}g_{22,4}^{5(1)}+\frac{24}{5}g_{22,5}^{5(1)}-g_{22,6}^{5(1)}+\frac{23}{3}g_{22,7}^{5(1)},\notag\\[.5em]
        g_{22,3}^{1(2)}=&-\frac{1}{2}g_{22,2}^{3(2)}-\frac{2}{3}g_{22,3}^{3(2)}+\frac{1}{2}g_{22,5}^{3(2)}-\frac{2}{3}g_{22,7}^{3(2)}+\frac{1}{3}g_{22,5}^{4(1)}-\frac{2}{3}g_{22,6}^{4(1)}-\frac{15}{2}g_{22,7}^{4(1)}-\frac{6}{5}g_{22,2}^{4(2)}-\frac{1}{2}g_{22,3}^{4(2)}+\frac{189}{4}g_{22,2}^{5(1)}+\frac{1}{3}g_{22,3}^{5(1)}\notag\\[.5em]
                    &+\frac{26}{3}g_{22,4}^{5(1)}+\frac{24}{5}g_{22,5}^{5(1)}-g_{22,6}^{5(1)}+\frac{23}{3}g_{22,7}^{5(1)},\notag\\[.5em]
        g_{22,1}^{2(1)}=&-\frac{1}{3}g_{22,5}^{4(1)}+\frac{1}{3}g_{22,6}^{4(1)}+\frac{3}{4}g_{22,7}^{4(1)}-16g_{22,2}^{5(1)}-2g_{22,3}^{5(1)}-2g_{22,4}^{5(1)}-\frac{10}{3}g_{22,5}^{5(1)}-\frac{8}{3}g_{22,7}^{5(1)},\notag\\[.5em]
        g_{22,2}^{2(1)}=&-g_{22,4}^{2(1)}-g_{22,6}^{3(1)}+\frac{1}{7}g_{22,5}^{4(1)}-\frac{1}{3}g_{22,6}^{4(1)}-\frac{4}{3}g_{22,7}^{4(1)}+16g_{22,2}^{5(1)}+\frac{1}{3}g_{22,3}^{5(1)}-\frac{1}{2}g_{22,4}^{5(1)}+\frac{25}{8}g_{22,5}^{5(1)}-\frac{1}{2}g_{22,6}^{5(1)}+\frac{5}{2}g_{22,7}^{5(1)},\notag\\[.5em]
        g_{22,3}^{2(1)}=&\frac{5}{4}g_{22,5}^{4(1)}-\frac{6}{5}g_{22,6}^{4(1)}+\frac{1}{6}g_{22,7}^{4(1)}+\frac{82}{3}g_{22,2}^{5(1)}-\frac{7}{2}g_{22,3}^{5(1)}-\frac{3}{2}g_{22,4}^{5(1)}+6g_{22,5}^{5(1)}+\frac{1}{3}g_{22,6}^{5(1)}+4g_{22,7}^{5(1)},\notag\\[.5em]
        g_{22,1}^{2(2)}=&-\frac{1}{6}g_{22,3}^{3(2)}+\frac{1}{2}g_{22,5}^{4(1)}-\frac{2}{3}g_{22,6}^{4(1)}-\frac{11}{3}g_{22,7}^{4(1)}-\frac{1}{3}g_{22,3}^{4(2)}+45g_{22,2}^{5(1)}+\frac{20}{3}g_{22,3}^{5(1)}+\frac{15}{2}g_{22,4}^{5(1)}+8g_{22,5}^{5(1)}-\frac{1}{3}g_{22,6}^{5(1)}+\frac{27}{4}g_{22,7}^{5(1)},\notag\\[.5em]
        g_{22,2}^{2(2)}=&\frac{7}{4}g_{22,3}^{2(2)}-g_{22,4}^{2(2)}+\frac{1}{2}g_{22,2}^{3(2)}-\frac{23}{3}g_{22,3}^{3(2)}-\frac{3}{2}g_{22,5}^{3(2)}-g_{22,6}^{3(2)}+5g_{22,7}^{3(2)}-\frac{5}{3}g_{22,5}^{4(1)}+\frac{11}{3}g_{22,6}^{4(1)}+\frac{171}{4}g_{22,7}^{4(1)}+\frac{33}{4}g_{22,2}^{4(2)}\notag\\[.5em]
                    &-\frac{5}{4}g_{22,3}^{4(2)}-248g_{22,2}^{5(1)}+\frac{29}{5}g_{22,3}^{5(1)}-46g_{22,4}^{5(1)}-22g_{22,5}^{5(1)}+\frac{43}{7}g_{22,6}^{5(1)}-\frac{83}{2}g_{22,7}^{5(1)},\notag\\[.5em]
        g_{22,1}^{3(1)}=&\frac{1}{7}g_{22,7}^{4(1)}-2g_{22,2}^{5(1)}+\frac{3}{4}g_{22,3}^{5(1)}+\frac{1}{3}g_{22,4}^{5(1)}-\frac{1}{2}g_{22,5}^{5(1)}-\frac{1}{2}g_{22,7}^{5(1)},\notag\\[.5em]
        g_{22,2}^{3(1)}=&2g_{22,2}^{5(1)}, \quad\quad g_{22,3}^{3(1)}=-2g_{22,3}^{5(1)},\\[.5em]
        g_{22,4}^{3(1)}=&-g_{22,6}^{3(1)}-\frac{1}{9}g_{22,6}^{4(1)}-\frac{2}{3}g_{22,7}^{4(1)}-\frac{1}{2}g_{22,2}^{5(1)}+\frac{1}{5}g_{22,3}^{5(1)}+\frac{1}{5}g_{22,4}^{5(1)}-\frac{1}{3}g_{22,5}^{5(1)}-\frac{1}{2}g_{22,7}^{5(1)},\notag\\[.5em]
        g_{22,5}^{3(1)}=&\frac{1}{2}g_{22,5}^{4(1)}-\frac{2}{3}g_{22,6}^{4(1)}-\frac{2}{3}g_{22,7}^{4(1)}+\frac{39}{2}g_{22,2}^{5(1)}+\frac{1}{5}g_{22,3}^{5(1)}+\frac{1}{5}g_{22,4}^{5(1)}+\frac{33}{8}g_{22,5}^{5(1)}+\frac{7}{2}g_{22,7}^{5(1)},\notag\\[.5em]
        g_{22,7}^{3(1)}=&-\frac{1}{2}g_{22,5}^{4(1)}+\frac{1}{2}g_{22,6}^{4(1)}-20g_{22,2}^{5(1)}-\frac{9}{2}g_{22,5}^{5(1)}-4g_{22,7}^{5(1)},\notag\\[.5em]
        g_{22,1}^{3(2)}=&-\frac{1}{6}g_{22,3}^{3(2)}-\frac{3}{2}g_{22,7}^{4(1)}-\frac{1}{3}g_{22,3}^{4(2)}+3g_{22,2}^{5(1)}-2g_{22,3}^{5(1)}+\frac{2}{3}g_{22,4}^{5(1)}-\frac{3}{4}g_{22,5}^{5(1)}-\frac{1}{3}g_{22,6}^{5(1)}+\frac{2}{3}g_{22,7}^{5(1)},\notag\\[.5em]
        g_{22,4}^{3(2)}=&g_{22,5}^{3(2)}-g_{22,6}^{3(2)}+g_{22,7}^{3(2)},\notag\\[.5em]
        g_{22,1}^{4(1)}=&\frac{7}{2}g_{22,2}^{5(1)}+g_{22,5}^{5(1)}+\frac{2}{3}g_{22,7}^{5(1)},\notag\\[.5em]
        g_{22,2}^{4(1)}=&\frac{1}{9}g_{22,5}^{4(1)}-\frac{1}{12}g_{22,6}^{4(1)}+\frac{19}{12}g_{22,7}^{4(1)}+\frac{77}{36}g_{22,2}^{5(1)}-\frac{13}{18}g_{22,3}^{5(1)}-\frac{16}{9}g_{22,4}^{5(1)}+2g_{22,5}^{5(1)}+\frac{17}{108}g_{22,6}^{5(1)}+g_{22,7}^{5(1)},\notag\\[.5em]
        g_{22,3}^{4(1)}=&\frac{2}{9}g_{22,5}^{4(1)}-\frac{1}{6}g_{22,6}^{4(1)}+\frac{7}{6}g_{22,7}^{4(1)}+\frac{59}{18}g_{22,2}^{5(1)}+\frac{5}{9}g_{22,3}^{5(1)}-\frac{14}{9}g_{22,4}^{5(1)}+2g_{22,5}^{5(1)}+\frac{17}{54}g_{22,6}^{5(1)}+g_{22,7}^{5(1)},\notag\\[.5em]
        g_{22,4}^{4(1)}=&4g_{00,1}^{3(3)}-\frac{1}{3}g_{22,6}^{4(1)}+\frac{7}{3}g_{22,7}^{4(1)}+\frac{50}{9}g_{22,2}^{5(1)}-\frac{8}{9}g_{22,3}^{5(1)}-\frac{28}{9}g_{22,4}^{5(1)}+3g_{22,5}^{5(1)}-\frac{10}{27}g_{22,6}^{5(1)}+2g_{22,7}^{5(1)},\notag\\[.5em]
        g_{22,1}^{4(2)}=&\frac{1}{6}g_{22,3}^{3(2)}+\frac{1}{9}g_{22,6}^{4(1)}+g_{22,7}^{4(1)}+\frac{1}{3}g_{22,3}^{4(2)}-\frac{22}{3}g_{22,2}^{5(1)}-\frac{1}{3}g_{22,3}^{5(1)}-\frac{4}{3}g_{22,4}^{5(1)}-g_{22,5}^{5(1)}+\frac{1}{8}g_{22,6}^{5(1)}-\frac{6}{5}g_{22,7}^{5(1)},\notag\\[.5em]
        g_{22,1}^{5(1)}=&-\frac{1}{7}g_{22,7}^{4(1)}+3g_{22,2}^{5(1)}+\frac{1}{5}g_{22,4}^{5(1)}+\frac{2}{3}g_{22,5}^{5(1)}+\frac{1}{2}g_{22,7}^{5(1)},\notag\\[.5em]
        g_{22,1}^{5(2)}=&\frac{1}{9}g_{22,6}^{4(1)}+g_{22,7}^{4(1)}+\frac{1}{6}g_{22,3}^{4(2)}-\frac{22}{3}g_{22,2}^{5(1)}-\frac{1}{3}g_{22,3}^{5(1)}-\frac{4}{3}g_{22,4}^{5(1)}-g_{22,5}^{5(1)}+\frac{1}{8}g_{22,6}^{5(1)}-\frac{6}{5}g_{22,7}^{5(1)}.\notag
    \end{alignat}}
\textcolor{black}{
For decuplet-decuplet sector, we can establish 26 sum rules in terms of these free parameters, $g_{00,1}^{3(3)}$, $g_{22,5}^{4(1)}$, $g_{22,6}^{4(1)}$, $g_{22,7}^{4(1)}$, $g_{22,2}^{5(1)}$, $g_{22,3}^{5(1)}$, $g_{22,4}^{5(1)}$, $g_{22,5}^{5(1)}$, $g_{22,6}^{5(1)}$, $g_{22,7}^{5(1)}$, $g_{22,2}^{3(2)}$, $g_{22,3}^{3(2)}$, $g_{22,5}^{3(2)}$, $g_{22,6}^{3(2)}$, $g_{22,7}^{3(2)}$, $g_{22,2}^{4(2)}$, $g_{22,3}^{4(2)}$, $g_{22,6}^{3(1)}$.}\\

\indent Next, \textcolor{black}{the NNLO sum rules of the $1/N_c$ Hartree potential} are provided in terms of the coupling constant. the sum rules of couplings are given as follows:\\
\textcolor{black}{
\indent $\bullet$ Octet-Octet (\scalebox{1}{$BBBB$})
    \begin{equation}
    \begin{split}
        g_{00,1}^{1(1)} =& -2g_{00,1}^{1(2)}-\frac{3}{8}g_{00,1}^{2(3)}-\frac{3}{4}g_{00,1}^{3(2)}-\frac{27}{4}g_{00,1}^{5(3)}, \quad\quad g_{00,1}^{2(1)} = -\frac{11}{4}g_{00,1}^{2(3)}+4g_{00,1}^{3(2)}-\frac{11}{2}g_{00,1}^{3(3)}-\frac{9}{2}g_{00,1}^{5(3)}, \\[.5em]
        g_{00,1}^{2(2)} =& g_{00,1}^{2(3)}-2g_{00,1}^{3(2)}+2g_{00,1}^{3(3)}+9g_{00,1}^{5(3)}, \hspace{5em} g_{00,1}^{3(1)} =\frac{3}{8}g_{00,1}^{2(3)}-2g_{00,1}^{3(2)}+\frac{3}{4}g_{00,1}^{3(3)}+\frac{27}{4}g_{00,1}^{5(3)},\\[.5em]
        g_{00,1}^{4(1)} =& - g_{00,1}^{5(3)}, \quad\quad g_{00,1}^{4(2)} = 5g_{00,1}^{5(3)},\quad\quad  g_{00,1}^{4(3)} = 2g_{00,1}^{5(3)}, \quad\quad g_{00,1}^{5(1)} = -\frac{1}{2}g_{00,1}^{5(3)}, \quad\quad g_{00,1}^{5(2)} = \frac{5}{2}g_{00,1}^{5(3)}.
    \end{split}
\end{equation}
For the Lagrangian of octet-octet sector, using  $1/N_c$ operator analysis method, we can establish 9 couplings sum rules , which express in terms of free parameters $g_{00,1}^{1(2)}$, $g_{00,1}^{2(3)}$,  $g_{00,1}^{3(2)}$ , $g_{00,1}^{3(3)}$, $g_{00,1}^{5(3)}$.}\\[.5em]
\textcolor{black}{
\indent $\bullet$ Mixed Octet-Decuplet (\scalebox{1}{$BBDB$})\\
    \begin{equation}
        g_{01,1}^{4(1)}=-5g_{01,1}^{4(2)}.
    \end{equation}
We also use the $1/N_c$ operator analysis on the $BB$ to $DB$ to form the sum rule with free parameter, $g_{01,1}^{4(2)}$.}\\[.5em]
\textcolor{black}{
\indent $\bullet$ Mixed Octet-Decuplet (\scalebox{1}{$DBDB$})
    \begin{equation}
        \begin{split}
            &g_{11,1}^{1(1)} = g_{00,1}^{1(3)}+g_{00,1}^{3(3)}, \quad\quad g_{11,1}^{1(2)} = -g_{11,1}^{1(4)}, \quad\quad\quad g_{11,1}^{1(3)} = 0, \\[.5em]
            &g_{11,1}^{2(1)} = 0, \quad\quad\quad\quad\qquad\,\, g_{11,1}^{2(2)} =-g_{11,1}^{2(4)}, \quad\quad\quad  g_{11,1}^{2(3)} = 0,\\[.5em]
            &g_{11,1}^{3(1)} = 0, \quad\quad\quad\quad\qquad\,\, g_{11,1}^{3(2)} = -g_{11,1}^{3(4)}, \quad\quad\quad g_{11,1}^{3(3)} = 0,\\[.5em]
            &g_{11,1}^{4(1)} = 0, \quad\quad\quad\quad\qquad\,\, g_{11,1}^{4(2)} = -g_{11,1}^{4(4)},  \quad\quad\quad g_{11,1}^{4(3)} = 0,\\[.5em]
            &g_{11,1}^{5(1)} = 0, \quad\quad\quad\quad\qquad\,\, g_{11,1}^{5(2)} = - g_{11,1}^{5(4)},\quad\quad\quad g_{11,1}^{5(3)} = 0.
        \end{split}
    \end{equation}
For the sector of $DB$ to $DB$, our $1/N_c$ operator analysis constructs 15 sum rules with these free parameters , $g_{11,1}^{1(4)}$, $g_{11,1}^{2(4)}$, $g_{11,1}^{3(4)}$, $g_{11,1}^{4(4)}$, $g_{11,1}^{5(4)}$, and $g_{00,1}^{1(3)}$, $g_{00,1}^{3(3)}$ from $BBBB$ sector .}\\[.5em]
\textcolor{black}{
\indent $\bullet$  Mixed Octet-Decuplet (\scalebox{1}{$BBDD$})
    \begin{equation}
    \begin{split}
        &g_{02,1}^{1(1)} = -\frac{333}{13}g_{00,1}^{5(3)}, \quad\quad g_{02,1}^{2(1)} =  -\frac{2295}{52}g_{00,1}^{5(3)},\\[.5em]
        &g_{02,1}^{3(1)} = \frac{513}{52}g_{00,1}^{5(3)}, \hspace{2.7em} g_{02,2}^{3(1)} = \frac{9315}{104}g_{00,1}^{5(3)},\\[.5em]
        &g_{02,1}^{4(1)}= 0, \hspace{6em} g_{02,2}^{41}=-27g_{00,1}^{5(3)}, \\[.5em]
         &g_{02,1}^{5(1)} = 54g_{00,1}^{5(3)}, \hspace{3.7em} g_{02,2}^{5(1)}= \frac{81}{2}g_{00,1}^{5(3)}.\\[.5em]
    \end{split}
\end{equation}
In $BB$ to $DD$ sector, we can summarize that 8 sum rules are described by single octet-octet free parameters, $g_{00,1}^{5(3)}$.}\\[.5em]
\textcolor{black}{
\indent $\bullet$  Mixed Octet-Decuplet (\scalebox{1}{$DBDD$})
    \begin{equation}
    g_{12,1}^{4(1)} = \frac{729}{116}g_{00,1}^{5(3)}, \quad\quad  g_{12,2}^{4(1)} = -\frac{2679}{58}g_{00,1}^{5(3)}, \quad\quad  g_{12,3}^{4(1)} = -\frac{243}{58}g_{00,1}^{5(3)}.
\end{equation}
The 3  couplings sum rules in sector $DB$ to $DD$ are described by single octet-octet free parameters, $g_{00,1}^{5(3)}$}\\[.1em]
\textcolor{black}{
\indent $\bullet$  Decuplet-Decuplet (\scalebox{1}{$DDDD$})
    \begin{alignat}{2}
        g_{22,2}^{1(1)}=&\frac{1}{5}g_{22,7}^{4(1)}+\frac{1}{9}g_{22,7}^{4(2)}-6g_{22,2}^{5(1)}-\frac{1}{2}g_{22,4}^{5(1)}-\frac{5}{6}g_{22,5}^{5(1)}-\frac{1}{7}g_{22,6}^{5(1)}-\frac{2}{3}g_{22,7}^{5(1)}-\frac{4}{3}g_{22,2}^{5(2)}-\frac{1}{5}g_{22,4}^{5(2)}-\frac{1}{5}g_{22,5}^{5(2)}-\frac{1}{4}g_{22,7}^{5(2)},\notag\\[.5em]
        g_{22,1}^{1(2)}=&-3g_{22,1}^{1(1)}-\frac{1}{5}g_{22,7}^{4(1)}-\frac{41}{3}g_{22,2}^{5(1)}-\frac{7}{3}g_{22,3}^{5(1)}-\frac{5}{4}g_{22,4}^{5(1)}-\frac{11}{3}g_{22,5}^{5(1)}+\frac{1}{3}g_{22,6}^{5(1)}-\frac{5}{2}g_{22,7}^{5(1)}-\frac{9}{2}g_{22,2}^{5(2)}-\frac{3}{4}g_{22,3}^{5(2)} \notag\\[.5em]
                    &-\frac{1}{2}g_{22,4}^{5(2)}-\frac{5}{4}g_{22,5}^{5(2)}+\frac{1}{7}g_{22,6}^{5(2)}-\frac{5}{6}g_{22,7}^{5(2)},\notag\\[.5em]
        g_{22,2}^{1(2)}=&-\frac{1}{16}g_{22,4}^{4(2)}+\frac{1}{8}g_{22,5}^{4(2)}-\frac{1}{8}g_{22,6}^{4(2)}-\frac{1}{8}g_{22,7}^{4(2)}-\frac{33}{16}g_{22,2}^{5(2)}+\frac{1}{8}g_{22,3}^{5(2)}-\frac{5}{16}g_{22,5}^{5(2)}-\frac{1}{16}g_{22,6}^{5(2)}-\frac{1}{8}g_{22,7}^{5(2)},\notag\\[.5em]
        g_{22,3}^{1(2)}=&-\frac{1}{16}g_{22,4}^{4(2)}+\frac{1}{8}g_{22,5}^{4(2)}-\frac{1}{8}g_{22,6}^{4(2)}-\frac{1}{8}g_{22,7}^{4(2)}-\frac{1}{16}g_{22,2}^{5(2)}+\frac{17}{8}g_{22,3}^{5(2)}-\frac{5}{16}g_{22,5}^{5(2)}-\frac{1}{16}g_{22,6}^{5(2)}-\frac{1}{8}g_{22,7}^{5(2)},\notag\\[.5em]
        g_{22,1}^{2(1)}=&\frac{25}{3}g_{00,1}^{2(3)}-\frac{57}{4}g_{00,1}^{3(2)}+\frac{33}{2}g_{00,1}^{3(3)}+\frac{43}{2}g_{00,1}^{5(3)}-\frac{2}{3}g_{22,1}^{1(1)}-\frac{1}{2}g_{22,7}^{4(1)}-\frac{1}{3}g_{22,7}^{4(2)}-\frac{29}{3}g_{22,2}^{5(1)}-\frac{13}{6}g_{22,3}^{5(1)}-\frac{1}{4}g_{22,4}^{5(1)}\notag\\[.5em]
                    &-\frac{13}{4}g_{22,5}^{5(1)}+\frac{1}{3}g_{22,6}^{5(1)}-2g_{22,7}^{5(1)}-\frac{1}{6}g_{22,2}^{5(2)}-\frac{1}{2}g_{22,3}^{5(2)}+\frac{1}{4}g_{22,4}^{5(2)}-\frac{1}{2}g_{22,5}^{5(2)}+\frac{1}{8}g_{22,6}^{5(2)},\notag\\[.5em]
        g_{22,2}^{2(1)}=&-g_{22,4}^{2(1)}-g_{22,6}^{3(1)}-\frac{1}{8}g_{22,6}^{4(1)}+\frac{1}{2}g_{22,7}^{4(1)}+\frac{1}{3}g_{22,4}^{4(2)}-\frac{1}{6}g_{22,5}^{4(2)}+\frac{1}{5}g_{22,6}^{4(2)}+\frac{3}{2}g_{22,2}^{5(1)}-\frac{1}{3}g_{22,3}^{5(1)}-3g_{22,4}^{5(1)}+\frac{4}{3}g_{22,5}^{5(1)}\notag\\[.5em]
                    &-\frac{1}{3}g_{22,6}^{5(1)}+\frac{1}{5}g_{22,7}^{5(1)}-\frac{19}{3}g_{22,2}^{5(2)}-\frac{3}{2}g_{22,5}^{5(2)}+\frac{1}{5}g_{22,6}^{5(2)}-\frac{4}{3}g_{22,7}^{5(2)},\notag\\[.5em]
        g_{22,3}^{2(1)}=&\frac{1}{3}g_{22,6}^{4(1)}+\frac{11}{3}g_{22,7}^{4(1)}-\frac{1}{6}g_{22,4}^{4(2)}-\frac{1}{3}g_{22,5}^{4(2)}+\frac{1}{2}g_{22,6}^{4(2)}+\frac{3}{2}g_{22,7}^{4(2)}-\frac{6}{5}g_{22,2}^{5(1)}-\frac{26}{5}g_{22,3}^{5(1)}-\frac{26}{3}g_{22,4}^{5(1)}+\frac{15}{4}g_{22,5}^{5(1)}\notag\\[.5em]
                    &-\frac{4}{3}g_{22,6}^{5(1)}+\frac{3}{4}g_{22,7}^{5(1)}-\frac{17}{2}g_{22,2}^{5(2)}-\frac{2}{3}g_{22,3}^{5(2)}-\frac{14}{5}g_{22,4}^{5(2)}-\frac{1}{4}g_{22,5}^{5(2)}-\frac{2}{3}g_{22,6}^{5(2)}-\frac{2}{3}g_{22,7}^{5(2)},\notag\\[.5em]
        g_{22,1}^{2(2)}=&-\frac{99}{4}g_{00,1}^{2(3)}+\frac{171}{4}g_{00,1}^{3(2)}-\frac{149}{3}g_{00,1}^{3(3)}-\frac{194}{3}g_{00,1}^{5(3)}+\frac{9}{4}g_{22,1}^{1(1)}-\frac{3}{4}g_{22,7}^{4(1)}-\frac{1}{6}g_{22,5}^{4(2)}+\frac{1}{6}g_{22,6}^{4(2)}+17g_{22,2}^{5(1)}\notag\\[.5em]
                    &+\frac{19}{9}g_{22,3}^{5(1)}+2g_{22,4}^{5(1)}+\frac{11}{3}g_{22,5}^{5(1)}-\frac{1}{2}g_{22,6}^{5(1)}+\frac{28}{9}g_{22,7}^{5(1)}-\frac{7}{2}g_{22,2}^{5(2)}-\frac{1}{4}g_{22,4}^{5(2)}-\frac{3}{4}g_{22,5}^{5(2)}-\frac{1}{4}g_{22,6}^{5(2)}-\frac{1}{2}g_{22,7}^{5(2)},\notag\\[.5em]
        g_{22,2}^{2(2)}=&-g_{22,4}^{2(2)}-g_{22,6}^{3(2)}-\frac{13}{16}g_{22,4}^{4(2)}+\frac{1}{2}g_{22,5}^{4(2)}-\frac{5}{8}g_{22,6}^{4(2)}+\frac{1}{2}g_{22,7}^{4(2)}+\frac{329}{16}g_{22,2}^{5(2)}-\frac{3}{8}g_{22,3}^{5(2)}-3g_{22,4}^{5(2)}+\frac{89}{16}g_{22,5}^{5(2)}\notag\\[.5em]
                    &-\frac{13}{16}g_{22,6}^{5(2)}+\frac{33}{8}g_{22,7}^{5(2)},\notag\\[.5em]
        g_{22,3}^{2(2)}=&\frac{1}{2}g_{22,4}^{4(2)}+g_{22,5}^{4(2)}-g_{22,6}^{4(2)}-g_{22,7}^{4(2)}+\frac{49}{2}g_{22,2}^{5(2)}-3g_{22,3}^{5(2)}+\frac{9}{2}g_{22,5}^{5(2)}+\frac{1}{2}g_{22,6}^{5(2)}+3g_{22,7}^{5(2)},\notag\\[.5em]
        g_{22,1}^{3(1)}=&-\frac{1}{2}g_{22,1}^{1(1)}-3g_{22,2}^{5(1)}+\frac{1}{2}g_{22,3}^{5(1)}+\frac{1}{2}g_{22,4}^{5(1)}-g_{22,5}^{5(1)}+\frac{1}{7}g_{22,6}^{5(1)}-\frac{3}{4}g_{22,7}^{5(1)}+\frac{1}{6}g_{22,2}^{5(2)}+\frac{1}{5}g_{22,4}^{5(2)},\notag\\[.5em]
        g_{22,2}^{3(1)}=&2g_{22,2}^{5(1)} \qquad g_{22,3}^{3(1)}=-2g_{22,3}^{5(1)},\notag\\[.5em]
        g_{22,4}^{3(1)}=&-g_{22,6}^{3(1)}-\frac{1}{7}g_{22,6}^{4(1)}-\frac{2}{3}g_{22,7}^{4(1)}-\frac{1}{2}g_{22,2}^{5(1)}+\frac{1}{5}g_{22,3}^{5(1)}+\frac{1}{4}g_{22,4}^{5(1)}-\frac{1}{2}g_{22,5}^{5(1)}-\frac{1}{2}g_{22,7}^{5(1)}-\frac{1}{8}g_{22,2}^{5(2)},\notag\\[.5em]
        g_{22,5}^{3(1)}=&\frac{6}{5}g_{22,7}^{4(1)}-\frac{1}{6}g_{22,5}^{4(2)}+\frac{1}{4}g_{22,6}^{4(2)}+\frac{1}{2}g_{22,7}^{4(2)}+\frac{36}{7}g_{22,2}^{5(1)}-\frac{2}{3}g_{22,3}^{5(1)}-3g_{22,4}^{5(1)}+\frac{11}{4}g_{22,5}^{5(1)}-\frac{1}{2}g_{22,6}^{5(1)}+\frac{5}{3}g_{22,7}^{5(1)}\notag\\[.5em]
                    &-\frac{24}{5}g_{22,2}^{5(2)}-\frac{1}{3}g_{22,3}^{5(2)}-g_{22,4}^{5(2)}-\frac{1}{2}g_{22,5}^{5(2)}-\frac{1}{5}g_{22,6}^{5(2)}-\frac{2}{3}g_{22,7}^{5(2)},\notag\\[.5em]
        g_{22,7}^{3(1)}=&-\frac{1}{7}g_{22,6}^{4(1)}-\frac{7}{4}g_{22,7}^{4(1)}+\frac{1}{6}g_{22,5}^{4(2)}-\frac{1}{4}g_{22,6}^{4(2)}-\frac{2}{3}g_{22,7}^{4(2)}-\frac{17}{3}g_{22,2}^{5(1)}+\frac{5}{6}g_{22,3}^{5(1)}+\frac{10}{3}g_{22,4}^{5(1)}-\frac{13}{4}g_{22,5}^{5(1)}+\frac{2}{3}g_{22,6}^{5(1)}\\[.5em]
                    &-\frac{9}{4}g_{22,7}^{5(1)}+\frac{19}{4}g_{22,2}^{5(2)}+\frac{1}{3}g_{22,3}^{5(2)}+\frac{9}{8}g_{22,4}^{5(2)}+\frac{1}{2}g_{22,5}^{5(2)}+\frac{1}{4}g_{22,6}^{5(2)}+\frac{1}{2}g_{22,7}^{5(2)},\notag\\[.5em]
        g_{22,1}^{3(2)}=&\frac{3}{2}g_{22,1}^{1(1)}-\frac{1}{3}g_{22,7}^{4(1)}+\frac{39}{4}g_{22,2}^{5(1)}+\frac{4}{3}g_{22,3}^{5(1)}+g_{22,4}^{5(1)}+\frac{7}{3}g_{22,5}^{5(1)}-\frac{1}{3}g_{22,6}^{5(1)}+\frac{7}{4}g_{22,7}^{5(1)}-\frac{1}{7}g_{22,2}^{5(2)}+\frac{3}{4}g_{22,3}^{5(2)}+\frac{1}{4}g_{22,4}^{5(2)}\notag\\[.5em]
                    &-\frac{1}{7}g_{22,6}^{5(2)}-\frac{1}{8}g_{22,7}^{5(2)},\notag\\[.5em]
        g_{22,2}^{3(2)}=&2g_{22,2}^{5(2)} \qquad g_{22,3}^{3(2)}=-2g_{22,3}^{5(2)},\notag\\[.5em]
        g_{22,4}^{3(2)}=&-g_{22,6}^{3(2)}-\frac{1}{16}g_{22,4}^{4(2)}-\frac{1}{8}g_{22,6}^{4(2)}-\frac{1}{2}g_{22,7}^{4(2)}-\frac{3}{16}g_{22,2}^{5(2)}+\frac{1}{8}g_{22,3}^{5(2)}-\frac{3}{16}g_{22,5}^{5(2)}-\frac{1}{16}g_{22,6}^{5(2)}-\frac{3}{8}g_{22,7}^{5(2)},\notag\\[.5em]
        g_{22,5}^{3(2)}=&-\frac{1}{16}g_{22,4}^{4(2)}+\frac{1}{2}g_{22,5}^{4(2)}-\frac{5}{8}g_{22,6}^{4(2)}-\frac{1}{2}g_{22,7}^{4(2)}+\frac{317}{16}g_{22,2}^{5(2)}+\frac{1}{8}g_{22,3}^{5(2)}+\frac{69}{16}g_{22,5}^{5(2)}-\frac{1}{16}g_{22,6}^{5(2)}+\frac{29}{8}g_{22,7}^{5(2)},\notag\\[.5em]
        g_{22,7}^{3(2)}=&-\frac{1}{2}g_{22,5}^{4(2)}+\frac{1}{2}g_{22,6}^{4(2)}-20g_{22,2}^{5(2)}-\frac{9}{2}g_{22,5}^{5(2)}-4g_{22,7}^{5(2)},\notag\\[.5em]
        g_{22,1}^{4(1)}=&\frac{2}{3}g_{22,1}^{1(1)}+\frac{1}{7}g_{22,7}^{4(2)}+\frac{27}{4}g_{22,2}^{5(1)}+\frac{1}{2}g_{22,3}^{5(1)}+2g_{22,5}^{5(1)}-\frac{1}{4}g_{22,6}^{5(1)}+\frac{3}{2}g_{22,7}^{5(1)}-\frac{3}{4}g_{22,2}^{5(2)}-\frac{1}{3}g_{22,4}^{5(2)},\notag\\[.5em]
        g_{22,2}^{4(1)}=&\frac{3}{2}g_{22,7}^{4(1)}+\frac{1}{5}g_{22,7}^{4(2)}+\frac{9}{4}g_{22,2}^{5(1)}-\frac{3}{4}g_{22,3}^{5(1)}-2g_{22,4}^{5(1)}+\frac{13}{6}g_{22,5}^{5(1)}+\frac{6}{5}g_{22,7}^{5(1)}+\frac{1}{2}g_{22,2}^{5(2)}-\frac{1}{3}g_{22,4}^{5(2)}+\frac{1}{3}g_{22,5}^{5(2)}\notag\\[.5em]
                    &-\frac{1}{8}g_{22,6}^{5(2)}+\frac{1}{4}g_{22,7}^{5(2)},\notag\\[.5em]
        g_{22,3}^{4(1)}=&g_{22,7}^{4(1)}-\frac{1}{6}g_{22,4}^{4(2)}+\frac{1}{3}g_{22,7}^{4(2)}+\frac{7}{2}g_{22,2}^{5(1)}+\frac{1}{2}g_{22,3}^{5(1)}-2g_{22,4}^{5(1)}+\frac{7}{3}g_{22,5}^{5(1)}-\frac{1}{5}g_{22,6}^{5(1)}+\frac{4}{3}g_{22,7}^{5(1)}+g_{22,2}^{5(2)}-\frac{1}{6}g_{22,3}^{5(2)}\notag\\[.5em]
                    &-\frac{2}{3}g_{22,4}^{5(2)}+\frac{2}{3}g_{22,5}^{5(2)}-\frac{1}{4}g_{22,6}^{5(2)}+\frac{1}{2}g_{22,7}^{5(2)},\notag\\[.5em]
        g_{22,4}^{4(1)}=&\frac{1}{6}g_{22,6}^{4(1)}+\frac{13}{6}g_{22,7}^{4(1)}-\frac{1}{3}g_{22,4}^{4(2)}+\frac{2}{3}g_{22,7}^{4(2)}+\frac{23}{4}g_{22,2}^{5(1)}-g_{22,3}^{5(1)}-4g_{22,4}^{5(1)}+\frac{11}{3}g_{22,5}^{5(1)}-\frac{4}{3}g_{22,6}^{5(1)}+\frac{8}{3}g_{22,7}^{5(1)}+2g_{22,2}^{5(2)}\notag\\[.5em]
                    &-\frac{1}{3}g_{22,3}^{5(2)}-\frac{4}{3}g_{22,4}^{5(2)}+\frac{5}{4}g_{22,5}^{5(2)}-\frac{1}{2}g_{22,6}^{5(2)}+\frac{4}{5}g_{22,7}^{5(2)},\notag\\[.5em]
        g_{22,5}^{4(1)}=&\frac{23}{18}g_{22,6}^{4(1)}+\frac{65}{18}g_{22,7}^{4(1)}-\frac{1}{3}g_{22,5}^{4(2)}+\frac{23}{54}g_{22,6}^{4(2)}+\frac{65}{54}g_{22,7}^{4(2)}-\frac{515}{18}g_{22,2}^{5(1)}-\frac{5}{3}g_{22,3}^{5(1)}-\frac{20}{3}g_{22,4}^{5(1)}-\frac{23}{9}g_{22,5}^{5(1)}-\frac{65}{54}g_{22,6}^{5(1)}\notag\\[.5em]
                    &-\frac{32}{9}g_{22,7}^{5(1)}-\frac{515}{54}g_{22,2}^{5(2)}-\frac{5}{9}g_{22,3}^{5(2)}-\frac{20}{9}g_{22,4}^{5(2)}-\frac{23}{27}g_{22,5}^{5(2)}-\frac{65}{162}g_{22,6}^{5(2)}-\frac{32}{27}g_{22,7}^{5(2)},\notag\\[.5em]
        g_{22,1}^{5(1)}=&\frac{1}{2}g_{22,1}^{1(1)}+\frac{15}{4}g_{22,2}^{5(1)}+\frac{1}{3}g_{22,3}^{5(1)}+\frac{7}{6}g_{22,5}^{5(1)}-\frac{1}{7}g_{22,6}^{5(1)}+\frac{3}{4}g_{22,7}^{5(1)}-\frac{1}{6}g_{22,2}^{5(2)}-\frac{1}{5}g_{22,4}^{5(2)},\notag\\[.5em]
        g_{22,1}^{5(2)}=&-\frac{5}{3}g_{22,1}^{1(1)}+\frac{1}{3}g_{22,7}^{4(1)}-\frac{39}{4}g_{22,2}^{5(1)}-\frac{4}{3}g_{22,3}^{5(1)}-g_{22,4}^{5(1)}-\frac{7}{3}g_{22,5}^{5(1)}+\frac{1}{3}g_{22,6}^{5(1)}-\frac{7}{4}g_{22,7}^{5(1)}+g_{22,2}^{5(2)}+\frac{1}{4}g_{22,4}^{5(2)}+\frac{1}{5}g_{22,5}^{5(2)}\notag\\[.5em]
                    &+\frac{1}{7}g_{22,6}^{5(2)}+\frac{1}{8}g_{22,7}^{5(2)},\notag\\[.5em]
        g_{22,1}^{4(2)}=&-\frac{795}{32}g_{00,1}^{2(3)}+\frac{171}{4}g_{00,1}^{3(2)}-\frac{795}{16}g_{00,1}^{3(3)}-\frac{1035}{16}g_{00,1}^{5(3)}-g_{22,1}^{2(2)},\notag\\[.5em]
        g_{22,2}^{4(2)}=&\frac{1}{4}g_{22,4}^{4(2)}+g_{22,7}^{4(2)}+\frac{3}{4}g_{22,2}^{5(2)}-\frac{1}{2}g_{22,3}^{5(2)}-g_{22,4}^{5(2)}+\frac{5}{4}g_{22,5}^{5(2)}+\frac{1}{4}g_{22,6}^{5(2)}+\frac{1}{2}g_{22,7}^{5(2)},\notag\\[.5em]
        g_{22,3}^{4(2)}=&\frac{1}{2}g_{22,4}^{4(2)}+\frac{1}{2}g_{22,2}^{5(2)}+g_{22,3}^{5(2)}+\frac{1}{2}g_{22,5}^{5(2)}+\frac{1}{2}g_{22,6}^{5(2)}.\notag
    \end{alignat}}
\textcolor{black}{
\noindent For decuplet-decuplet sector, we can establish 32 sum rules in terms of these free parameters, $g_{00,1}^{2(3)}$,  $g_{00,1}^{3(2)}$ , $g_{00,1}^{3(3)}$, $g_{00,1}^{5(3)}$, $g_{22,1}^{1(1)}$, $g_{22,4}^{2(1)}$, $g_{22,6}^{3(1)}$, $g_{22,4}^{2(2)}$, $g_{22,6}^{3(2)}$, $g_{22,6}^{4(1)}$, $g_{22,7}^{4(1)}$, $g_{22,4}^{4(2)}$, $g_{22,5}^{4(2)}$, $g_{22,6}^{4(2)}$, $g_{22,7}^{4(2)}$, $g_{22,2}^{5(1)}$, $g_{22,3}^{5(1)}$, $g_{22,4}^{5(1)}$, $g_{22,5}^{5(1)}$, $g_{22,6}^{5(1)}$, $g_{22,7}^{5(1)}$, $g_{22,2}^{5(2)}$, $g_{22,3}^{5(2)}$, $g_{22,4}^{5(2)}$, $g_{22,5}^{5(2)}$, $g_{22,6}^{5(2)}$, $g_{22,7}^{5(2)}$.}\\
\newpage
\subsection{The partial-wave LECs at leading order}\label{A-5-PartialWave}
In the following sections, we present the explicit form of the four-point contact potential, denoted as $V_{ct}$. This potential involves the general two-body spin operator and accounts for the non-vanishing transitions between partial-waves $^{2S+1}L_j$ at leading order (LO). Following the approach outlined in \cite{Skibi_ski_2011,haiden2017scattdecupletbary}, we can express it as follows:\\
\\
\indent $\bullet$ Octet-Octet (\scalebox{1}{$BBBB$})
    \begin{equation}
        \begin{split}
            &V_{ct} = a_1\cdot\mathbb{1}+a_2\vec{\sigma}_1\cdot\vec{\sigma}_2,
            \\
            &V_{^1S_0}=\braket{^1S_0|V_{ct}|^1S_0} = a_1-3a_2,
            \\
            &V_{^3S_1}=\braket{^3S_1|V_{ct}|^3S_1} = a_1+a_2,
        \end{split}
    \end{equation}
\textcolor{black}{where $a_1 = \mathcal{F}_1^{(BBBB)} c_{S}^{(1)} + \mathcal{F}_2^{(BBBB)} c_{S}^{(2)}+\mathcal{F}_3^{(BBBB)} c_{S}^{(3)}$ and $a_2 = \mathcal{F}_1^{(BBBB)} c_{T}^{(1)} + \mathcal{F}_2^{(BBBB)} c_{T}^{(2)}+\mathcal{F}_3^{(BBBB)} c_{T}^{(3)}$. The $\mathcal{F}_i^{(BBBB)}$ are the flavor factors of the potential in $BBBB$ sector.}
\\
\\
\indent $\bullet$ Mixed Octet-Decuplet (\scalebox{1}{$BBDB$})
    \begin{equation}
        \begin{split}
            &V_{ct} =a_1\vec{S}_{1}^{\dagger}\cdot\vec{\sigma}_{2},\\
            &V_{^3S_1}=\braket{^3S_1|V_{ct}|^3S_1} = -2\sqrt{\frac{2}{3}}a_1,
        \end{split}
    \end{equation}
\textcolor{black}{where $a_1 = \mathcal{F}_1^{(BBDB)} c_{1,BBDB}^{(1)} + \mathcal{F}_2^{(BBDB)} c_{2,BBDB}^{(2)}$.  The $\mathcal{F}_i^{(BBDB)}$ are the flavor factors of the potential in $BBDB$ sector.}
\\
\\
\indent $\bullet$ Mixed Octet-Decuplet (\scalebox{1}{$DBDB$})
    \begin{equation}
        \begin{split}
            &V_{ct} =a_1\cdot\mathbb{1}+a_2\vec{\Sigma}_1\cdot\vec{\sigma}_{2},\\
            &V_{^3S_1}=\braket{^3S_1|V_{ct}|^3S_1} = a_1-5a_2,\\
            &V_{^5S_2}=\braket{^5S_2|V_{ct}|^5S_2} = a_1+3a_2,
        \end{split}
    \end{equation}
\textcolor{black}{where $a_1 = \mathcal{F}_1^{(DBDB)} c_{1,DBDB}^{(1)} + \mathcal{F}_2^{(DBDB)} c_{1,DBDB}^{(2)}+\mathcal{F}_3^{(DBDB)} c_{1,DBDB}^{(3)}+\mathcal{F}_4^{(DBDB)} c_{1,DBDB}^{(4)}$ and $a_2 = \mathcal{F}_1^{(DBDB)} c_{2,DBDB}^{(1)} + \mathcal{F}_2^{(DBDB)} c_{2,DBDB}^{(2)}+\mathcal{F}_3^{(DBDB)} c_{2,DBDB}^{(3)}+\mathcal{F}_4^{(DBDB)} c_{2,DBDB}^{(4)}$.  The $\mathcal{F}_i^{(DBDB)}$ are the flavor factors of the potential in $DBDB$ sector.}
\\
\\
\indent $\bullet$ Mixed Octet-Decuplet (\scalebox{1}{$BBDD$})
    \begin{equation}
        \begin{split}
            &V_{ct} =a_1\vec{S}_{1}^{\dagger}\cdot\vec{S}_{2}^{\dagger}+a_2S_{1}^{ij\dagger}S_{2}^{ij\dagger},\\
            &V_{^1S_0}=\braket{^1S_0|V_{ct}|^1S_0} = -\sqrt{2}a_1-\frac{5}{3}\sqrt{2}a_2,\\
            &V_{^3S_1}=\braket{^3S_1|V_{ct}|^3S_1} = -\frac{1}{3}\sqrt{3}(a_1+a_2),
        \end{split}
    \end{equation}
\textcolor{black}{where $a_1 = \mathcal{F}_1^{(BBDD)} c_{2,BBDD}^{(1)} $ and $a_2 = \mathcal{F}_1^{(BBDD)} c_{3,BBDD}^{(1)} $. The $\mathcal{F}_i^{(BBDD)}$ are the flavor factors of the potential in $BBDD$ sector.}
\\
\\
\indent $\bullet$ Mixed Octet-Decuplet (\scalebox{1}{$DBDD$})
    \begin{equation}
        \begin{split}
             &V_{ct} =a_1\vec{\Sigma}_{1}\cdot\vec{S}_{2}^{\dagger}+a_2\Sigma_{1}^{ij}\Sigma_{2}^{ij}\\
             &V_{^3S_1}=\braket{^3S_1|V_{ct}|^3S_1} = 2\sqrt{\frac{5}{3}}a_1+\sqrt{10}a_2\\
            &V_{^5S_2}=\braket{^5S_2|V_{ct}|^5S_2} = 2\sqrt{3}a_1-\sqrt{2}a_2
        \end{split}
    \end{equation}
\textcolor{black}{where $a_1 = \mathcal{F}_1^{(DBDD)} c_{2,DBDD}^{(1)} $ and $a_2 = \mathcal{F}_1^{(DBDD)} c_{3,DBDD}^{(1)} $. The $\mathcal{F}_i^{(DBDD)}$ are the flavor factors of the potential in $DBDD$ sector.}
\\
\\
\indent $\bullet$ Decuplet-Decuplet (\scalebox{1}{$DDDD$})
    \begin{equation}
        \begin{split}
            &V_{ct} =a_1\cdot\mathbb{1}+a_2\vec{\Sigma}_{1}\cdot\vec{\Sigma}_{2}+a_3\Sigma_{1}^{ij}\Sigma_{2}^{ij}+a_4\Sigma_{1}^{ijk}\Sigma_{2}^{ijk},\\
            &V_{^1S_0}=\braket{^1S_0|V_{ct}|^1S_0} = a_1-15a_2+\frac{15}{2}a_3-\frac{350}{3}a_4,\\
            &V_{^3S_1}=\braket{^3S_1|V_{ct}|^3S_1} = a_1-11a_2+\frac{3}{2}a_3+70a_4,\\
            &V_{^5S_2}=\braket{^5S_2|V_{ct}|^5S_2} = a_1-3a_2-\frac{9}{2}a_3-\frac{70}{3}a_4,\\
            &V_{^7S_3}=\braket{^7S_3|V_{ct}|^7S_3} = a_1+9a_2+\frac{3}{2}a_3+\frac{10}{3}a_4,
        \end{split}
    \end{equation}
\textcolor{black}{where $a_1 = \mathcal{F}_1^{(DDDD)} c_{1,DDDD}^{(1)} +\mathcal{F}_2^{(DDDD)} c_{1,DDDD}^{(2)} $,  $a_2 = \mathcal{F}_1^{(DDDD)} c_{2,DDDD}^{(1)} +\mathcal{F}_2^{(DDDD)} c_{2,DDDD}^{(2)} $, $a_3 = \mathcal{F}_1^{(DDDD)} c_{3,DDDD}^{(1)} +\mathcal{F}_2^{(DDDD)} c_{3,DDDD}^{(2)} $ and $a_4 = \mathcal{F}_1^{(DDDD)} c_{4,DDDD}^{(1)} +\mathcal{F}_2^{(DDDD)} c_{4,DDDD}^{(2)} $.  The $\mathcal{F}_i^{(DDDD)}$ are the flavor factors of the potential in $DDDD$ sector.}
\\
\\
To obtain the various SU(3) relations for our approach, we have projected the contact terms from the chiral Lagrangian onto the partial-wave sector at LO. The following relations are obtained \cite{haiden2017scattdecupletbary},\\[1em]
\indent $\bullet$  Octet-Octet (\scalebox{1}{$BBBB$})
\\
\small{\begin{equation}
        \begin{split}
            &\mathcal{C}_{00}^{1}=\frac{2}{3}\bigl(c_{S,BBBB}^{(1)}-3c_{T,BBBB}^{(1)}-8c_{S,BBBB}^{(2)}+24c_{T,BBBB}^{(2)}-3c_{S,BBBB}^{(3)}+9c_{T,BBBB}^{(3)}\bigr),\\[1em]
            &\mathcal{C}_{00}^{8s}=\frac{4}{3}c_{S,BBBB}^{(1)}-4c_{T,BBBB}^{(1)}-\frac{5}{3}c_{S,BBBB}^{(2)}+5c_{T,BBBB}^{(2)}-2c_{S,BBBB}^{(3)}+6c_{T,BBBB}^{(3)},\\[1em]
            &\mathcal{C}_{00}^{8a}=3c_{S,BBBB}^{(2)}+3c_{T,BBBB}^{(2)}-2\bigl(c_{S,BBBB}^{(3)}+c_{T,BBBB}^{(3)}\bigr),\\[1em]
            &\mathcal{C}_{00}^{10}=2\bigl(c_{S,BBBB}^{(1)}+c_{T,BBBB}^{(1)}-c_{S,BBBB}^{(3)}-c_{T,BBBB}^{(3)}\bigr),\\[1em]
            &\mathcal{C}_{00}^{\overline{10}}=-2\bigl(c_{S,BBBB}^{(1)}+c_{T,BBBB}^{(1)}+c_{S,BBBB}^{(3)}+c_{T,BBBB}^{(3)}\bigr),\\[1em]
            &\mathcal{C}_{00}^{27}=-2\bigl(c_{S,BBBB}^{(1)}-3c_{T,BBBB}^{(1)}+c_{S,BBBB}^{(3)}-3c_{T,BBBB}^{(3)}\bigr).
        \end{split}
\end{equation}}
\\[1em]
\indent $\bullet$  Mixed Octet-Decuplet (\scalebox{1}{$DBDB$})\\
\small{\begin{alignat}{2}
    \mathcal{C}_{11}^{35,3S1}=&-c_{1,DBDB}^{(1)}+5c_{2,DBDB}^{(1)}-c_{1,DBDB}^{(3)}+5c_{2,DBDB}^{(3)}\notag,\\[1em]
            \mathcal{C}_{11}^{27,3S1}=&\frac{1}{3}\bigl(-3c_{1,DBDB}^{(1)}+15c_{2,DBDB}^{(1)}+c_{1,DBDB}^{(3)}-5c_{2,DBDB}^{(3)}\bigr)\notag,\\[1em]
            \mathcal{C}_{11}^{10,3S1}=&\frac{1}{3}\bigl(-3c_{1,DBDB}^{(1)}+15c_{2,DBDB}^{(1)}-4c_{1,DBDB}^{(2)}+20c_{2,DBDB}^{(2)}-c_{1,DBDB}^{(3)}+5c_{2,DBDB}^{(3)}-4c_{1,DBDB}^{(4)}+20c_{2,DBDB}^{(4)}\bigr)\notag,\\[1em]
            \mathcal{C}_{11}^{8,3S1}=&\frac{1}{6}\bigl(-6c_{1,DBDB}^{(1)}+30c_{2,DBDB}^{(1)}-10c_{1,DBDB}^{(2)}+50c_{2,DBDB}^{(2)}+2c_{1,DBDB}^{(3)}-10c_{2,DBDB}^{(3)}+5c_{1,DBDB}^{(4)}-25c_{2,DBDB}^{(4)}\bigr)\notag,\\[1em]
            \mathcal{C}_{11}^{35,5S2}=&-c_{1,DBDB}^{(1)}-3c_{2,DBDB}^{(1)}-c_{1,DBDB}^{(3)}-3c_{2,DBDB}^{(3)}\notag,\\[1em]
            \mathcal{C}_{11}^{27,5S2}=&-c_{1,DBDB}^{(1)}-3c_{2,DBDB}^{(1)}+\frac{1}{3}c_{1,DBDB}^{(3)}+c_{2,DBDB}^{(3)}\notag,\\[1em]
            \mathcal{C}_{11}^{10,5S2}=&\frac{1}{3}\bigl(-3c_{1,DBDB}^{(1)}-9c_{2,DBDB}^{(1)}-4c_{1,DBDB}^{(2)}-12c_{2,DBDB}^{(2)}-c_{1,DBDB}^{(3)}-3c_{2,DBDB}^{(3)}-4c_{1,DBDB}^{(4)}-12c_{2,DBDB}^{(4)}\bigr)\notag,\\[1em]
            \mathcal{C}_{11}^{8,5S2}=&-c_{1,DBDB}^{(1)}-3c_{2,DBDB}^{(1)}-\frac{5}{3}c_{1,DBDB}^{(2)}-5c_{2,DBDB}^{(2)}+\frac{1}{3}c_{1,DBDB}^{(3)}+c_{2,DBDB}^{(3)}+\frac{5}{6}c_{1,DBDB}^{(4)}+\frac{5}{2}c_{2,DBDB}^{(4)}.
\end{alignat}}
\\
\indent $\bullet$  Decuplet-Decuplet (\scalebox{1}{$DDDD$})
\\
\small{\begin{alignat}{2}
    &\mathcal{C}_{22}^{\overline{10},3S1}=\frac{1}{3}\bigl(-6c_{1,DDDD}^{(1)}+66c_{2,DDDD}^{(1)}-9c_{3,DDDD}^{(1)}-420c_{4,DDDD}^{(1)}+2c_{1,DDDD}^{(2)}-22c_{2,DDDD}^{(2)}+3c_{3,DDDD}^{(2)}+140c_{4,DDDD}^{(2)}\bigr)\notag,\\[1em]
            &\mathcal{C}_{22}^{\overline{10},7S3}=-2c_{1,DDDD}^{(1)}-18c_{2,DDDD}^{(1)}-3c_{3,DDDD}^{(1)}-\frac{20}{3}c_{4,DDDD}^{(1)}+\frac{2}{3}c_{1,DDDD}^{(2)}+6c_{2,DDDD}^{(2)}+c_{3,DDDD}^{(2)}+\frac{20}{9}c_{4,DDDD}^{(2)}\notag,\\[1em]
            &\mathcal{C}_{22}^{27,1S0}=\frac{1}{27}\bigl(-54c_{1,DDDD}^{(1)}+810c_{2,DDDD}^{(1)}-405c_{3,DDDD}^{(1)}+6300c_{4,DDDD}^{(1)}+6c_{1,DDDD}^{(2)}-90c_{2,DDDD}^{(2)}+45c_{3,DDDD}^{(2)}-700c_{4,DDDD}^{(2)}\bigr)\notag,\\[1em]
            &\mathcal{C}_{22}^{27,5S2}=-2c_{1,DDDD}^{(1)}+6c_{2,DDDD}^{(1)}+9c_{3,DDDD}^{(1)}+\frac{140}{3}c_{4,DDDD}^{(1)}+\frac{2}{9}c_{1,DDDD}^{(2)}-\frac{2}{3}c_{2,DDDD}^{(2)}-c_{3,DDDD}^{(2)}-\frac{140}{27}c_{4,DDDD}^{(2)}\notag,\\[1em]
            &\mathcal{C}_{22}^{35,3S1}=\frac{1}{3}\bigl(-6c_{1,DDDD}^{(1)}+66c_{2,DDDD}^{(1)}-9c_{3,DDDD}^{(1)}-420c_{4,DDDD}^{(1)}-2c_{1,DDDD}^{(2)}+22c_{2,DDDD}^{(2)}-3c_{3,DDDD}^{(2)}-140c_{4,DDDD}^{(2)}\bigr)\notag,\\[1em]
            &\mathcal{C}_{22}^{35,7S3}=-2c_{1,DDDD}^{(1)}-18c_{2,DDDD}^{(1)}-3c_{3,DDDD}^{(1)}-\frac{20}{3}c_{4,DDDD}^{(1)}-\frac{2}{3}c_{1,DDDD}^{(2)}-6c_{2,DDDD}^{(2)}-c_{3,DDDD}^{(2)}-\frac{20}{9}c_{4,DDDD}^{(2)}\notag,\\[1em]
            &\mathcal{C}_{22}^{28,1S0}=\frac{1}{3}\bigl(-6c_{1,DDDD}^{(1)}+90c_{2,DDDD}^{(1)}-45c_{3,DDDD}^{(1)}+700c_{4,DDDD}^{(1)}-6c_{1,DDDD}^{(2)}+90c_{2,DDDD}^{(2)}-45c_{3,DDDD}^{(2)}+700c_{4,DDDD}^{(2)}\bigr)\notag,\\[1em]
            &\mathcal{C}_{22}^{28,5S2}=-2c_{1,DDDD}^{(1)}+6c_{2,DDDD}^{(1)}+9c_{3,DDDD}^{(1)}+\frac{140}{3}c_{4,DDDD}^{(1)}-2c_{1,DDDD}^{(2)}+6c_{2,DDDD}^{(2)}+9c_{3,DDDD}^{(2)}+\frac{140}{3}c_{4,DDDD}^{(2)}.
\end{alignat}}
\bibliography{references.bib}

\end{document}